\documentclass[submission, Phys]{SciPost}
\usepackage[T1]{fontenc}
\usepackage{booktabs}
\usepackage{graphicx}
\usepackage{xspace}
\usepackage{multirow}
\usepackage{amssymb,amsmath}
\usepackage{mathtools}
\usepackage{cleveref}
\usepackage{xstring}
\usepackage[normalem]{ulem}
\usepackage{algorithm}
\usepackage{algpseudocode}
\usepackage{ulem}
\usepackage{mathrsfs}
\usepackage{calligra}
\hypersetup{colorlinks, linkcolor={red!50!black}, citecolor={blue!50!black},
  urlcolor={blue!80!black}}

\usepackage[many]{tcolorbox}

\definecolor{main}{HTML}{5989cf}    
\definecolor{sub}{HTML}{cde4ff}     

\tcbset{
    sharp corners,
    colback = white,
    before skip = 0.2cm,    
    after skip = 0.5cm      
}                           

\newtcolorbox[auto counter]{boxH}[2]{
    label = {#1},
    title = {#2},
    fonttitle=\bfseries, 
    coltitle=black,
    colback = sub,
    colframe = main,
    boxrule = 0pt,
    leftrule = 6pt 
}

\newtcolorbox{mybox}[1]{
  colback=blue!5!white,    
  colframe=blue!75!black,  
  fonttitle=\bfseries,     
  title=#1                 
}

\newcommand{\likelihood}[1]{\ensuremath{\mathcal{L}_{\mathrm{#1}}}\,}
\newcommand{\method}{\ensuremath{\mathrm{MRL}}\,}
\newcommand{\fracdiff}{\ensuremath{\Delta(\likelihood{\text{c}})}\,}
\newcommand{\estimatedfrac}[2]{\ensuremath{f(\likelihood{#1}|#2)}\,}
\newcommand{\trash}[1]{}

\begin{document}




\begin{center}

\begin{Large}
\textbf{\textsc{Defining a Minimum Resolution for Unbinned Analyses}}
\end{Large}

\vspace{1cm}

{\sc
Manuel~Szewc$^{a}$%
\footnote{{\tt \href{mailto:mszewc@unsam.edu.ar}{mszewc@unsam.edu.ar}}}%
}

\vspace*{.7cm}

{\sl
$^a$International Center for Advanced Studies (ICAS) and ICIFI-CONICET, UNSAM, \\
25 de Mayo y Francia, CP1650, San Mart\'{\i}n, Buenos Aires, Argentina

\vspace*{0.1cm}

\vspace*{0.1cm}

}

\end{center}

\vspace{0.1cm}

\begin{abstract}
\noindent 
Collider analyses combine rigorous statistical techniques with state-of-the-art Machine Learning models. However, when the latter are used directly to estimate the likelihood function of the background, hard to quantify systematic effects may bias the estimation of the relevant signal parameters. To address this problem, we present the Minimum Resolution Likelihood (\method) method, which defines a Fiducial Signal Region that aims to turn systematic effects into statistical uncertainties. We show with examples that the resulting signal strength estimation has a reduced bias, with the effect depending on the sample size and type of signal considered. We consider both toy examples and a realistic application based on the \texttt{HI-SIGMA} technique applied to di-Higgs searches.
\vskip 1cm
~
\end{abstract}

\tableofcontents
\clearpage

\section{Introduction}

High-energy physics experiments, and in particular collider analyses, have long benefitted from rigorous statistical techniques~\cite{Behnke:2013pga} that take advantage of the quality of the measurements to extract information about the fundamental parameters of our theories. A key part of any analysis is the proper definition of the background processes, from which we want to extract a signal (which in high-energy physics is not done in terms of outlier events but as estimations of parameters of interest that capture over or under-densities of the data with respect to the backgrounds). If available, state-of-the art generators with appropriate uncertainties are preferred, since they define the background unambigously. However, for some cases this is not possible, due to a combination of compute cost and simulator accuracy, and data-driven backgrounds are estimated. This is not a ``niche'' problem, as data-driven backgrounds are considered for ``bread and butter'' analyses at the LHC, e.g. to estimate multi-jet backgrounds~\cite{ATLAS:2018rnh,CMS:2024tdk,CMS:2026znu}.

Traditional data-driven background estimation techniques are confined to few dimensions in feature space, which, although already powerful, restrict the power of the analysis. With the advent of Machine Learning techniques the situation has changed, however, and data-driven backgrounds have been computed in large dimensions, both for specific searches~\cite{ATLAS:2018rnh,CMS:2024tdk,CMS:2026znu} and in the context of Anomaly Detection~\cite{Hallin:2021wme, Raine:2022hht, Golling:2022nkl, Hallin:2022eoq, Sengupta:2023xqy, Das:2023bcj}. Although this increases significantly the power of the analyses, so does the risk of bias and uncertainty underestimation, which have motivated several proposed solutions~\cite{ATLAS:2020ocz, Dauncey:2014xga,Haussmann:2026gbi}. However, this remains an open problem and it is hard to estimate the biases and uncertainties associated with the use of a data-driven model defined over many dimensions in an unbinned analysis.

Moreover, the issue is not confined to data-driven models. Even when trained on simulations, the use of unbinned multivariate models (which can originate from either classifiers or density estimation algorithms) to estimate parameters of interest in what is known as Simulation Based Inference (SBI)~\cite{Brehmer:2018kdj, Brehmer:2018eca, Cranmer_sbi_review,Ghosh_jrjc_proc, GomezAmbrosio:2022mpm, Bahl:2021dnc, Barrue:2023ysk, Schofbeck:2024zjo, Chai:2024zyl, nsbi_dihiggs, Benato:2025rgo} can lead to uncontrollable biases when the analysis is powerful enough that it is sensitive to the bias of the unbinned model with respect to the training distribution (or even to the bias of the original simulations themselves with respect to the true background model).

One possible failure mode originating from this model bias is the overestimation of the signal-to-background likelihood ratio in the regions where the signal is present, with a consequent overestimation of the number of signal events in the sample.\footnote{The converse case, where we overestimate the background and underestimate the signal, is of course worrisome since we would miss signals that are there, but is ``nicer'' in the sense that it requires daring, not conservative, solutions. Moreover, we are usually more concerned about overestimating signals, not underestimating them.} This may happen for example when the background model underestimates the probability density in the region where signal and background overlap due to the unbinned, high-dimensional background model wrongly ``spilling'' probability density into the tails due to training artifacts or architectural biases.

In this work, we propose a very simple solution to signal overestimation due to background mismodelling, which is not meant as a ``be-all, end-all patch'' but simply as a conservative band-aid that trades systematic uncertainties for statistical uncertainties. We claim that for some analyses this trade-off is worth it, since it allows the use of more powerful techniques. The method, which we call the Minimum Resolution Likelihood (\method) method, can be summed up very easily: for a given background model that is meant to be applied in a Signal Region, we consider an additional Calibration Region where there is only the background and which we use to define a minimum resolution in a smartly chosen feature below which we do not trust the model. Thus, any events in Signal Region that fall below that resolution are discarded, defining a Fiducial Signal Region where the statistical analysis is performed.\footnote{Similar issues, and the introduction of an analogue resolution concept, can be found in the generative model literature when discussing amplification~\cite{Diefenbacher:2026kki}.} In this way, we effectively turn systematic uncertainties over the background model itself into statistical uncertainties due to the more limited sample size of the data sample.  A more succinct list of requirement and assumptions for the \method method to be useful can be found in Box \ref{box}.

\begin{boxH}{box}{Box \thetcbcounter: Requirements and assumptions}

Requirements for applicability:
\begin{itemize}
    \item A background model that allows both to evaluate the likelihood and to generate samples.
    \item A Calibration Region with background-only distributions (or at least closer to the true background than the model).
\end{itemize}

Assumptions for usefulness:
\begin{itemize}
    \item The difference between the background model and the true background results in the underestimation of background in the Signal Region for the no-signal case. In other words, for a given signal model we assume the background model induces an overestimation of the signal-to-background likelihood ratio in regions where the signal is present.
    \item The Fiducial Signal Region is such that some fraction of the signal distribution is retained. Thus, there will be a minimum number of signal events for which we are still sensitive and the resulting drop in power due to the discarded events (trading uncontrollable systematic error for statistical error) is offset by the use of the unbinned likelihood-based techniques in the first place.
\end{itemize}

\end{boxH}

The work is structured as follows. In \cref{sec:prescription} we introduce the \method method, including how we obtain the minimum likelihood itself and how we use it to define a Fiducial Signal Region. In \cref{sec:toys} we apply the \method method to two toy examples to showcase its capabilities and possible drawbacks. In \cref{sec:hi-sigma} we show \method can improve a realistic di-Higgs study based on the \texttt{HI-SIGMA} technique introduced in Ref.~\cite{Amram:2025vqw}. We conclude and discuss future directions in \cref{sec:outlook}. Additional interpretations of the minimum likelihood can be found in the two appendices, \ref{app:eft} and \ref{app:derivation}.

\section{Prescription}
\label{sec:prescription}

In this section, we introduce the problem of model misspecification more explicitly (\cref{subsec:problem_statement}) and detail how we can obtain a minimum likelihood (\cref{subsec:how_to}) that we can use to define a Fiducial Signal Region and accommodate floating nuisance parameters (\cref{subsec:fiducial_signal_region}). 

\subsection{The problem of background model misspecification}
\label{subsec:problem_statement}

As a starting point, we define the usual extended likelihood considered in most collider analyses~\cite{Cranmer:2014lly} for $N$ total observed events in the Signal Region, with observed features $\{x_n\}_{n=1}^{N}$:
\begin{align}
    \label{eq:extended_likelihood}
    \mathcal{L}(\mathcal{D},\mu_s,\mu_b,\theta_s,\theta_b)&=\mathcal{P}(N|\mu_s \lambda_s+\mu_b \lambda_b)\nonumber\\
    &\prod_{n=1}^{N}\left(\frac{\mu_s \lambda_s}{\mu_s \lambda_s+\mu_b \lambda_b}p_s(x_n|\theta_s)+\frac{\mu_b \lambda_b}{\mu_s \lambda_s+\mu_b \lambda_b}p_b(x_n|\theta_b)\right)g(\theta_{s},\mu_{b},\theta_{b})\,,
\end{align}
where the parameter of interest is $\mu_s$, $\lambda_s$ and $\lambda_b$ are the expected signal and background rates and all others are nuisance parameters, for which we add a constraint term $g$ that stands for any additional measurements or theoretical systematic uncertainties that constrain said nuisance parameters. We then define the usual test statistic derived from the profiled log-likelihood,
\begin{align}
    t(\mu_s) = - 2 \ln \frac{\mathcal{L}(\mathcal{D},\mu_s,\hat{\hat{\mu}}_b,\hat{\hat{\theta}}_{s},\hat{\hat{\theta}}_{b})}{\mathcal{L}(\mathcal{D},\hat{\mu}_s,\hat{\mu}_b,\hat{\theta}_{s},\hat{\theta}_{b})}\equiv - 2 \ln \frac{\mathcal{L}(\mathcal{D},\mu_s,\hat{\hat{\nu}})}{\mathcal{L}(\mathcal{D},\hat{\mu}_s,\hat{\nu})}\,,
\end{align}
where we have grouped all nuisance parameters under $\nu$, and use the conventional notation where $\hat{\mu}_{s}$ and $\hat{\nu}$ are the maximum likelihood estimates of $\mu_{s}$ and $\nu$ and $\hat{\hat{\nu}}$ is the maximum likelihood estimate of $\nu$ for a fixed value of $\mu_s$. The test statistic can be used to build confidence intervals of size $\alpha \%$, obtaining the interval $(\mu^{\min}_{s}(\alpha,\mathcal{D}),\mu^{\max}_{s}(\alpha,\mathcal{D}))$\footnote{We are assuming monotonicity and certain smoothness of the likelihood function with respect to $\mu_{s}$.} such that
\begin{align}
P(\mu^{\min}_{s}(\alpha,\mathcal{D})\leq \mu_s < \mu^{\max}_{s}(\alpha,\mathcal{D})) = 1-\alpha\,,
\end{align}
where $\mu^{\min}_{s}(\alpha,\mathcal{D})$ and $\mu^{\max}_{s}(\alpha,\mathcal{D})$ are random variables and $\mu_{s}$ an unknown constant. This, of course, assumes that we have a correct model. If our likelihood is incorrect, the test loses statistical power and worse, we will obtain biased estimates of $\hat{\mu}_{s}$  and confidence intervals $(\mu^{\min}_{s},\mu^{\max}_{s})$ without proper coverage. The main issue is that in many, if not most or even all cases, the model is wrong. 

It may happen that the wrongness of the model is subleading, or accountable for by increasing the associated error via systematic uncertainties. In some cases, however, the wrongness of the model noticeably impacts the estimation of $\hat{\mu}_{s}$, and it is hard to assign a meaningful uncertainty that encapsulates model misspecification. We emphasize that this may not be because of imperfect model training, but simply a reflection of the power of the analysis and how under control other sources of uncertainty are. In that sense, the analysis may be a victim of its own success.

We are particularly interested in the case where the bias in the background model leads to the overestimation of the signal-to-background likelihood ratio in the regions where the signal is present, with a consequent overestimation of the number of signal events in the sample. This, for rare enough signals, leads to false discoveries due to underestimated background rates and overestimated signal rates. We detail the \method aimed to address this in the next section.

\subsection{The minimum likelihood determination}
\label{subsec:how_to}

We consider the background model $p(x|\theta_b)\neq p_{\text{true b}}(x)$ used in the statistical inference, and which we assume can be used to generate samples and to evaluate the likelihood of a given data point under said model. We can characterize such a model, even in the multidimensional case, by the hyper-volumes parameterized by the background-only likelihood $\mathcal{L}\equiv p(x|\theta_b)$\footnote{Not to be confused with the extended likelihood introduced in \cref{eq:extended_likelihood}.}, in a procedure similar in spirit to the philosophy behind Nested Sampling~\cite{Buchner_2023}.\footnote{As in Nested Sampling, an analogy of \method in terms of the energy density can be helpful. We introduce such an analogy in App. \ref{app:eft}.} In particular, we can compute the expected fraction of events above a certain likelihood value, $\estimatedfrac{\text{c}}{\text{model}}$
\begin{align}
    \estimatedfrac{c}{\text{model}} = \int_{\likelihood{c}}^{\likelihood{\max}} d\mathcal{\likelihood{}}\,p(\likelihood{}|\text{model}) = \mathbb{E}_{x\sim p(x|\text{model})}[\Theta(p(x|\theta_{b})-\likelihood{c})]\,,
\end{align}
where again $\likelihood{}$ is always computed with respect to the background model to be inspected. If we have a Calibration Region where to evaluate the model, and which we can assume shares the same background as the Signal Region (or is a better approximation to it than our model at least), we can compare the likelihood volumes for toys generated under the background model and under the correct background in terms of the following expectation values
\begin{align}
    \mathbb{E}_{x\sim p_{b}(x|\theta_{b})}[\Theta(p_{b}(x|\theta_{b})-\likelihood{c})] &\stackrel{?}{=} \mathbb{E}_{x\sim p_{\text{true b}}(x)}[\Theta(p_{b}(x|\theta_{b})-\likelihood{c})]\,,\nonumber\\
     \estimatedfrac{c}{\theta_{b}} &\stackrel{?}{=} \estimatedfrac{c}{\text{true b}}\,.
\end{align}

\begin{figure}[h!]
    \centering
    \includegraphics[width=\linewidth]{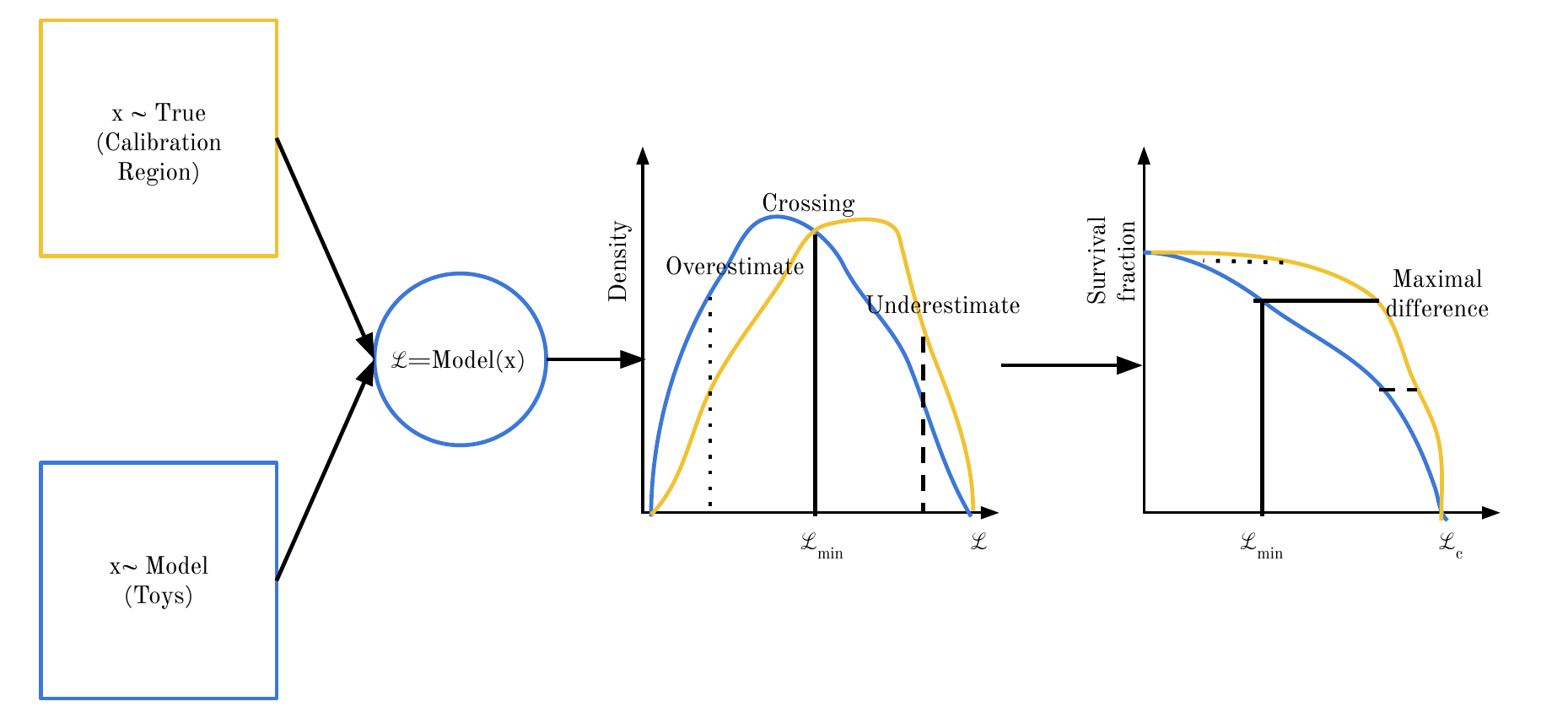}
    \caption{A diagrammatic description of the \likelihood{\text{min}} computation. See text for details.}
    \label{fig:diagram}
\end{figure}

We know that trivially these fractions coincide for $\likelihood{c}=0$, where they are both 1 and for $\likelihood{c}=\infty$, where they are both zero. However, we expect that the two models do not match perfectly, and thus there exists a critical point $\likelihood{c}$ where the difference between toys and Calibration Region expectation values is maximal. We can use this to define a minimum resolution likelihood below which we no longer trust our model

\begin{align}
    \label{eq:condition}
    \likelihood{\min}&\equiv\arg\max_{\likelihood{c}}|\estimatedfrac{\text{c}}{\theta_{b}} - \estimatedfrac{\text{c}}{\text{true b}}| \equiv\arg\max_{\likelihood{c}}|\Delta(\likelihood{c})|\,.
\end{align}
A diagrammatic description of this procedure is shown in \cref{fig:diagram}.

The definition of \likelihood{\text{min}} is not unique. Instead of the maximum, we could for example select the highest likelihood for which the difference is above a certain threshold. However, the definition of said threshold is arbitrary and may depend on the specific background and signal distributions considered, while the maximum prescription can be stated with no reference to specific distributions (although its uselfuness will depend on the particulars of the problem). 

Moreover, the definition of $|\fracdiff|$ itself is arbitrary and may be replaced by other quantifications of distribution disagreements based on different intuitions. \Cref{eq:condition} is attractive because it explicitly quantifies the disagreement in the survival function of the models using the likelihood itself as a summary statistic, which is particularly relevant for excluding the background-only hypothesis.  In App.~\ref{app:derivation}, we show that \cref{eq:condition} finds $\likelihood{\text{min}}$ that encodes the crossing between the regions where the background model underestimates and overestimates the background likelihood surfaces. We do highlight, however, that this will depend on the specific parameterization of the feature space and thus we assume a domain-expert defined choice has been made. 

In particular, we expect this choice to concentrate the signal in a region, which we may deem ``central''. Then, if the background model has heavier ``tails'' (ie a larger probability volume away from that central region) than the true background, then for higher likelihood hypersurfaces the background model under-predicts the expected the number of background events in that hypersurface, while for lower likelihood hypersurfaces the background model over-predicts. Thus, the choice of $\likelihood{\text{min}}$ is such that on average the two likelihoods match, corresponding to the ``crossing'' region where we go from under- to over-prediction. This is highly informed by the fact that to us underestimating the number of background events is worse than overestimating it. 

In practice, many local maxima may exist for complex likelihoods. However, these local maxima risk being spurious due to finite sample noise in the $|\Delta(\likelihood{c})|$  estimation. In particular, even if the two samples originate from the same distribution, \cref{eq:condition} will find some non-zero $\likelihood{\text{min}}$ due to finite sample sizes. To avoid this, which would be overly conservative since it penalizes correct background models, we always select the global maximum and set a dataset-dependent minimum threshold on the $|\Delta(\likelihood{c})|$ value, such that if no $\likelihood{c}$ crosses that threshold, $\likelihood{\text{min}}=0$.

\begin{align}
    \label{eq:condition_modified}
    \likelihood{\min}(\epsilon)=\begin{cases} 
      \arg\max_{\likelihood{c}}|\Delta(\likelihood{c})| & \text{if } \max_{\likelihood{c}}|\Delta(\likelihood{c})|>\epsilon\,,  \\
      0.0 & \text{if } \max_{\likelihood{c}}|\Delta(\likelihood{c})|\leq\epsilon\,.
    \end{cases}
\end{align}

In the presence of several comparable maximum candidates, we select the one with the largest likelihood. This implies that the method will be more conservative, possibly discarding well-modelled regions that could be contained between two maxima. The $\epsilon$ threshold may be computed by comparing the toy samples with an additional set of toy samples generated using the (in principle) biased background model which is of equal size to the Calibration Region. We should note that setting $\likelihood{\text{min}}=0$ does not imply that the background model is perfect, but simply that its bias is subdominant to the finite size effects in the calibration and toy datasets.

\subsection{The Fiducial Signal Region}
\label{subsec:fiducial_signal_region}

This minimum likelihood allows us to define a Fiducial Signal Region (FSR) given by $p_{b}(x|\theta)\geq \likelihood{\min}$. That is, if events are rare enough under our background model, we discard these events.\footnote{One could take less drastic measures, such as a clipping the likelihood to a constant value. We leave such explorations for future work.} If the signal does not lie completely below this likelihood cut, then the model will find the signal with reduced significance but with less bias. One does risk running into the case of no signal, and this is why this is a conservative technique.\footnote{This is an example of how there can be no free lunch once the model is estimated from data, since there is a complicated relationship between overfitting, look elsewhere effect and statistical power at play.} The extended likelihood for the statistical inference is thus modified to

\begin{align}
    \label{eq:extended_fid_likelihood}
    \mathcal{L}(\tilde{\mathcal{D}},\mu_{s},\nu)&=\mathcal{P}(\tilde{N}|\mu_{s} \lambda_s\epsilon_{\mathrm{sig}}(\theta_s)+\mu_{b} \lambda_b \epsilon_{\mathrm{CR}})\nonumber\\
    &\prod_{x \in \mathrm{FSR}}^{\tilde{N}}\left(\frac{\mu_{s}\lambda_s \epsilon_{\mathrm{sig}}(\theta_s)}{\mu_{s}\lambda_s \epsilon_{\mathrm{sig}}(\theta_s)+\mu_{b}\lambda_b \epsilon_{\mathrm{CR}}}\frac{p_s(x_n|\theta_s)}{\epsilon_{\mathrm{sig}}(\theta_s)}+\frac{\mu_{b}\lambda_b \epsilon_{\mathrm{CR}}}{\mu_{s}\lambda_s \epsilon_{\mathrm{sig}}(\theta_s)+\mu_{b}\lambda_b \epsilon_{\mathrm{CR}}}\frac{p_b(x_n|\theta_b)}{\epsilon_{\mathrm{toys}}(\theta_b)}\right)\,,
\end{align}
where $\tilde{N}$ is the number of events in the FSR and we have introduced the efficiencies estimated at the Calibration Region, additional signal simulations and using toys:
\begin{align}
    \epsilon_{i}(\nu_i) = \frac{\int dx\,p_{i}(x|\nu_i)\Theta(p_b(x|\theta_{b})-\likelihood{\min})}{\int dx\,p_{i}(x|\nu_i)}\,.
\end{align}

Although in this work we consider them as fixed, they could also be treated as additional nuisance parameters to be profiled over in the Fiducial Signal Region. We emphasize as well the difference in the efficiency used to correct the normalization for the expected background (Calibration Region) and the efficiency used to properly normalize the probability distribution (coming from the toys), to better increase the agreement between the background model and the true background.

This framework can easily accommodate the lack of a full model for the nuisance parameters, which is usually approximated by a Template Morphing based on two-point uncertainties~\cite{Read:1999kh, cranmer:2012sba}. Let us assume we have, a nominal estimate of the nuisance parameters, $\theta^{\text{nom.}}_{b}$. Then, we perform the $\likelihood{\text{min}}$ estimation using $p(x|\theta^{\text{nom.}}_{b})$ and estimate the different efficiencies using this cut on toys, simulation and Calibration Region. Since in the final analyses we still want to profile over $\theta_{b}$ to allow for better parameter inference with realistic errors, we can perform the trivial re-writing

\begin{align}
    p(x|\theta_b)&=\frac{p(x|\theta_b)}{p(x|\theta^{\text{nom.}}_{b})}p(x|\theta^{\text{nom.}}_{b})\nonumber\\
    &=w(x,\theta_b,\theta^{\text{nom.}}_{b})p(x|\theta^{\text{nom.}}_{b})\,,
\end{align}
which is very amenable to the use of parameterized nuisance parameter dependence, if needed. We can compute the modified efficiency using these weights as

\begin{align}
    \epsilon_{\mathrm{toys}}(\theta_b)&=\mathbb{E}_{x\sim\theta_{b}}[\Theta(p(x|\theta^{\text{nom.}}_{b})-\likelihood{\min})]\,,\nonumber\\
    &=\mathbb{E}_{x\sim\theta^{\text{nom.}}_{b}}[w(x,\theta_b,\theta^{\text{nom.}}_{b})\Theta(p(x|\theta^{\text{nom.}}_{b})-\likelihood{\min})]\nonumber\\
    &= \epsilon_{\mathrm{toys}}(\theta^{\text{nom.}}_b)\mathbb{E}_{x\sim\theta^{\text{nom.}}_{b}|\text{FSR}}[w(x,\theta_b,\theta^{\text{nom.}}_{b})]\,,
\end{align}
with the resulting extended likelihood being
\begin{align}
    \mathcal{L}(\tilde{\mathcal{D}},\theta)&=\mathcal{P}(\tilde{N}|\mu_{s} \lambda_s \epsilon_{\mathrm{sig}}(\theta_s)+\mu_{b} \lambda_b \epsilon_{\mathrm{CR}})\nonumber\\
    &\prod_{x_n \in \mathrm{FSR}}^{\tilde{N}}\Bigg(\frac{\mu_{s}\lambda_s \epsilon_{\mathrm{sig}}(\theta_s)}{\mu_{s}\lambda_s \epsilon_{\mathrm{sig}}(\theta_s)+\mu_{b}\lambda_b \epsilon_{\mathrm{CR}}}\frac{p_s(x_n|\theta_s)}{\epsilon_{\mathrm{sig}}(\theta_s)}\nonumber\\
    &+\frac{\mu_{b}\lambda_b \epsilon_{\mathrm{CR}}}{\mu_{s}\lambda_s \epsilon_{\mathrm{sig}}(\theta_s)+\mu_{b}\lambda_b \epsilon_{\mathrm{CR}}}\frac{w(x,\theta_b,\theta^{\text{nom.}}_{b})}{\mathbb{E}_{x\sim\theta^{\text{CR}}_{b}|\text{FSR}}[w(x,\theta_b,\theta^{\text{nom.}}_{b})]}\frac{p_b(x_n|\theta^{\text{nom.}}_b)}{\epsilon_{\mathrm{toys}}(\theta^{\text{nom.}}_b)}\Bigg)\,.
\end{align}
where, as in \cref{eq:extended_fid_likelihood}, we have used different efficiencies to correct the normalization for the expected background and to properly normalize the probability distribution, to better increase the agreement between the background model and the true background, and thus limit the impact of the nuisance parameters to the latter.

\section{Toy data}
\label{sec:toys}

To illustrate the method, we explore two toy examples, chosen for their simplicity and their ability to highlight the power and limitations of the method. In these examples, shown in \cref{subsec:toy_gauss,subsec:toy_exp}, we define the true signal and background distributions and a background model in terms of parametric functions, and run pseudo-experiments to characterize the impact of the method in the statistical power of the analysis. In each pseudo-experiment, the parametric functions are used both to generate the datasets and to perform a statistical fit of the data. We assume the number of signal and background number of events is Poisson distributed, with known expected rates $\{\lambda_s,\lambda_b\}$, and fit signal and background strength modifiers $\{\mu_s,\mu_b\}$ so that the nominal result corresponds to $\mu_{s,b}=1$. All parameter fits and uncertainty estimation are done with \texttt{iminuit}~\cite{iminuit}. 

Each pseudo-experiment consists of the generation of a Signal Region dataset using the true signal and background distributions and fixed Poisson rates and a fit to this dataset using either the true background distribution or the background model. Then, a Calibration Region and a ``toy'' dataset are generated using the true background and the background model, respectively, to calibrate and obtain $\likelihood{\text{min}}$ (for the toy examples, $\epsilon$ is set arbitrarily to $10^{-4}$ although we find no dependence on its value due to the simplicity of the problems). The estimated $\likelihood{\text{min}}$ is used to define a Fiducial Signal Region, where a new fit with the background model is performed. The size of the Calibration Region and the toy dataset is five times larger than the Signal Region, to avoid statistical fluctuations and disentangle the problems of estimating the Fiducial Signal Region from its impact on the fit itself.\footnote{Although because of the increased statistics, one could share the Calibration Region among pseudo-experiments since \likelihood{\text{min}} is almost unchanged, the generation cost is negliglible and self-consistency is preferred.}

\subsection{Gaussian case}
\label{subsec:toy_gauss}

The simplest case is when the signal and background are both gaussians, and we mismodel the standard deviation of the background.
\begin{align}
    p(x|\theta_s)&=\mathcal{N}(\bar{x}_s,\sigma_s)\nonumber\\
    p(x|\text{true b})&=\mathcal{N}(\bar{x}_b,\sigma_b)\nonumber\\
    p(x|\theta_b)&=\mathcal{N}(\bar{x}_b,\tilde{\sigma}_{b})\nonumber\\
\end{align}
where $\bar{x}_{s,b}$ are the signal and background means, and $\sigma_{s,b}$ ($\tilde{\sigma}_{b}$) the (mismodelled) standard deviations. To capture the range of applications of the method, we chose the parameters so that the signal-to-background is overestimated for values of $x$ where the signal is present but also the difference is small enough that the definition of the fiducial region does not remove all the signal. We consider two models, to show how the degree of mismodelling affects both the bias and the resulting \method performance. We select
\begin{align}
    (\bar{x}_s,\sigma_s) &= (0.5,0.1)\nonumber\\
    (\bar{x}_b,\sigma_b) &= (0.0,1.0)\nonumber\\
    \tilde{\sigma}_{b}&= 
    \begin{cases} 
      1.1 & \text{Nominal} \\
      1.025 & \text{Better model }
    \end{cases}
\end{align}

\begin{figure}[h!]
    \centering
    \includegraphics[width=0.32\linewidth]{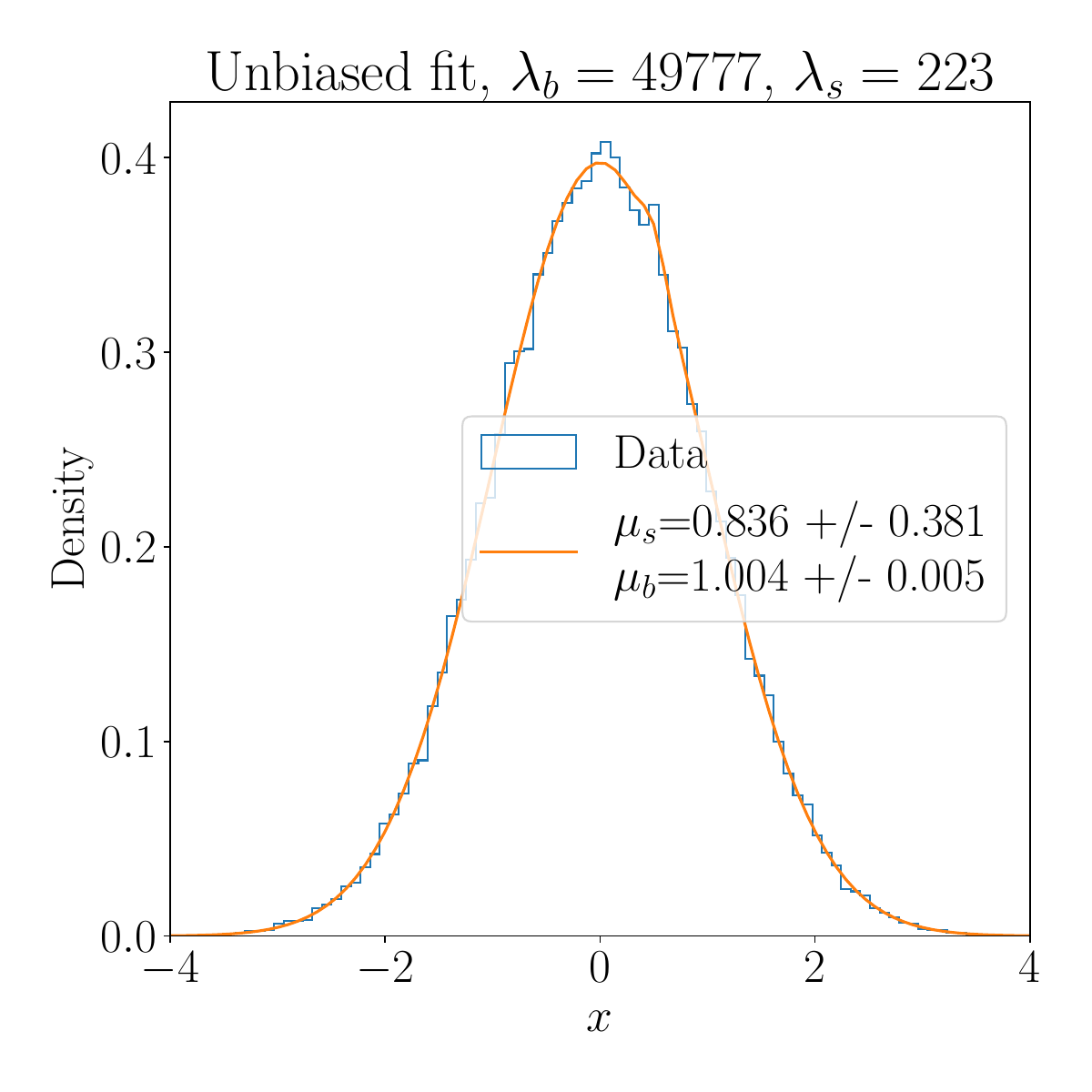}
    \includegraphics[width=0.32\linewidth]{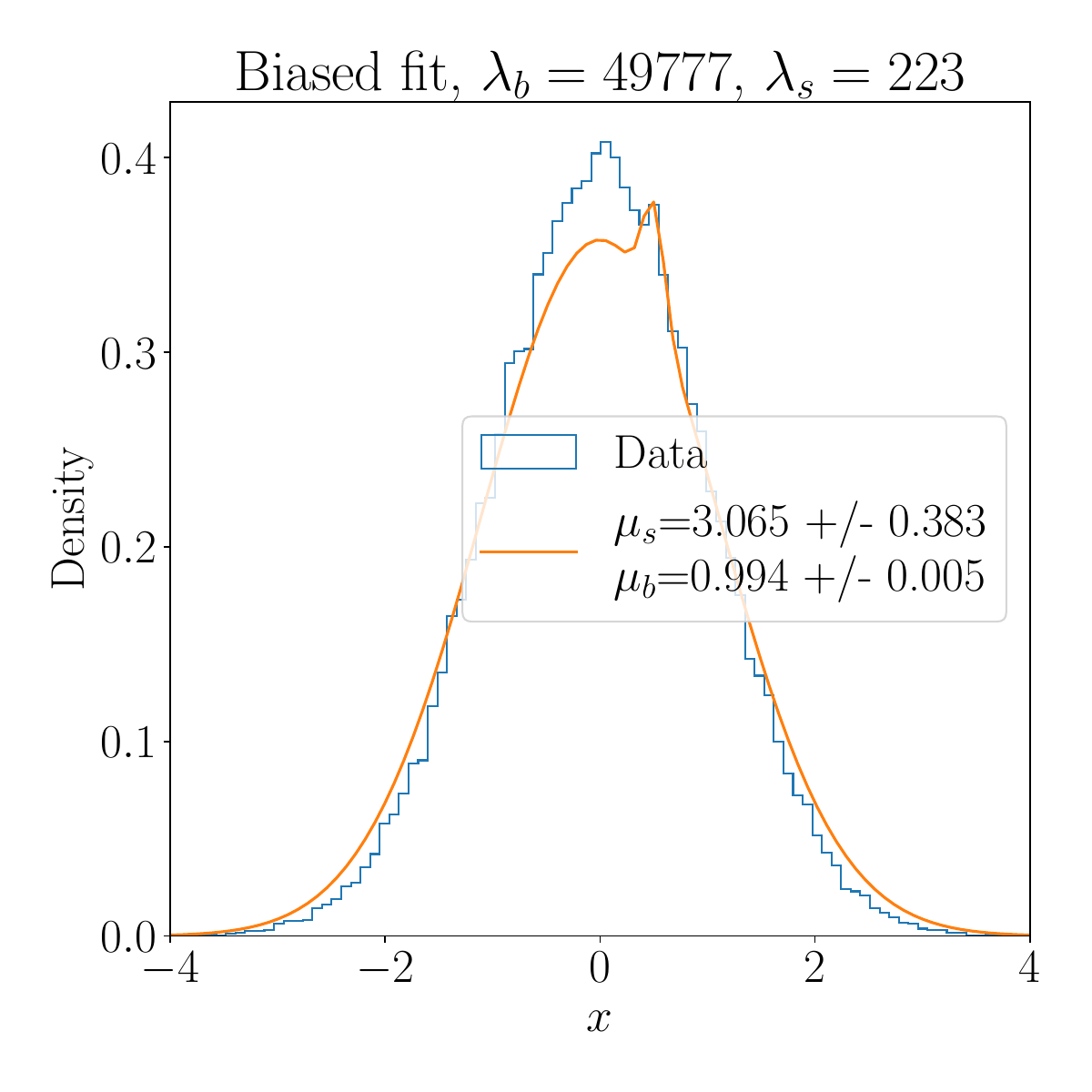}
    \includegraphics[width=0.32\linewidth]{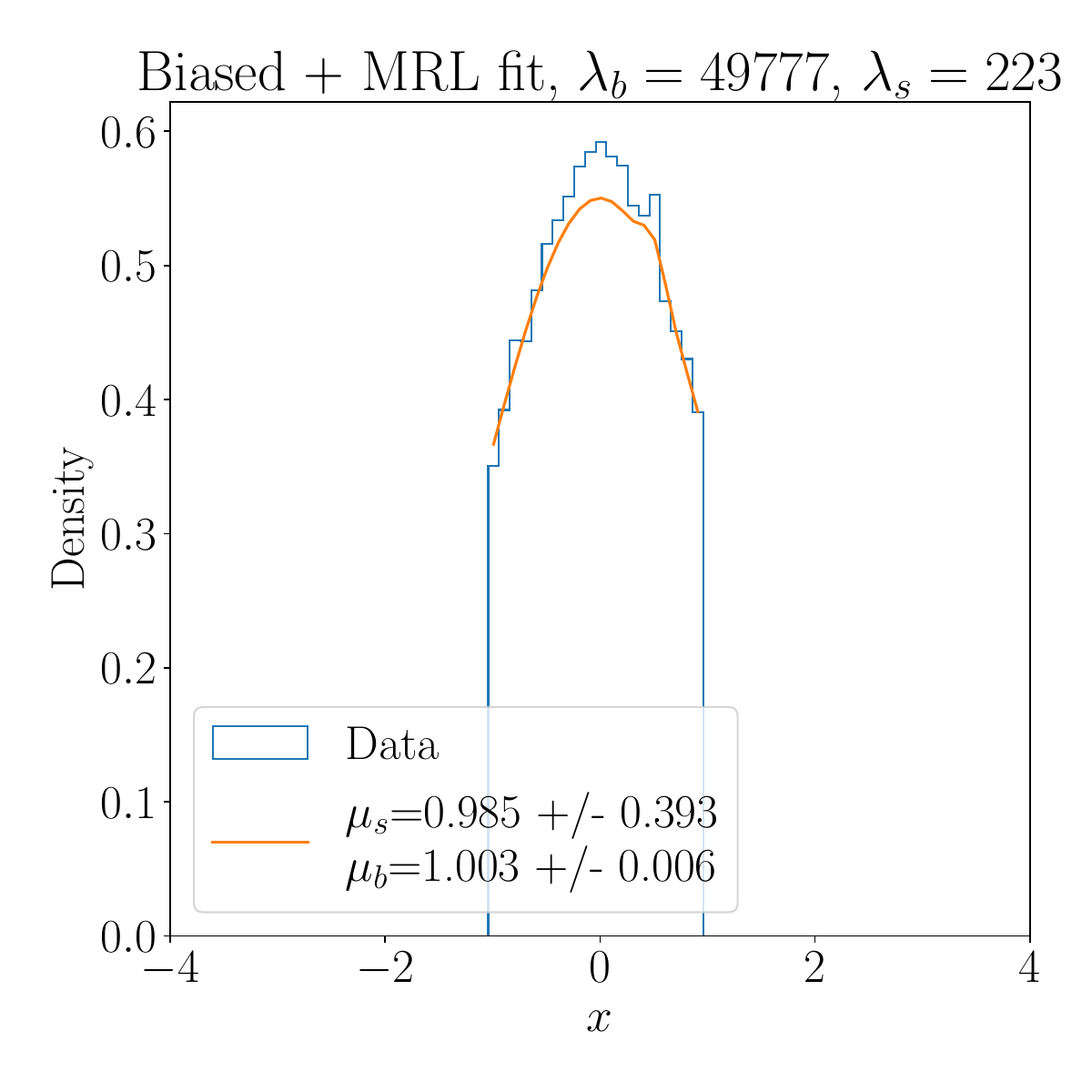}\\
    \includegraphics[width=0.32\linewidth]{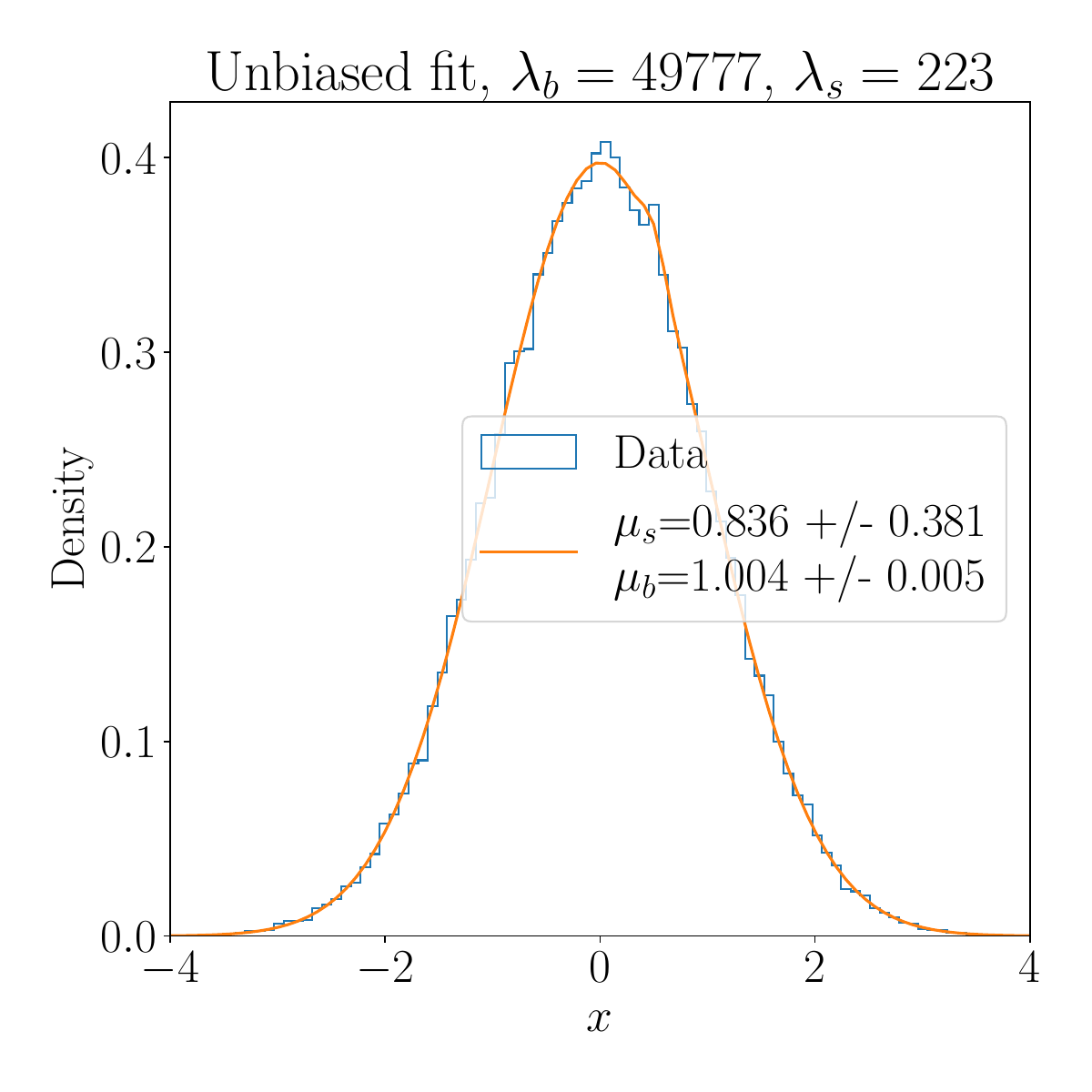}
    \includegraphics[width=0.32\linewidth]{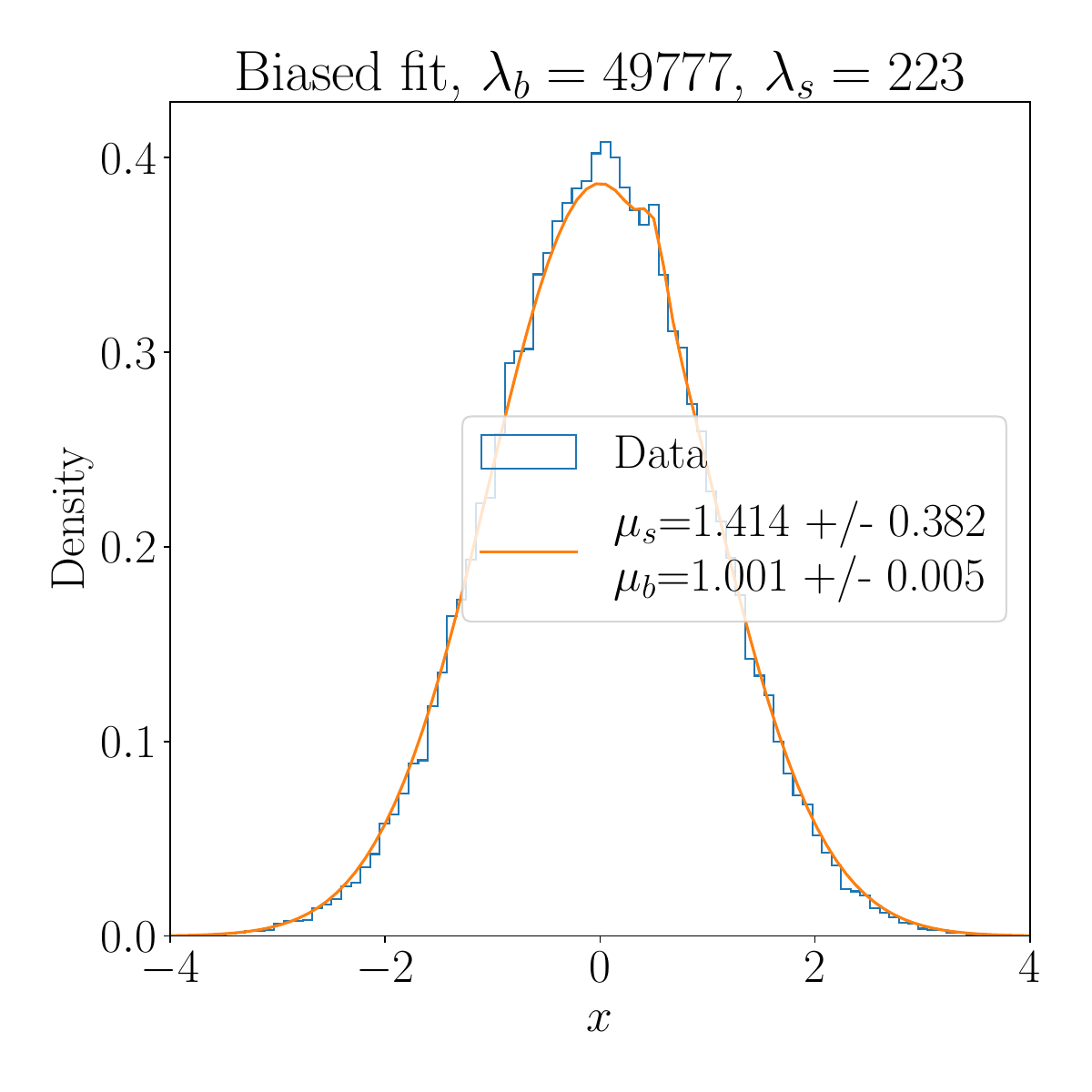}
    \includegraphics[width=0.32\linewidth]{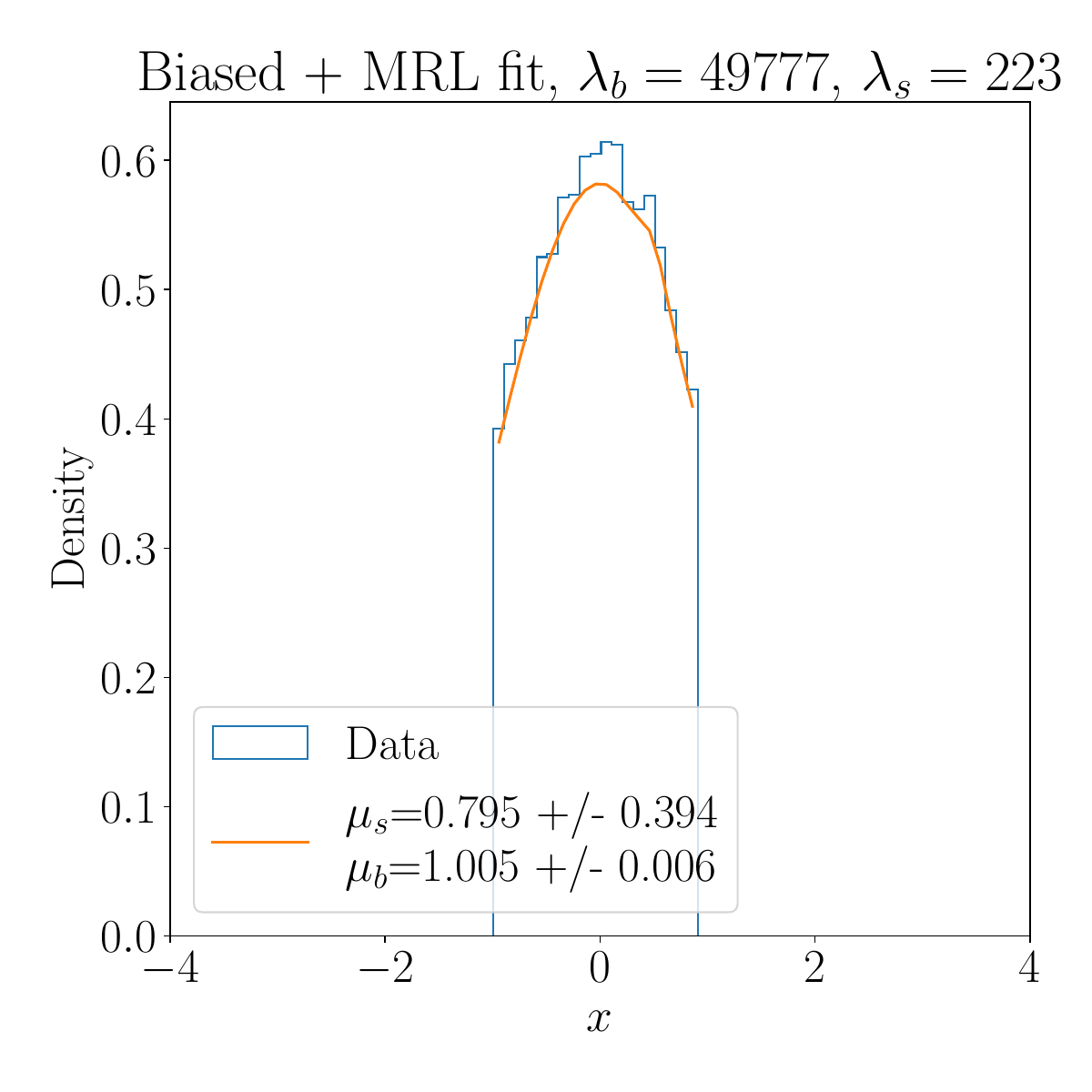}\\    
    \caption{The three fits done on a single pseudo-experiment for the 1D Gaussian example. We show the result of the fit with the correct model, with the biased background model, and with the biased background model but on the Fiducial Signal Region defined with the \method method. Upper (lower) row considers the nominal (better) background model.}
    \label{fig:toy_results_gaussian_1d}
\end{figure}

\begin{figure}[t!]
    \centering
    \includegraphics[width=0.45\linewidth]{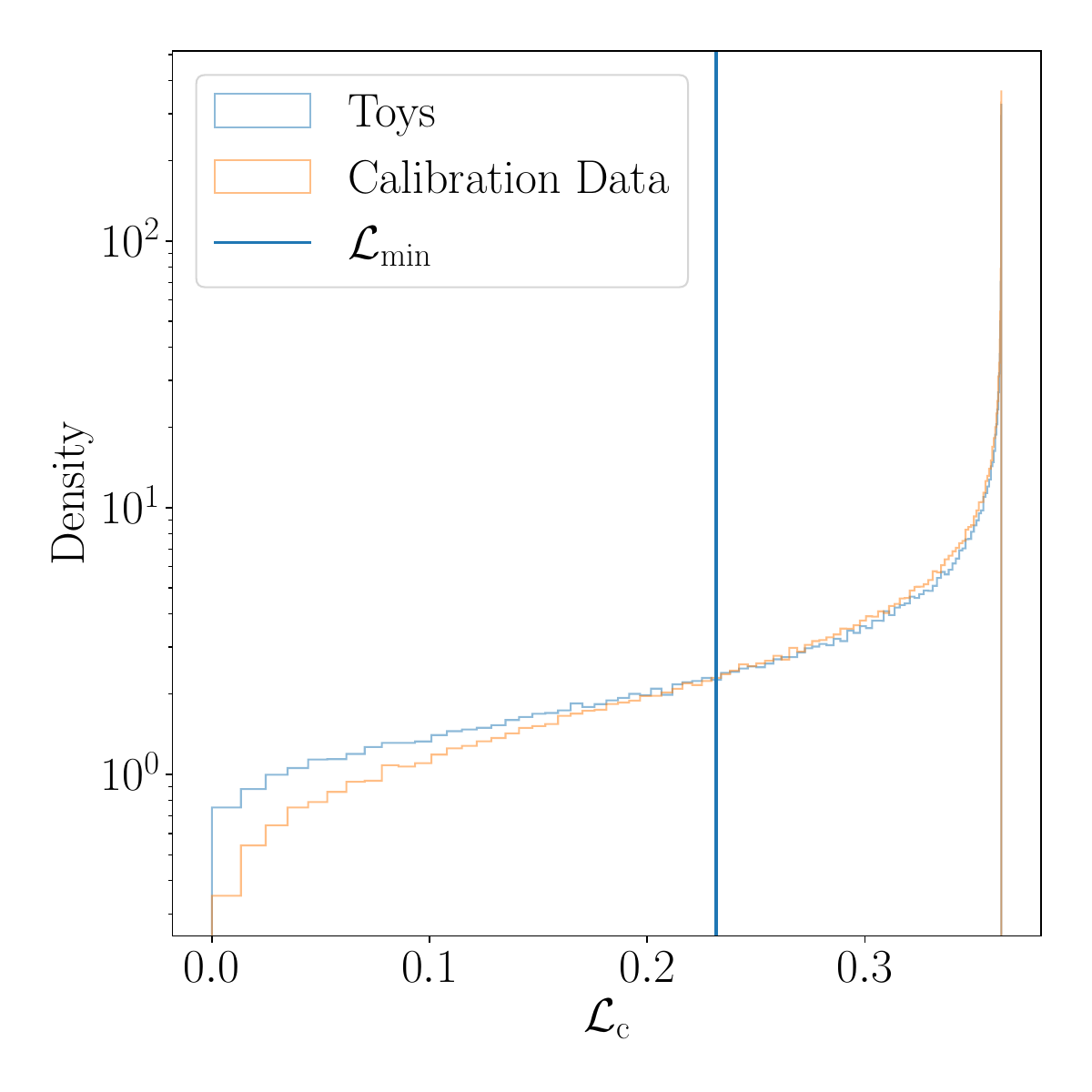}      
    \includegraphics[width=0.45\linewidth]{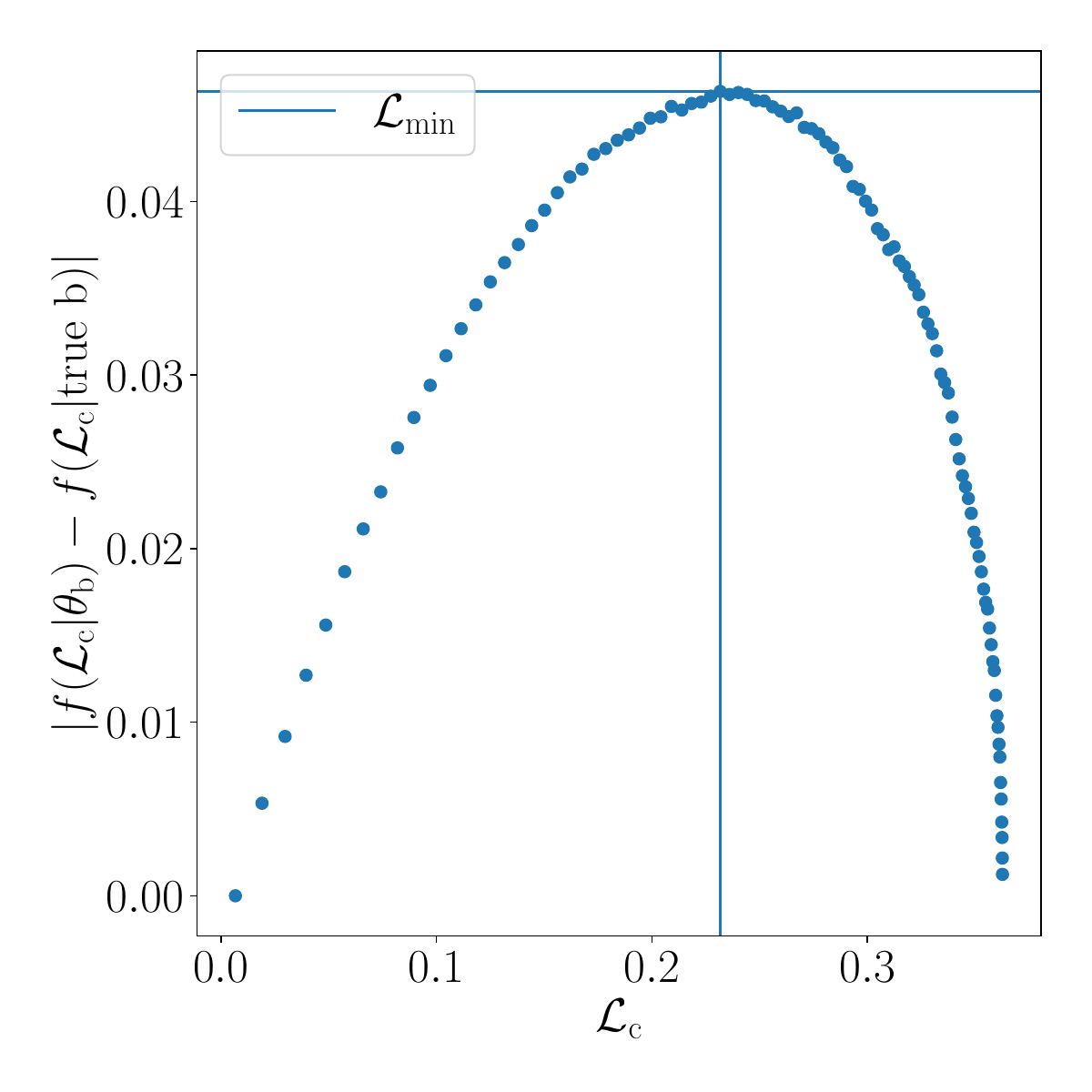}\\
    \includegraphics[width=0.45\linewidth]{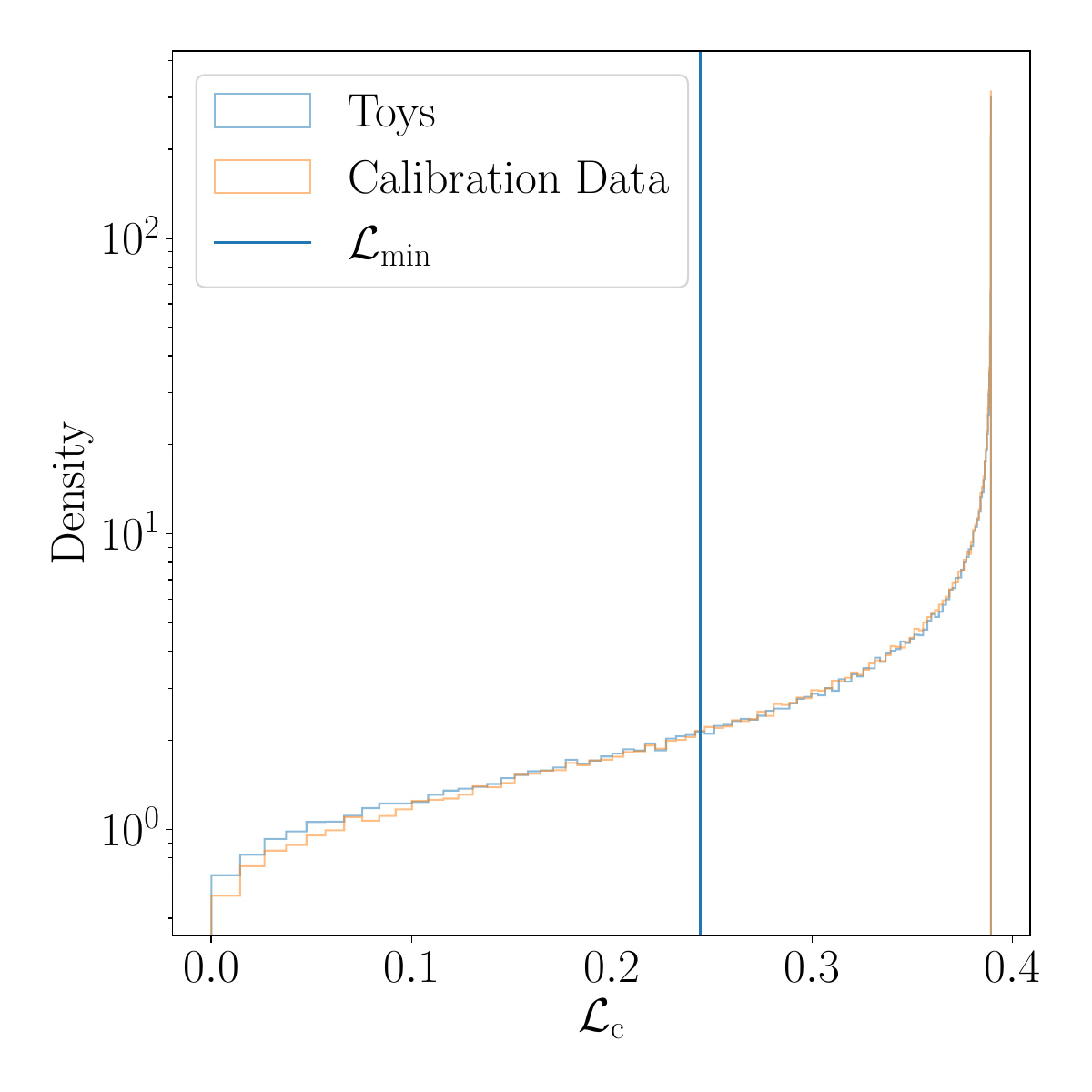}      
    \includegraphics[width=0.45\linewidth]{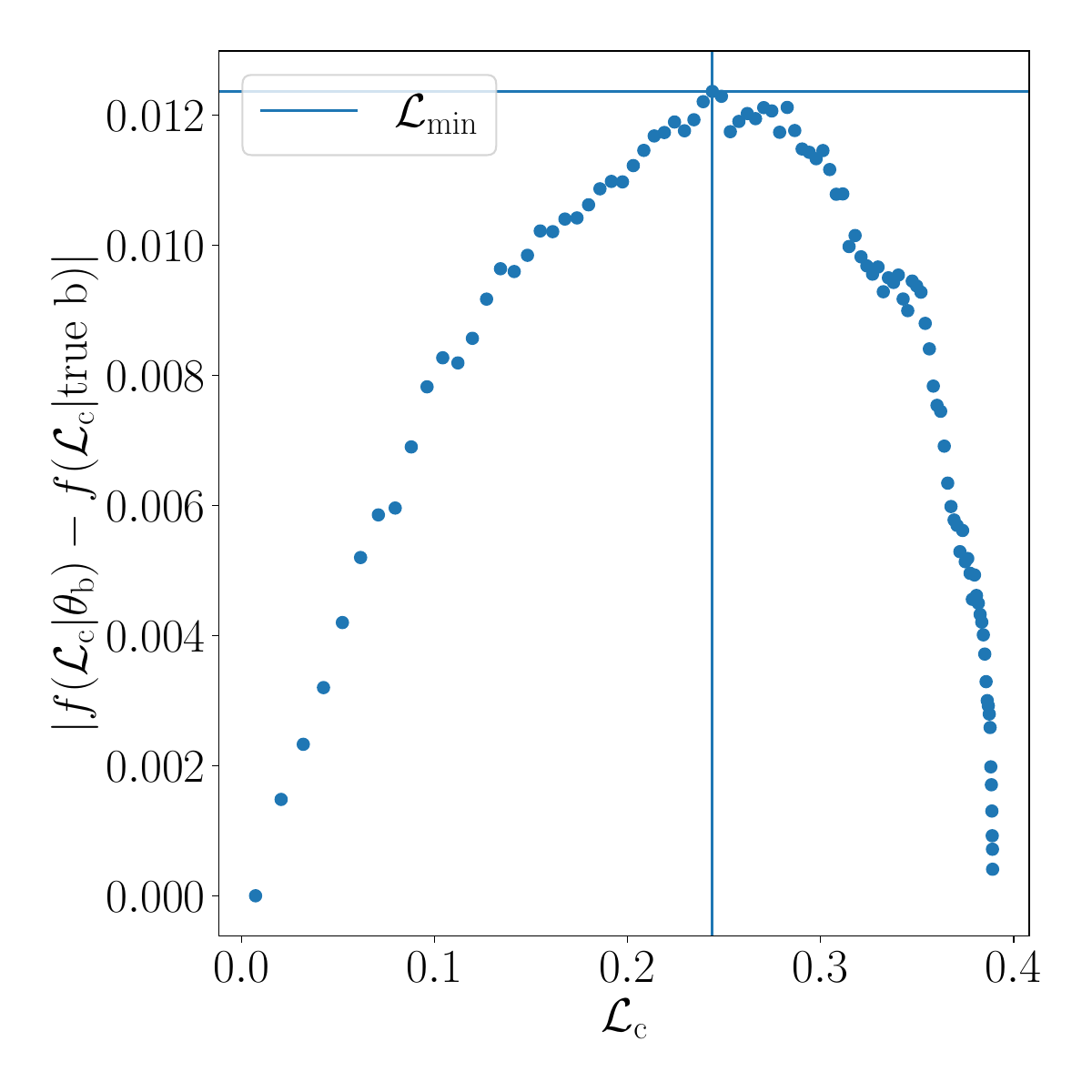}\\
    \caption{$\likelihood{\text{min}}$ determination obtained by comparing a Calibration Region and a set of toys for the 1D Gaussian example. Left: The probability distribution of the model likelihood under the background model and the true background. Right: The fraction difference distribution as a function of the critical likelihood \likelihood{c} obtained from comparing the toy and Calibration datasets. We observe how $\likelihood{\text{min}}$ captures the crossing between background overestimation and background underestimation. Upper (lower) row considers the nominal (better) background model.}
    \label{fig:toy_distribution_comparison_gaussian_1d}
\end{figure}

To highlight the impact of the \method method, we show the fitted values of a single pseudo-experiment in Fig.~\ref{fig:toy_results_gaussian_1d}. We observe how the biased model overestimates the number of signal events in order to increase the total number of events in the bulk of the distribution and suppress the events in the tail. The Fiducial Signal Region definition clips the tails, and unbiases the estimation of the signal and background strengths, at the expense of an increased uncertainty. The fit is not perfect, but is consistent with the true values for the reduced statistics. The $\likelihood{\text{min}}$ estimation, shown in Fig.~\ref{fig:toy_distribution_comparison_gaussian_1d}, highlights how there is a clear global maximum in the deviation between the calibration and toy region, that it corresponds to the crossing between background over and underestimation and that its value depends on the quality of the background model. 

\begin{figure}[h!]
    \centering
    \includegraphics[width=0.45\linewidth]{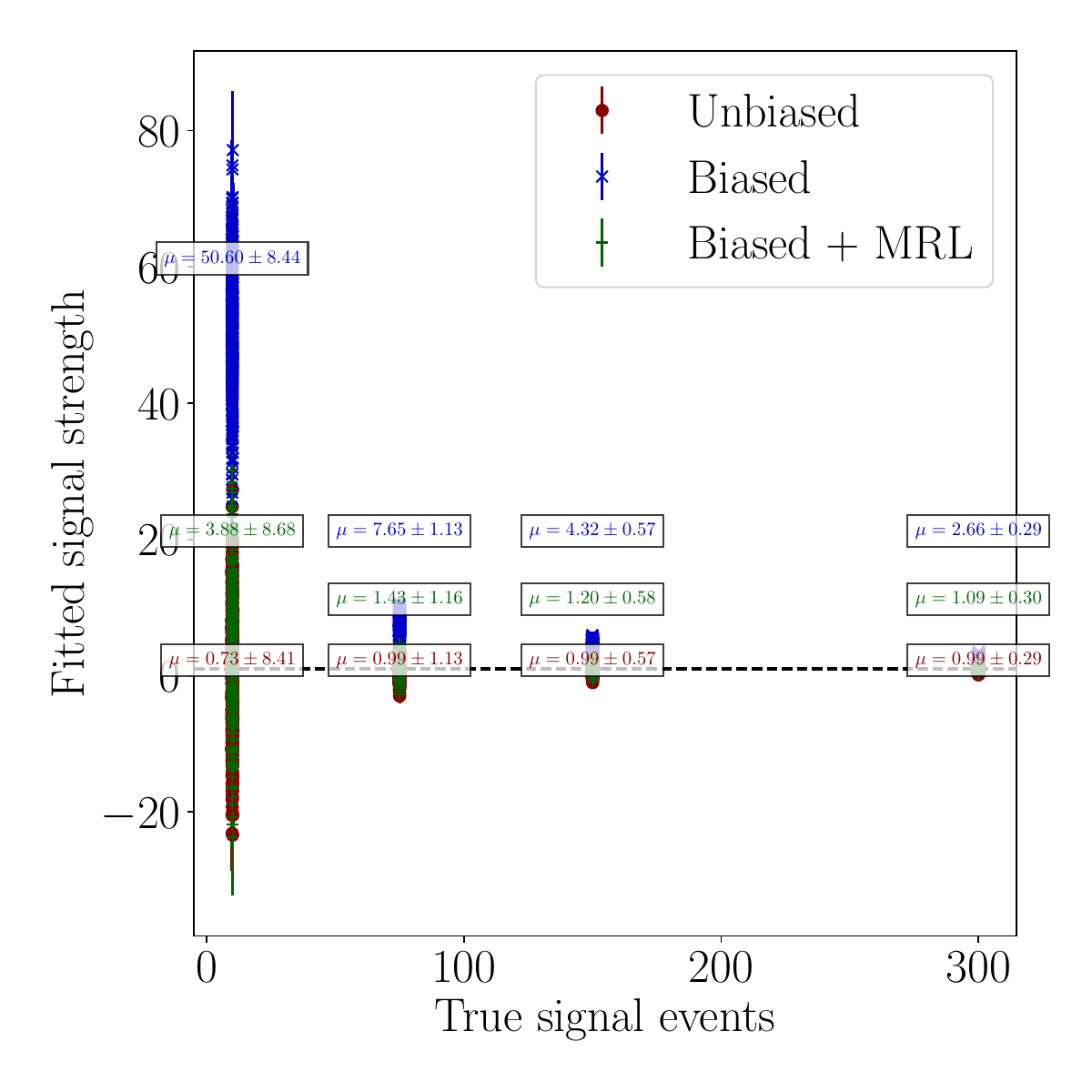}
    \includegraphics[width=0.45\linewidth]{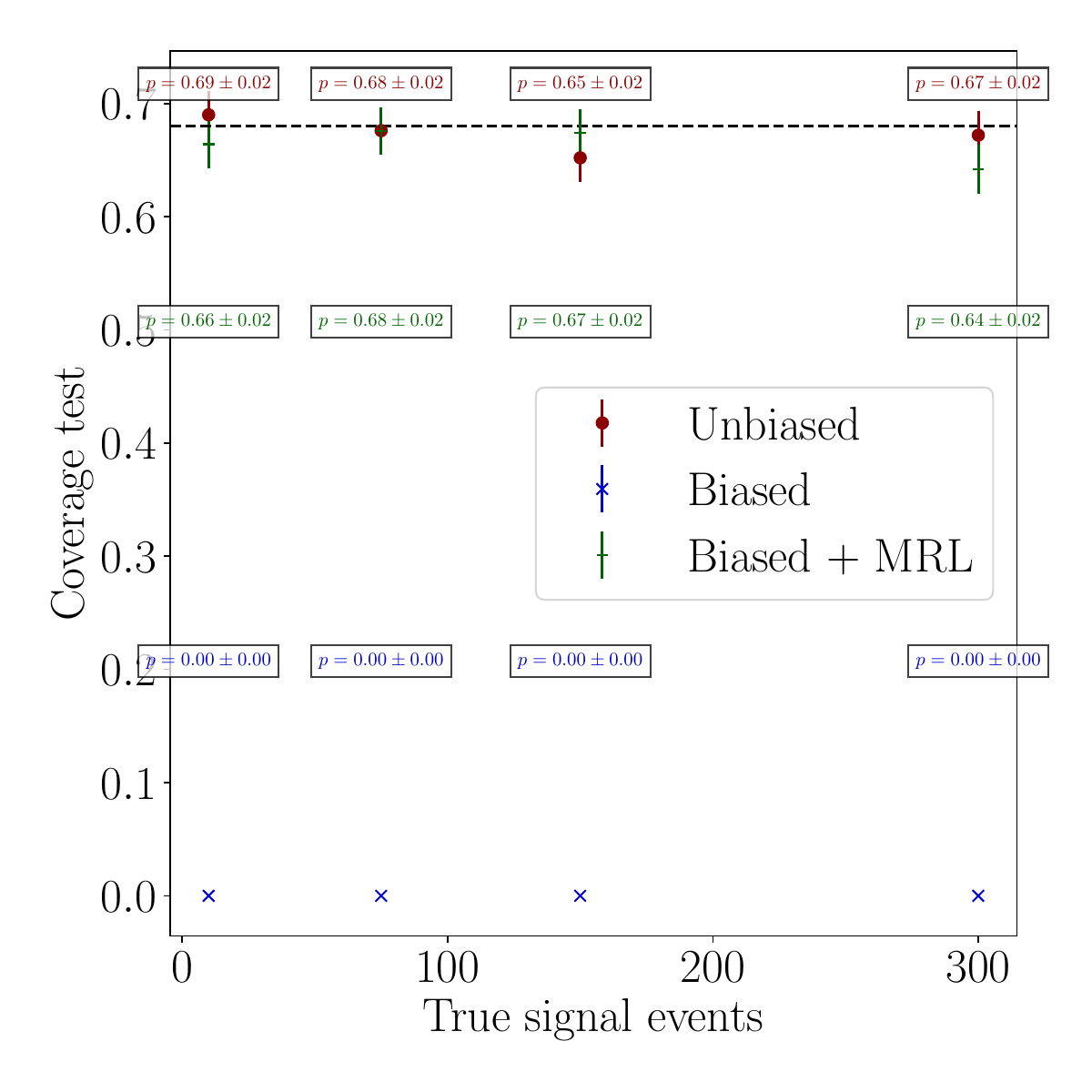}
    \includegraphics[width=0.45\linewidth]{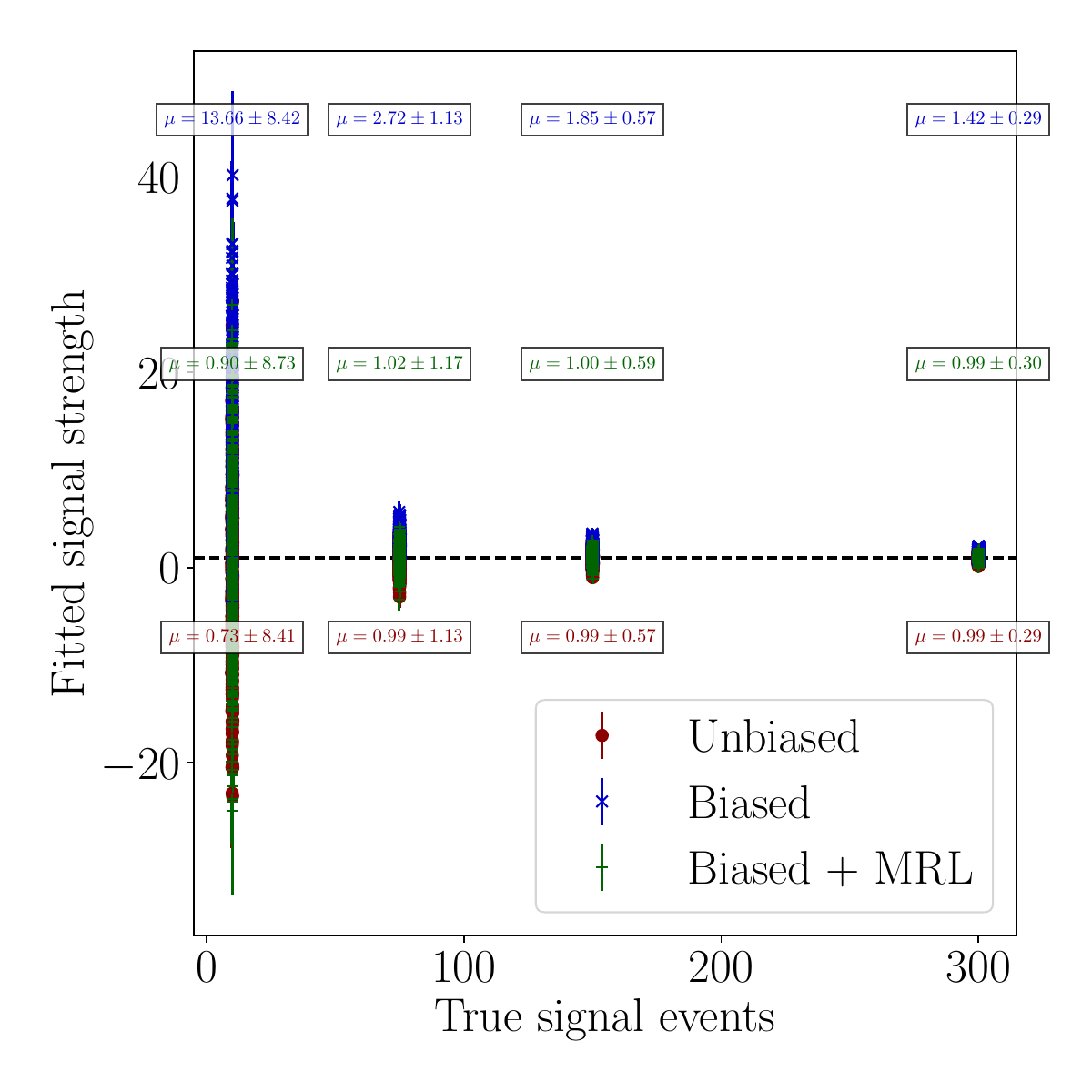}
    \includegraphics[width=0.45\linewidth]{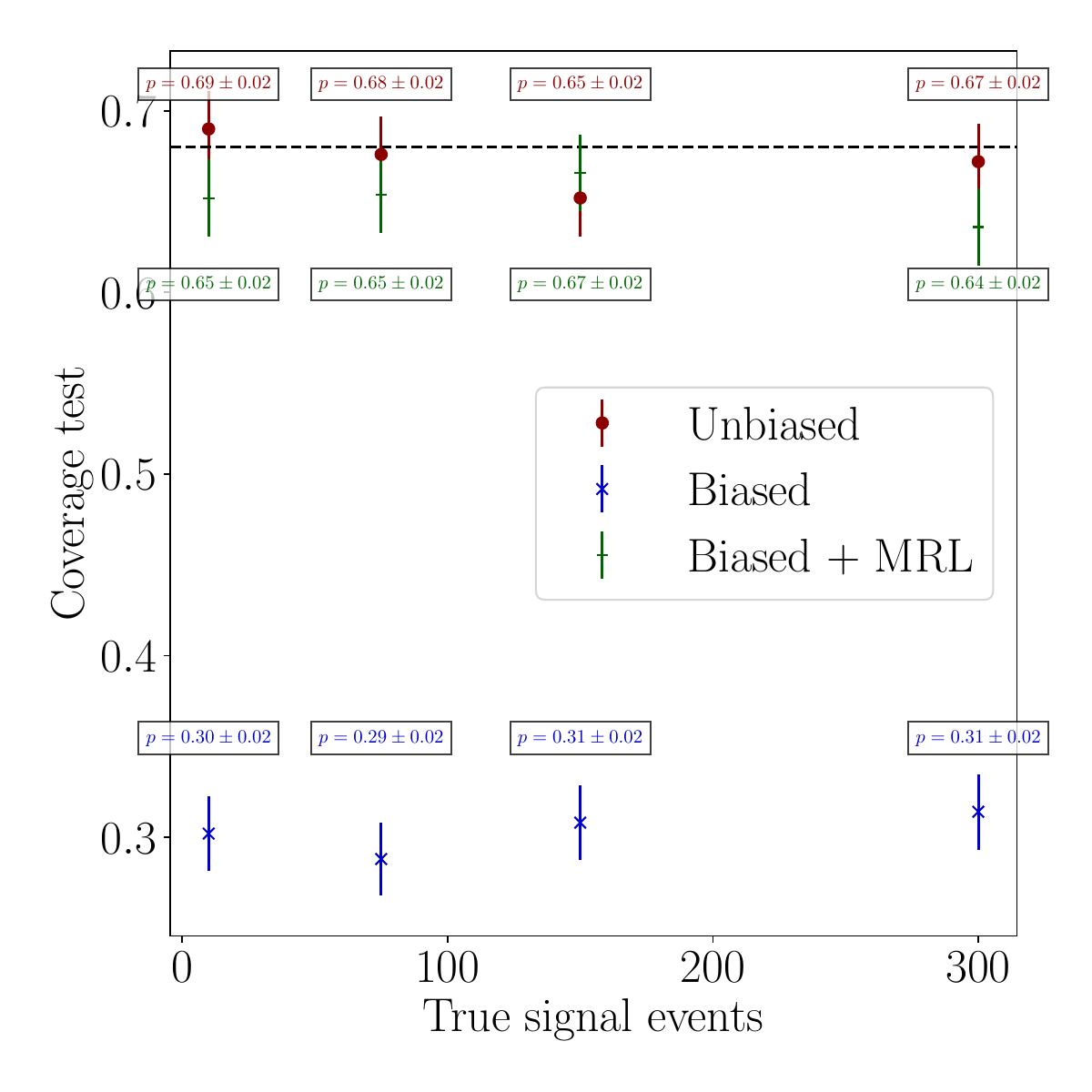}
    \caption{500 pseudo-experiment runs for different signal injection for the 1D Gaussian example. Left:  Maximum Likelihood Estimates of the signal strength as a function of the true expected signal events, with uncertainties. Right: Coverage derived from the confidence interval as a function of the true expected signal events. 
    Upper (lower) row considers the nominal (better) background model.}
    \label{fig:toy_signal_injection_study_gaussian_1d}
\end{figure}

Although illustrative, a single pseudo-experiment is not sufficient to quantify the effectiveness of the method. To do so, we run multiple pseudo-experiments for different signal injections.  We show the results in Fig.~\ref{fig:toy_signal_injection_study_gaussian_1d}. 

For the upper row corresponding to the nominal background model, we observe in the left plot the distribution of the Maximum Likelihood Estimates (MLE) of the signal strength with its uncertainties,\footnote{The uncertainties are estimated using the \texttt{HESSE} method in \texttt{iminuit}, which uses a Hessian approximation to compute symmetric uncertainties.} and see how the \method method greatly improves the results, yielding less biased estimations at the expense of slightly larger errors. This is further reinforced by looking at the coverage in the right plot, where we observe how the coverage becomes much closer to the desired value. The biased signal strength estimators arise from the same bias in the background model that the method aims to solve, and is unavoidable for large enough statistics.
This highlights that the \method method can only do so much towards fixing the bias for such misspecified models, specially where the central value of the MLE deviates more strongly from one. 
However, although it does not completely fix the bias, defining a FSR certainly reduces it and the corresponding danger of a false discovery.

This is further reinforced by the lower row, where the improvement in the background model results in a better \method performance, providing reasonable Maximum
Likelihood Estimates that yield good coverage for all considered signal strengths, effectively correcting the analysis. This also shows that if the bias of the background model is small enough for the statistical size of the sample, \method will not over-penalize the signal strength estimates.

\subsection{2D Exponential and Gaussian}
\label{subsec:toy_exp}

Although easily interpretable, the 1D example is too simple. To further study the model, and thinking of the realistic example introduced in Section~\ref{sec:hi-sigma}, we consider a 2D example where the distributions are inspired by the di-Higgs search.

We consider the case where the signal and background distributions are bounded by a two dimensional box of side $1$. The background distribution is a factorized product of truncated exponentials, while the signal is a truncated diagonal two-dimensional gaussian. The background model is also a product of truncated exponentials, but with misspecificied parameters.

\begin{align}
    p(x,y|\theta_s)&=\text{TruncNorm}(x;\bar{x}_s,\sigma_s)\text{TruncNorm}(y;\bar{x}_s,\sigma_s)\,,\nonumber\\
    p(x,y|\text{true b})&=\text{TruncExp}(x;r_b)\text{TruncExp}(y;r_b)\,,\nonumber\\
    p(x,y|\theta_b)&=\text{TruncExp}(x;\tilde{r}_b)\text{TruncExp}(y;\tilde{r}_b)\,.
\end{align}

We consider a single background model, and two possible signals, defined by the parameters
\begin{align}
    \bar{x}_s,\sigma_s &= 
    \begin{cases} 
      (0.3,0.1) & \text{Nominal}\,, \\
      (0.6,0.2) & \text{Unlikely}\,,
    \end{cases}\nonumber\\
    r_b &= 1.0\,,\nonumber\\
    \tilde{r}_b&=1.075\,,
\end{align}
where ``Unlikely'' reflects the fact that under that set of parameters, the signal will tend to populate lower background model probability regions than the ``Nominal'', although both models have the same total event yield. Using these models, we consider three benchmarks to highlight the power but also the limitations of the method. The first benchmark uses the nominal signal model to test the method in multiple dimensions. The second benchmark considers the nominal signal model as well but takes reduced values of $\lambda_s$ and $\lambda_b$ to show the impact of reduced statistics. The third uses the ``Unlikely'' signal to show how the method can be over-conservative if the Fiducial Signal Region definition is too restrictive.

\begin{figure}[h!]
    \centering
    \includegraphics[width=0.32\linewidth]{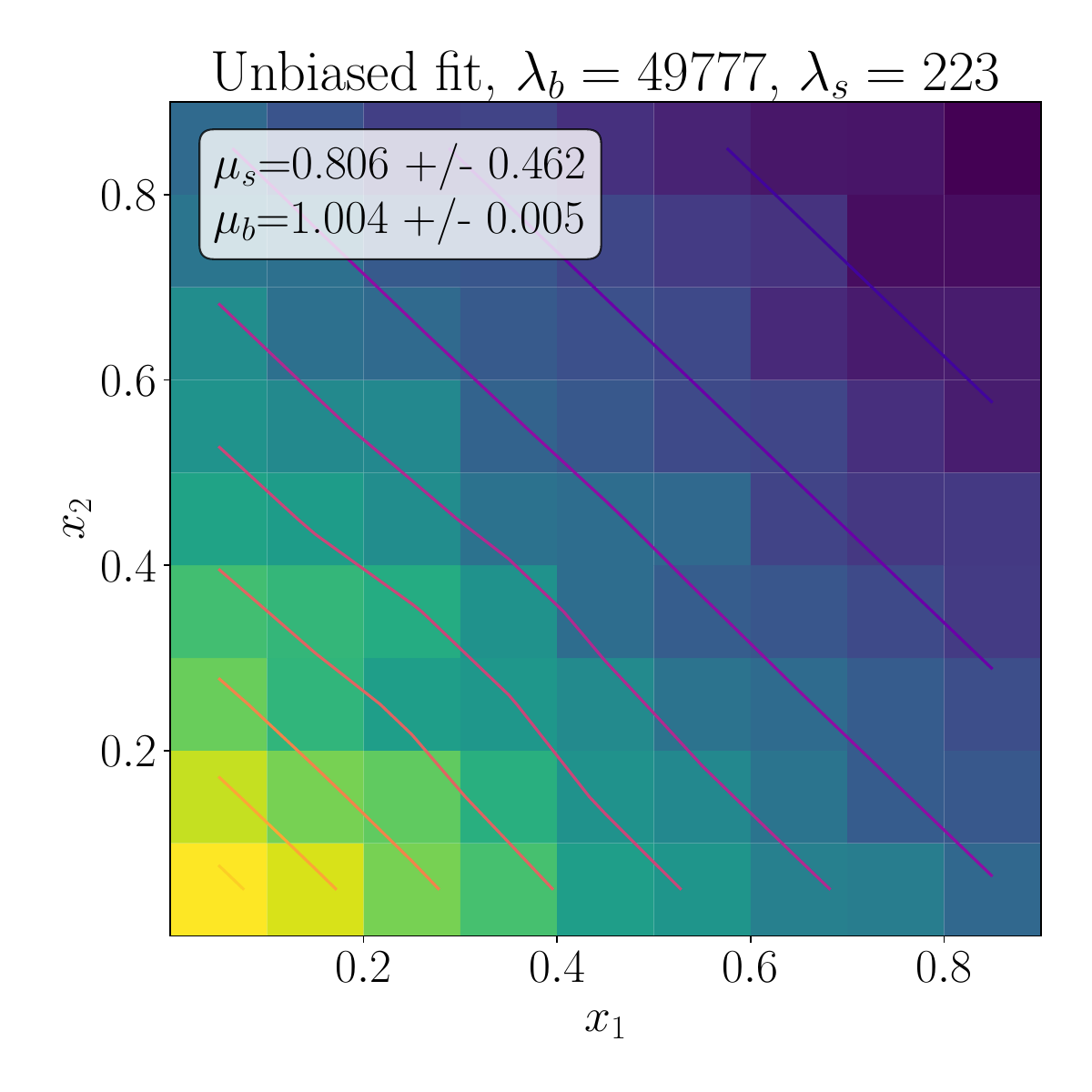}
    \includegraphics[width=0.32\linewidth]{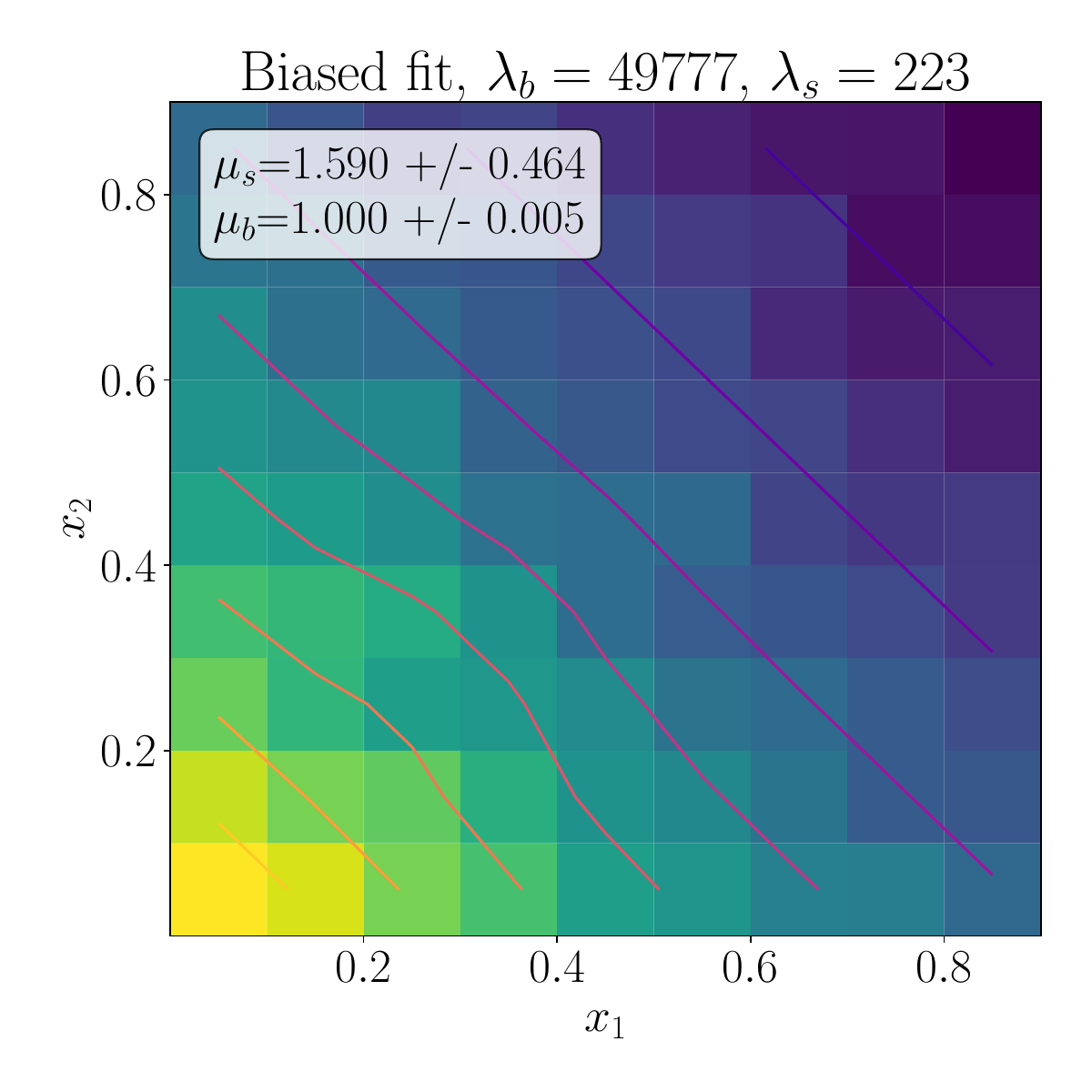}
    \includegraphics[width=0.32\linewidth]{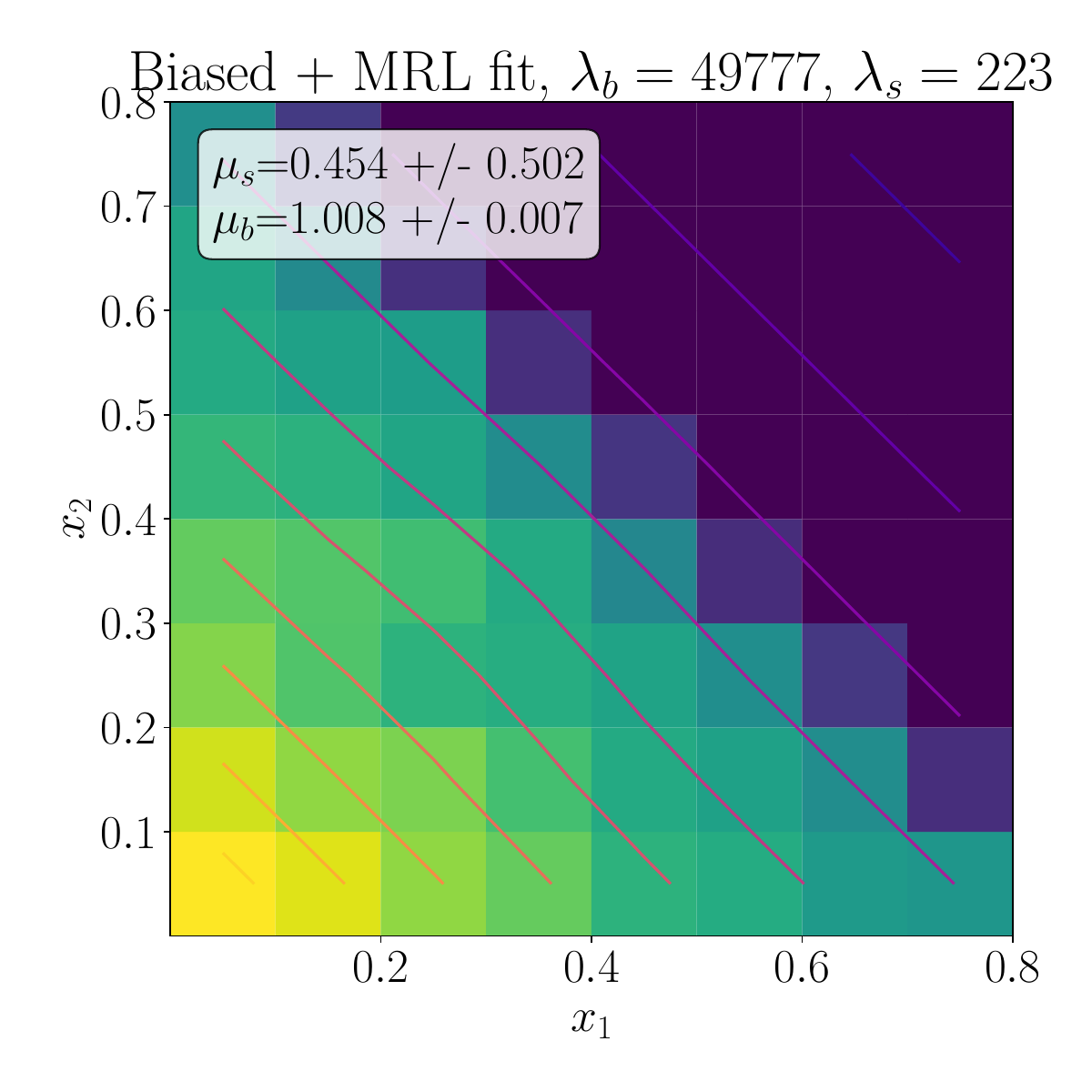}    
    \\
    \includegraphics[width=0.32\linewidth]{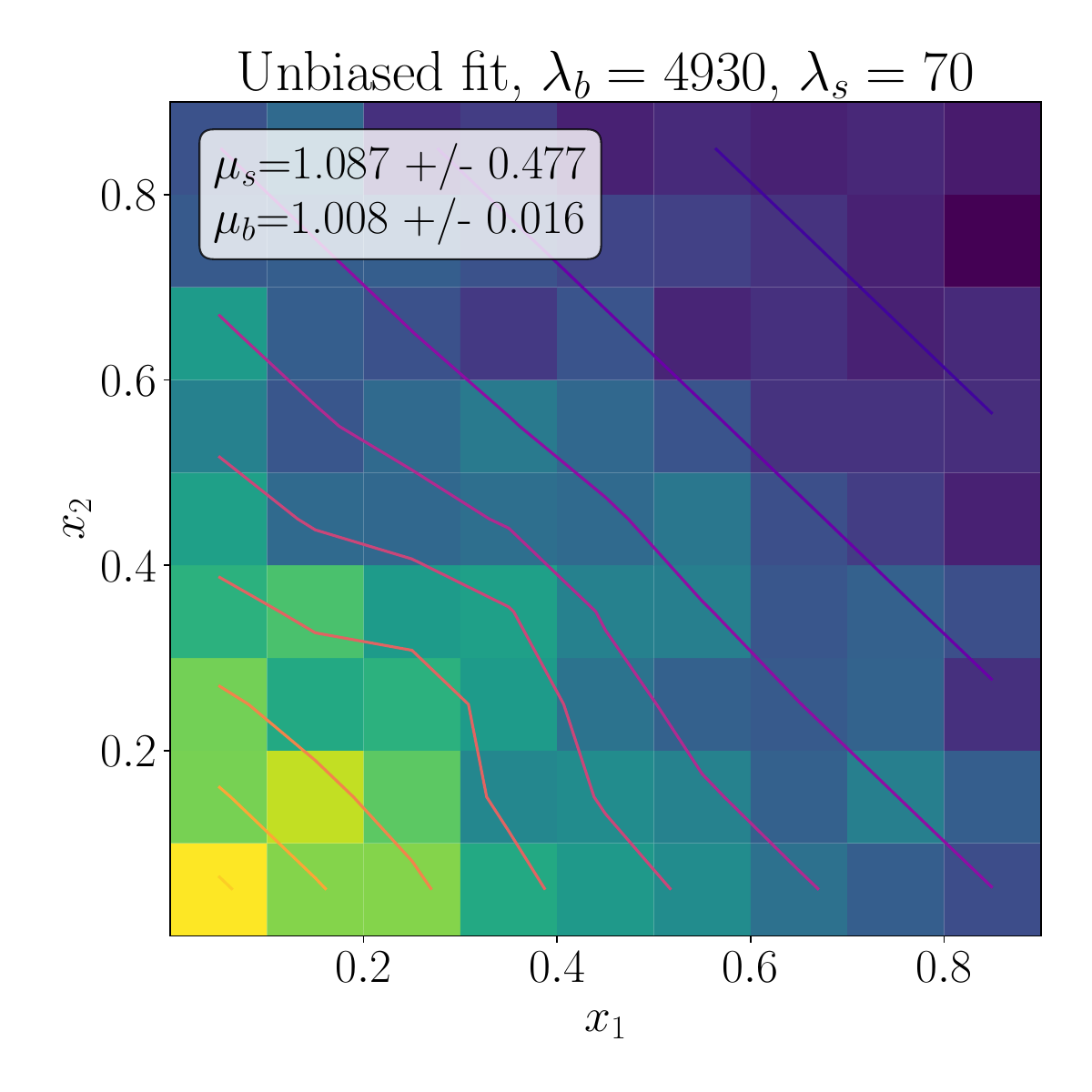}
    \includegraphics[width=0.32\linewidth]{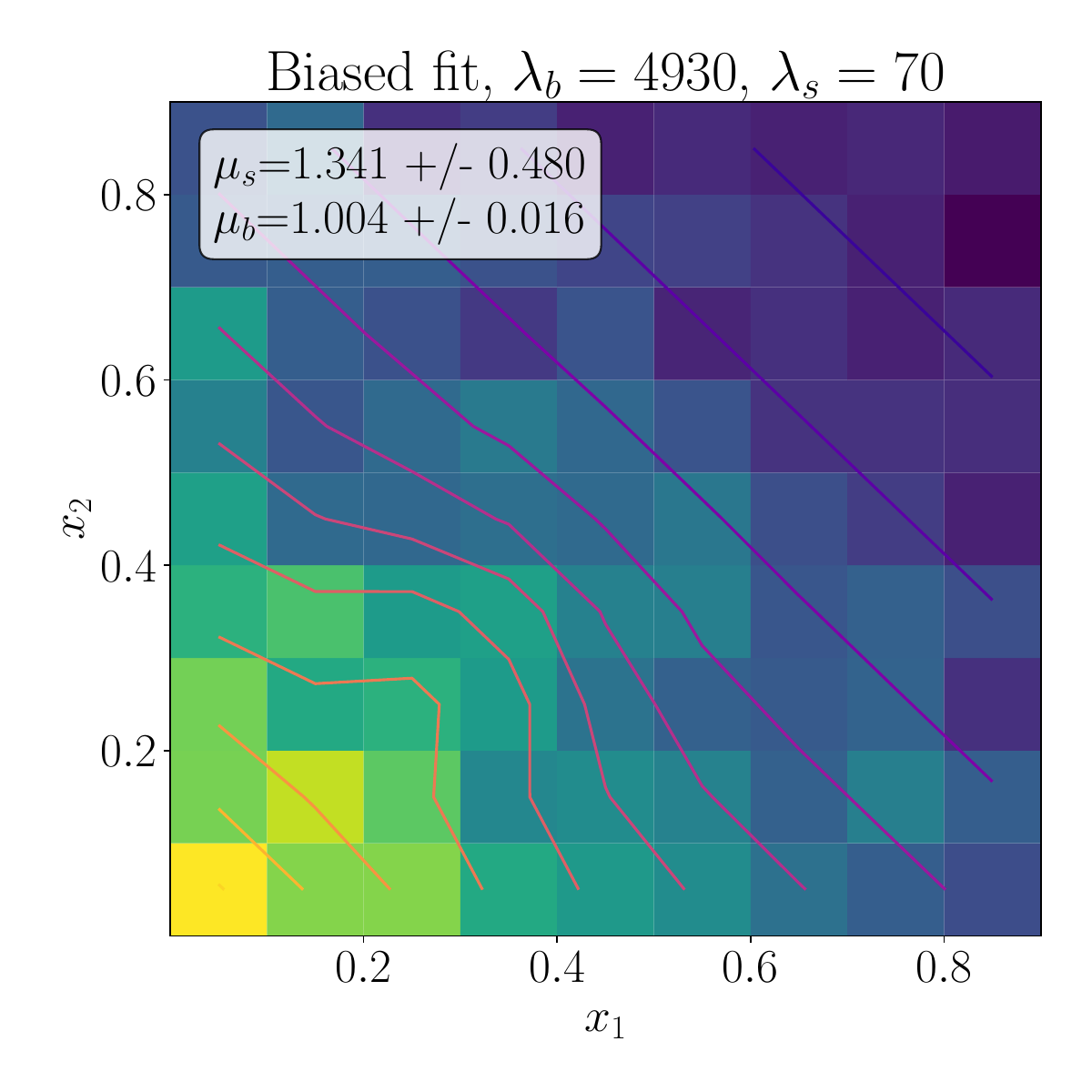}
    \includegraphics[width=0.32\linewidth]{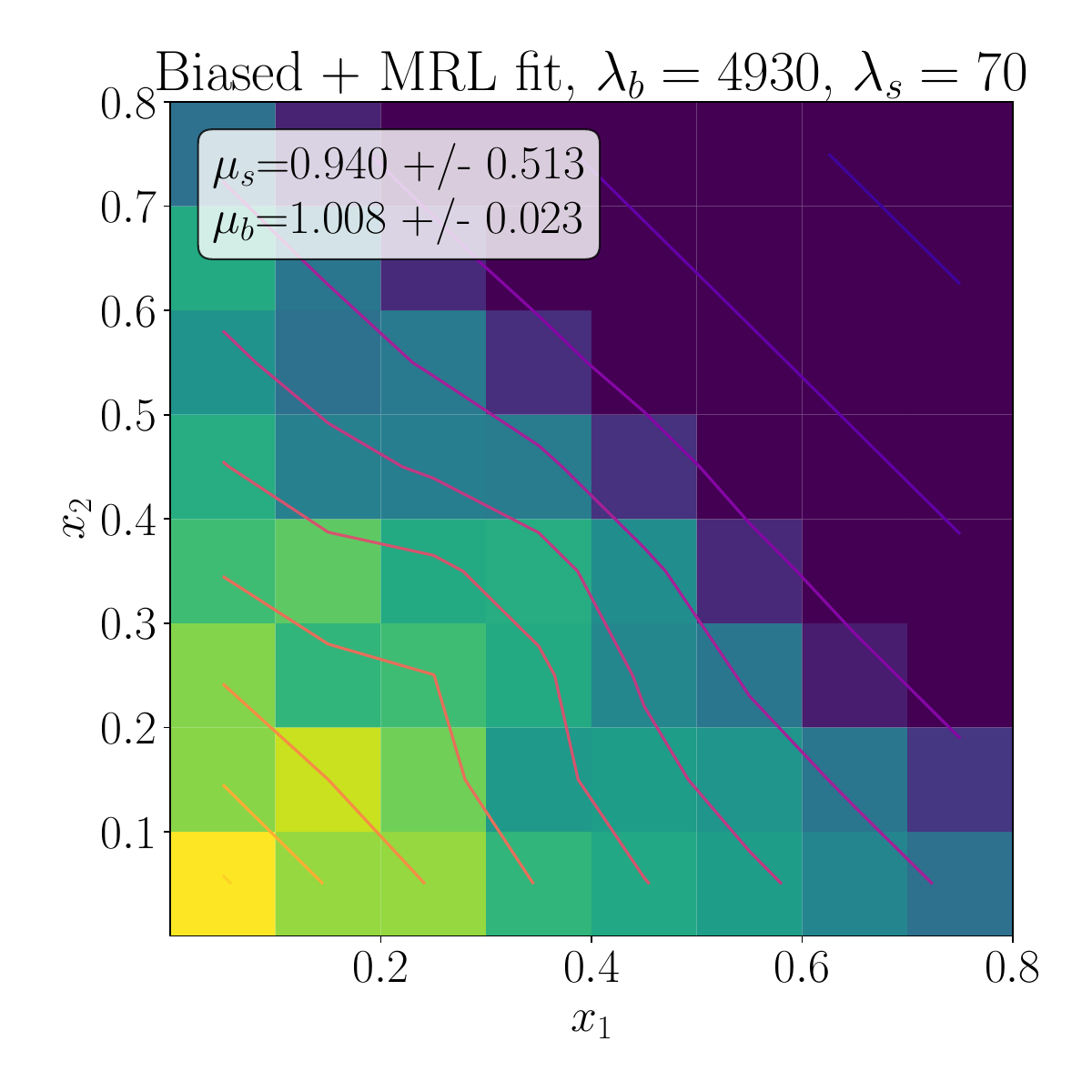}    
    \\
    \includegraphics[width=0.32\linewidth]{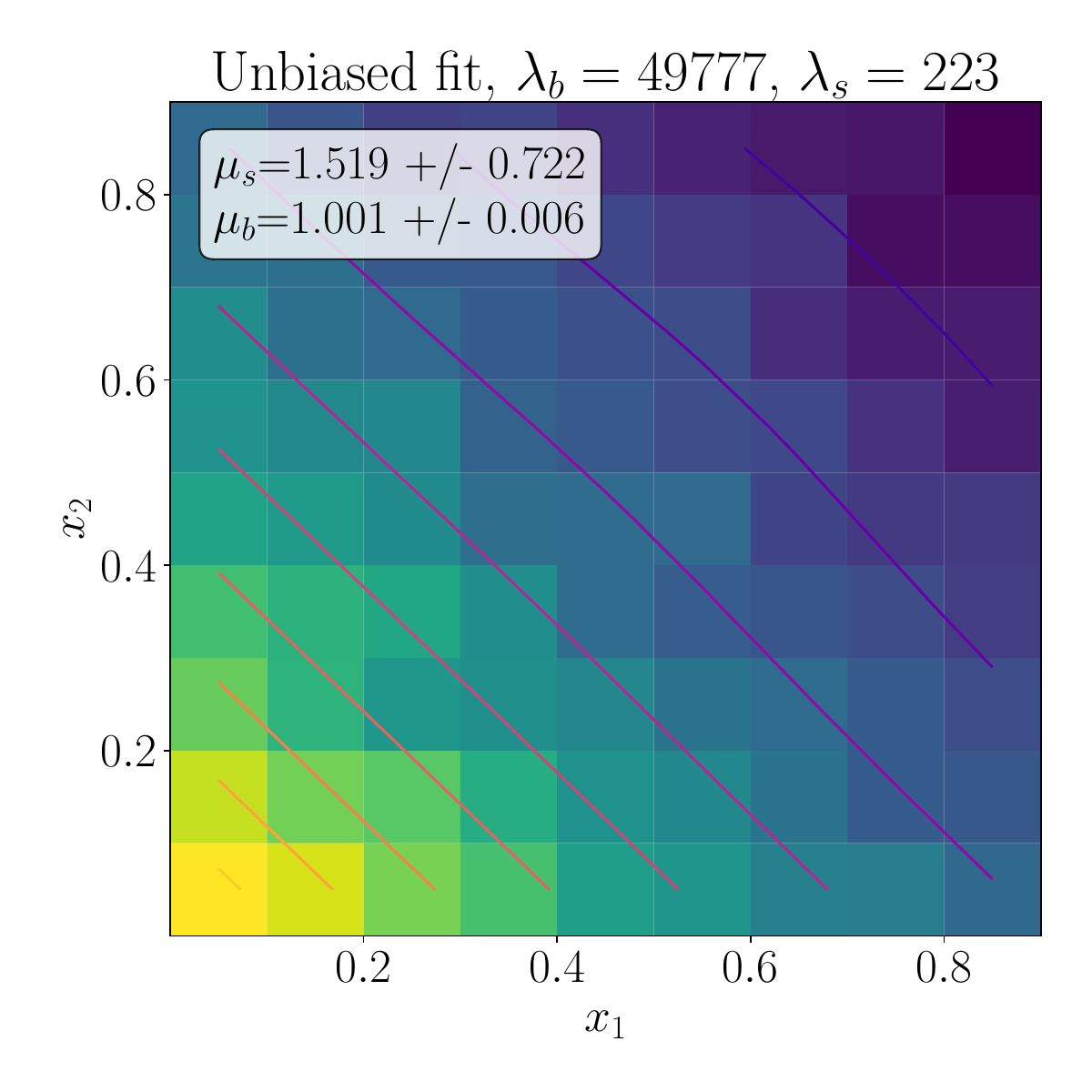}
    \includegraphics[width=0.32\linewidth]{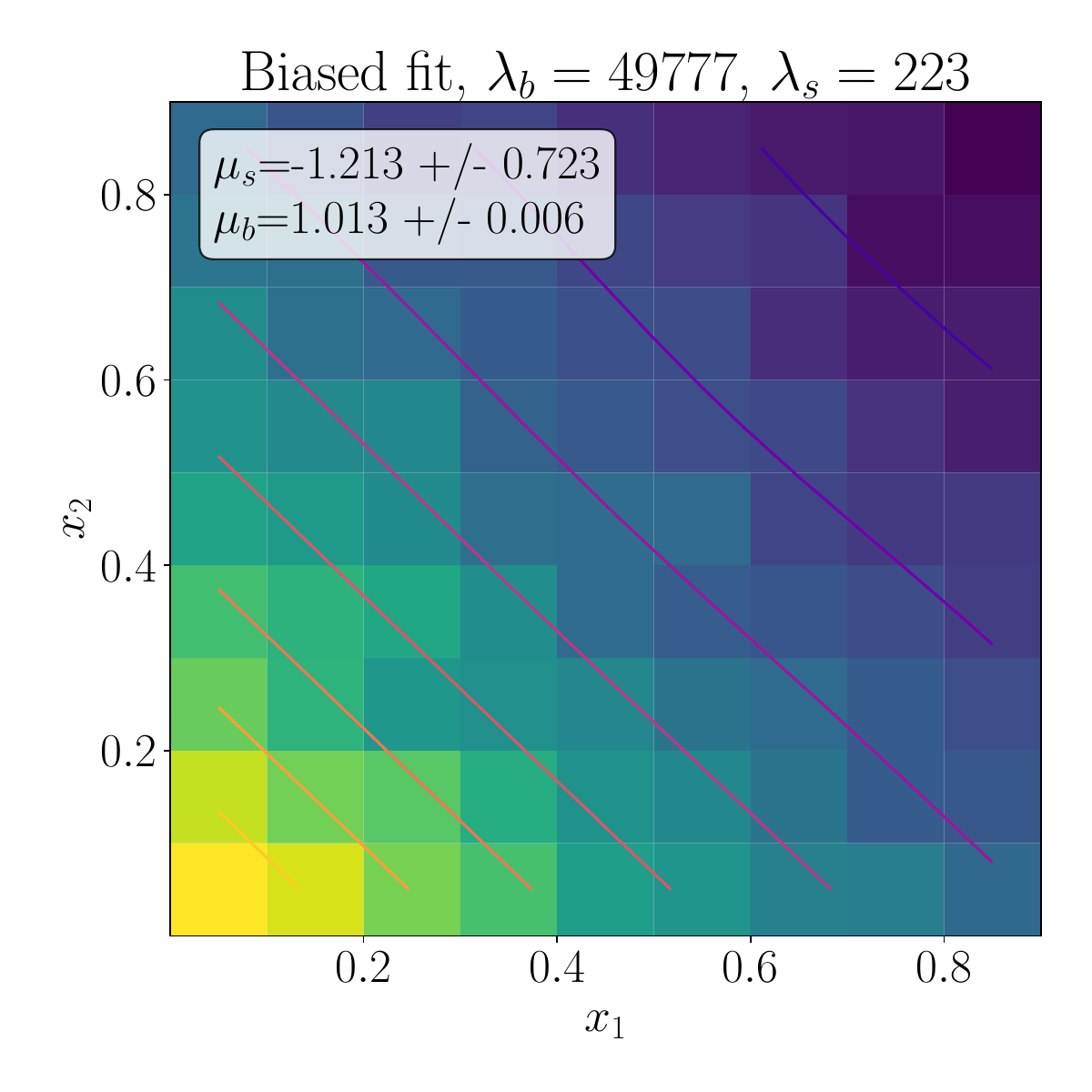}
    \includegraphics[width=0.32\linewidth]{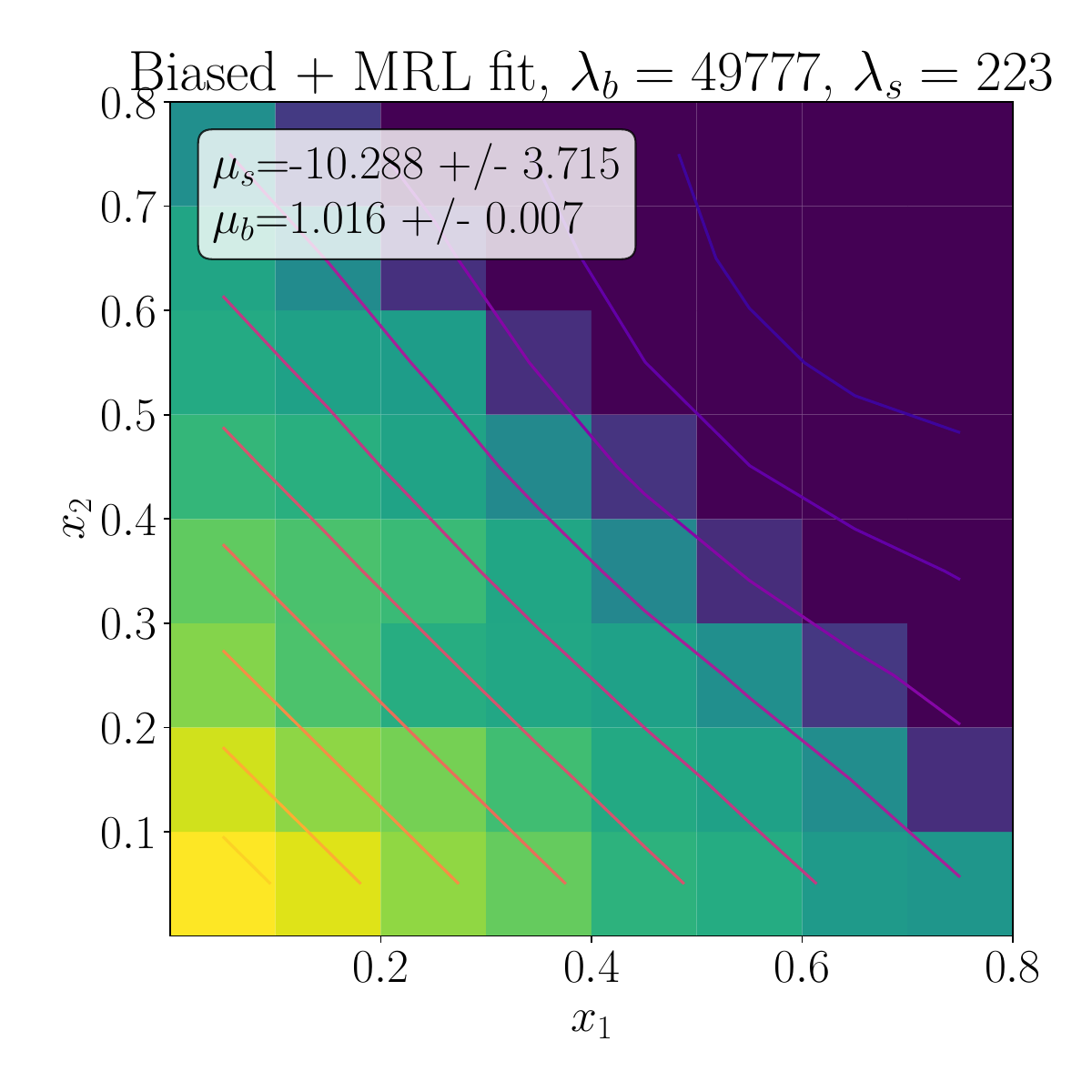}
    \caption{The three fits done on a single pseudo-experiment for the 2D example under the three different benchmarks. We show the result of the fit with the correct model, with the biased background model, and with the biased background model but on the Fiducial Signal Region defined with the \method method. The lines correspond to constant likelihood surfaces for the fitted values.}
    \label{fig:toy_results_exp_2d}
\end{figure}

\begin{figure}[h!]
    \centering
    \includegraphics[width=0.45\linewidth]{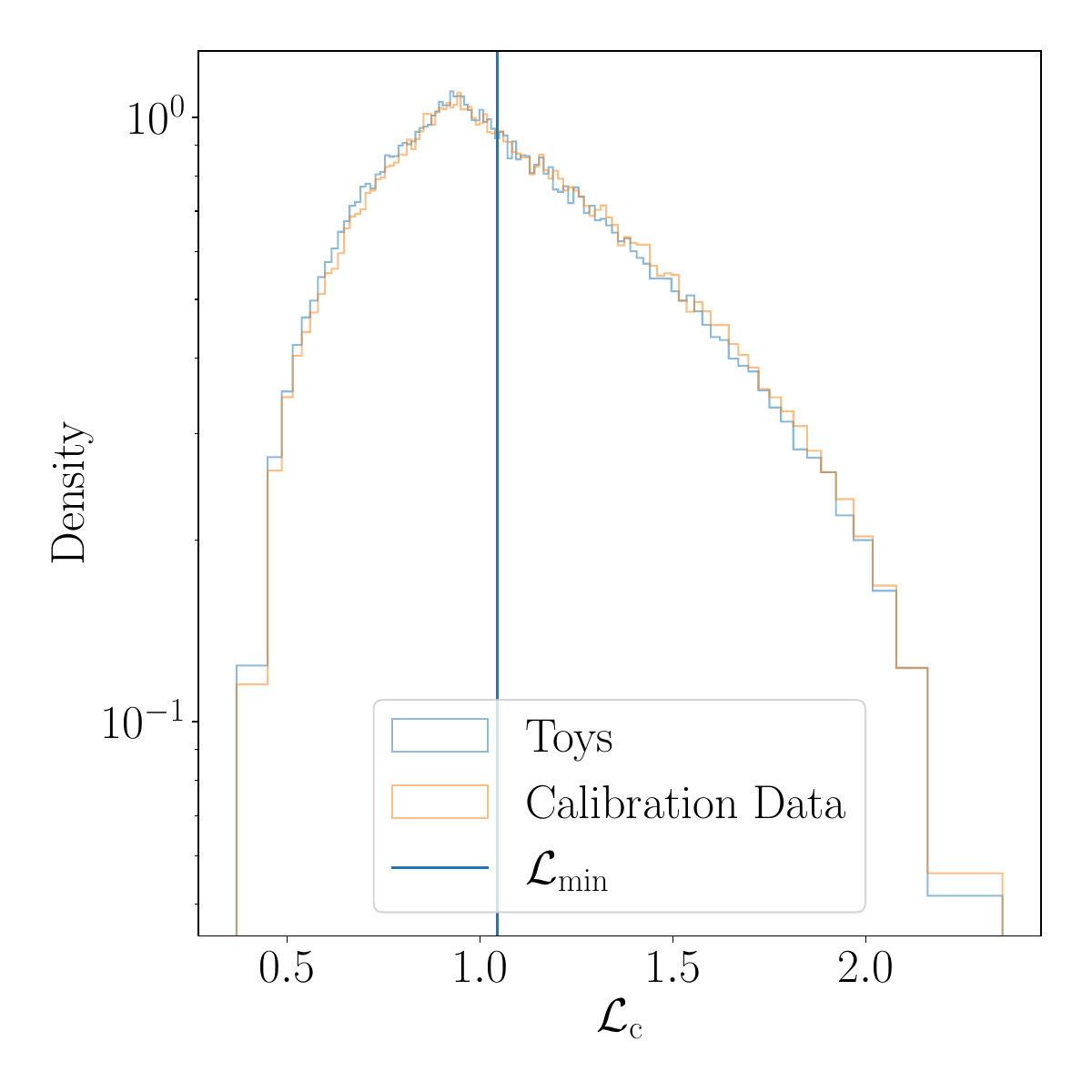}      
    \includegraphics[width=0.45\linewidth]{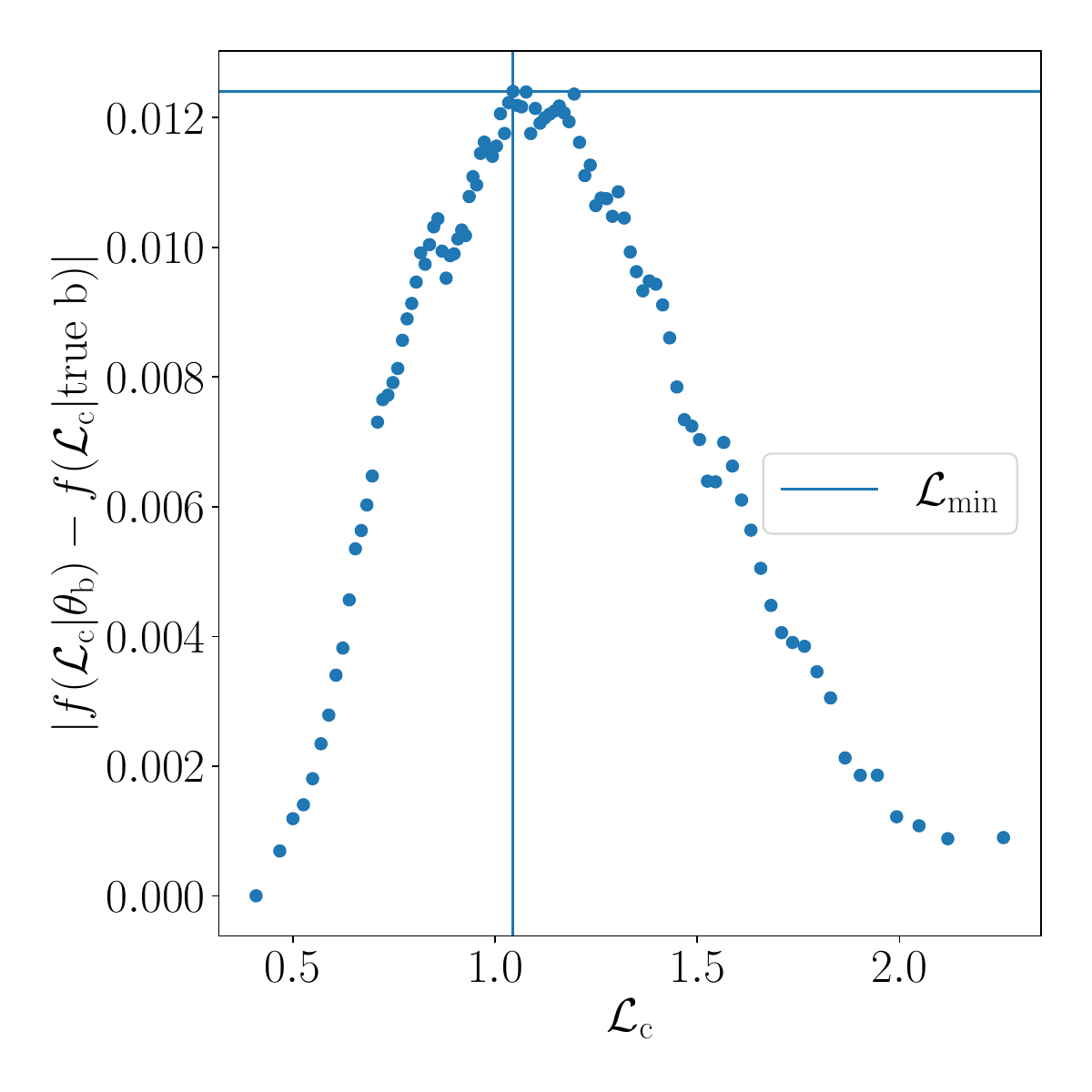}  
    \caption{$\likelihood{\text{min}}$ determination obtained by comparing a Calibration Region and a set of toys for the 2D Gaussian example. Left: The probability distribution of the model likelihood under the background model and the true background. Right: The fraction difference distribution as a function of the critical likelihood \likelihood{c} obtained from comparing the toy and Calibration datasets. We observe how $\likelihood{\text{min}}$ captures the crossing between background overestimation and background underestimation.}
    \label{fig:toy_distribution_comparison_exp_2d}
\end{figure}

We show the results for a single pseudo-experiment for each benchmark in Fig.~\ref{fig:toy_results_exp_2d}. For each fit, we show the constant likelihood lines for the fitted model. We observe how the Biased model degrades significantly with respect to the Unbiased fit. The Fiducial Signal Region, obtained via Fig.~\ref{fig:toy_distribution_comparison_exp_2d}, is almost identical for all three sets of parameters since the true background and background models are the same, and the Calibration Region and the generated toys have very large statistics. It also corresponds to a clipping of the background tails, albeit now defined in 2D.

\begin{figure}[h!]
    \centering
    \includegraphics[width=0.45\linewidth]{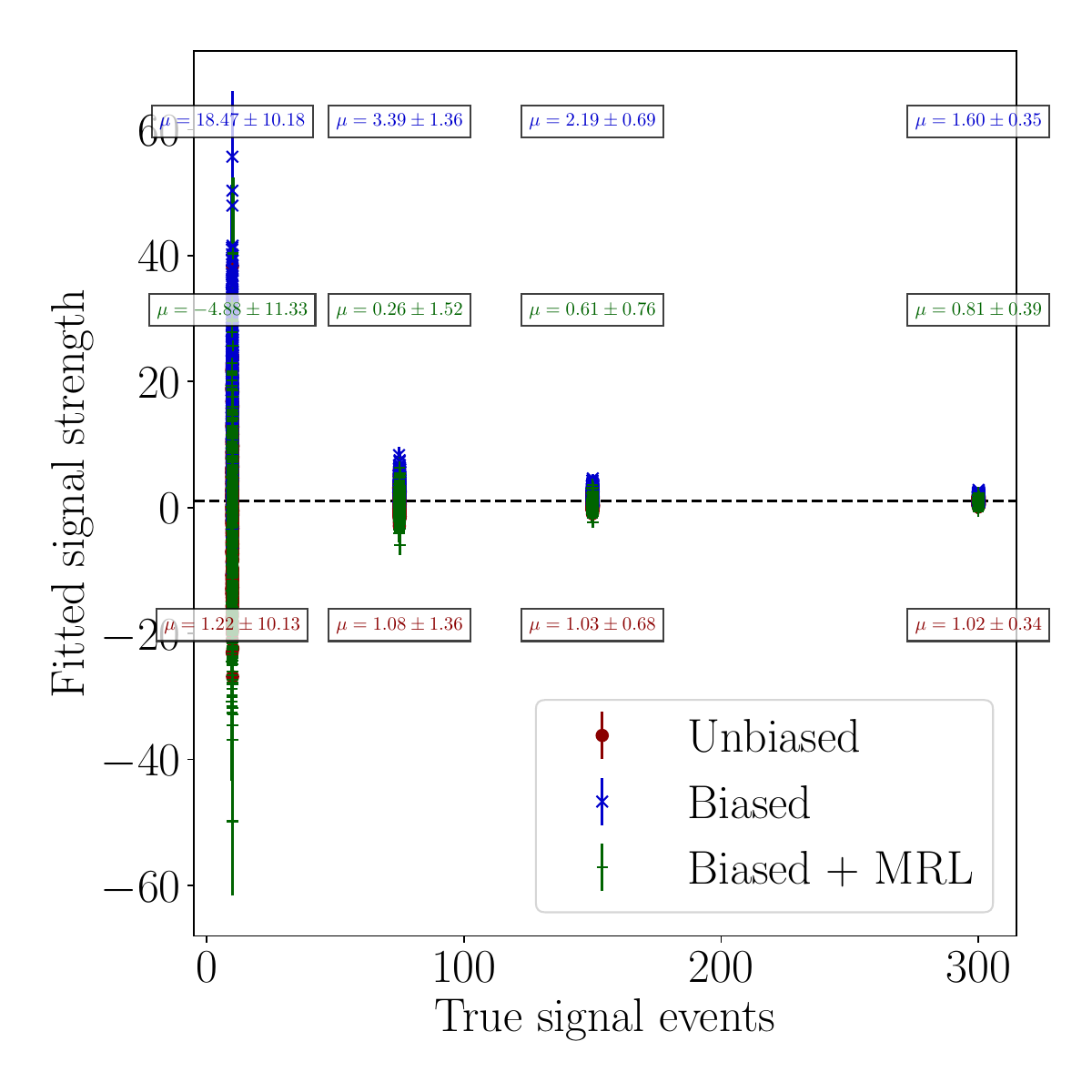}
    \includegraphics[width=0.45\linewidth]{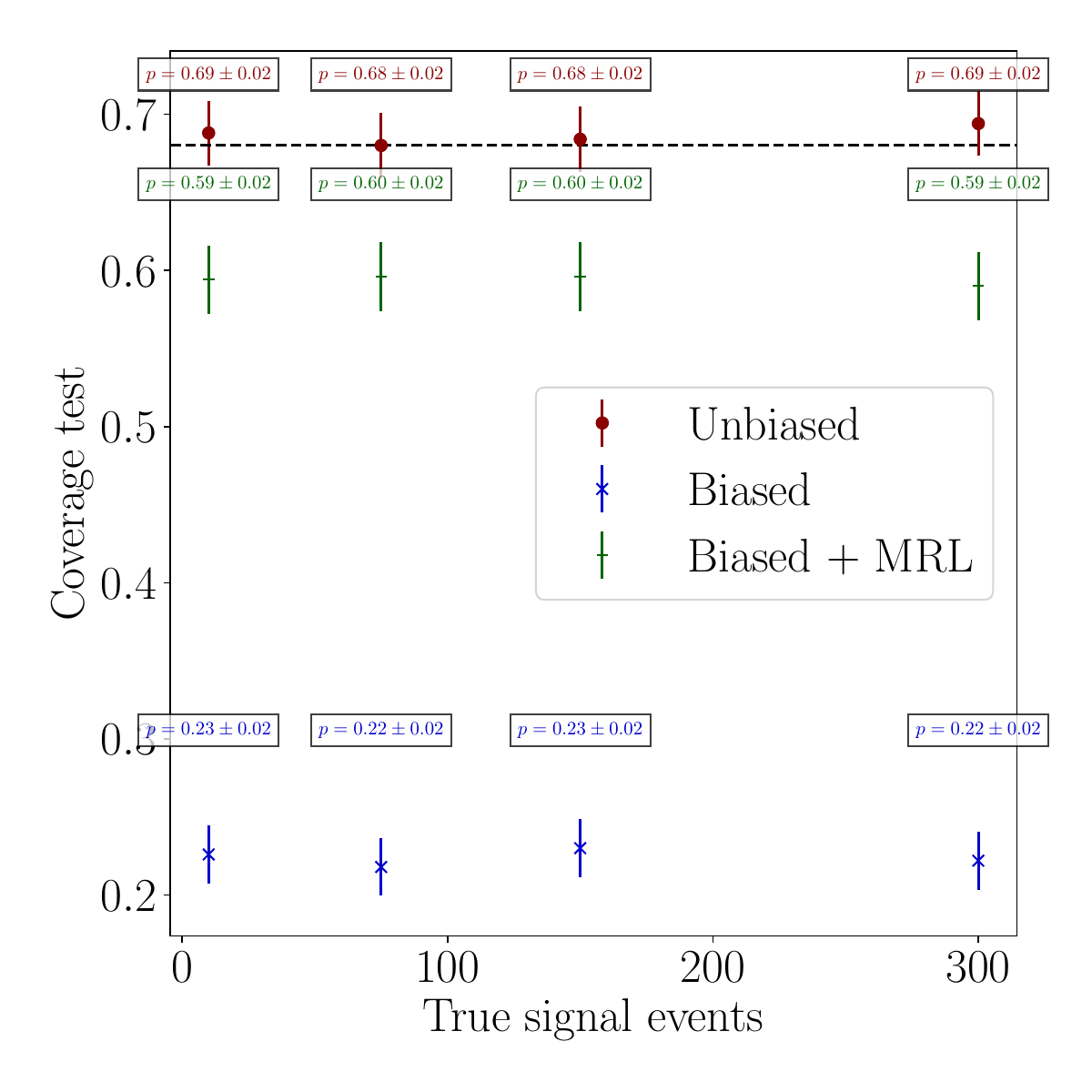}
    \\
    \includegraphics[width=0.45\linewidth]{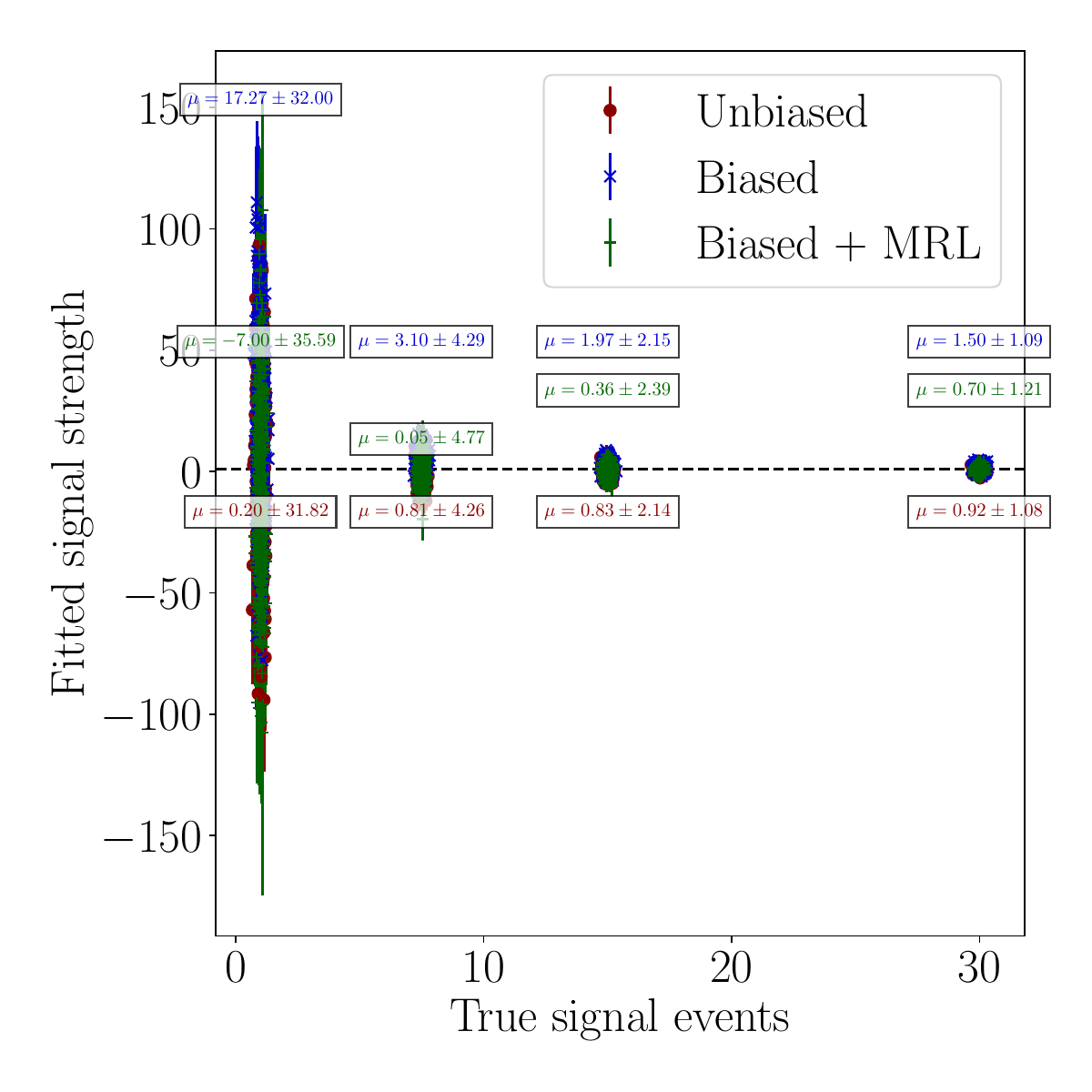}
    \includegraphics[width=0.45\linewidth]{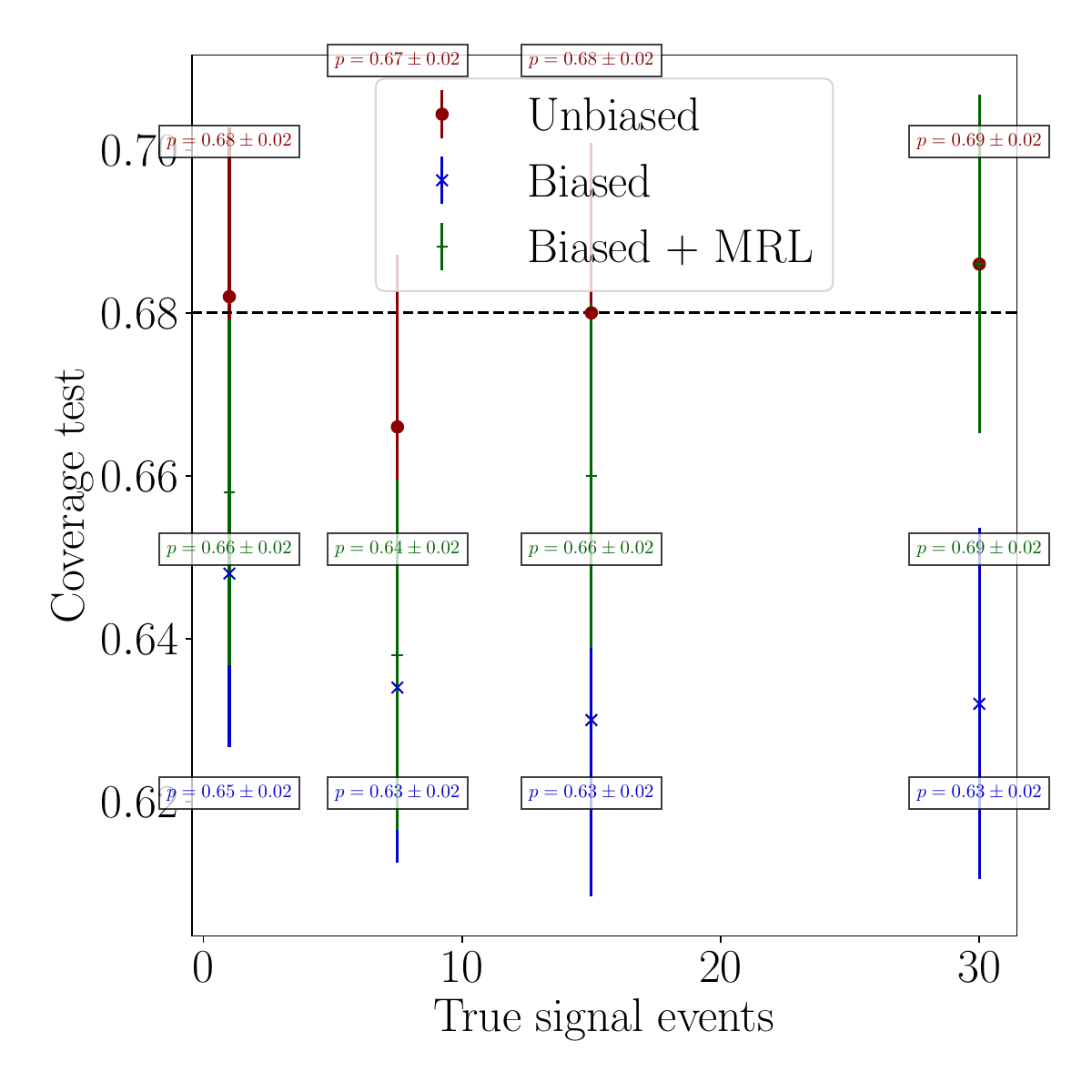}
    \\
    \includegraphics[width=0.45\linewidth]{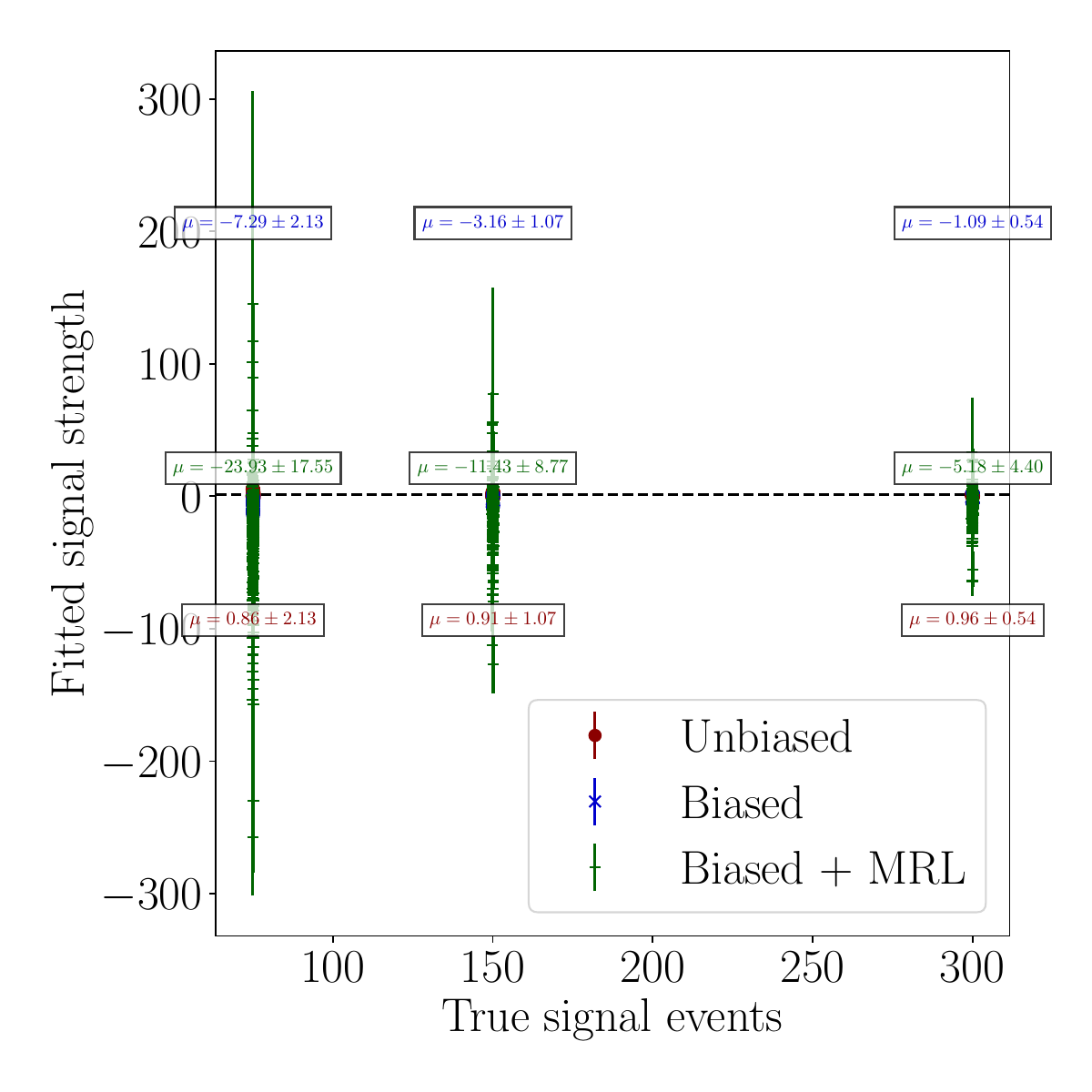}
    \includegraphics[width=0.45\linewidth]{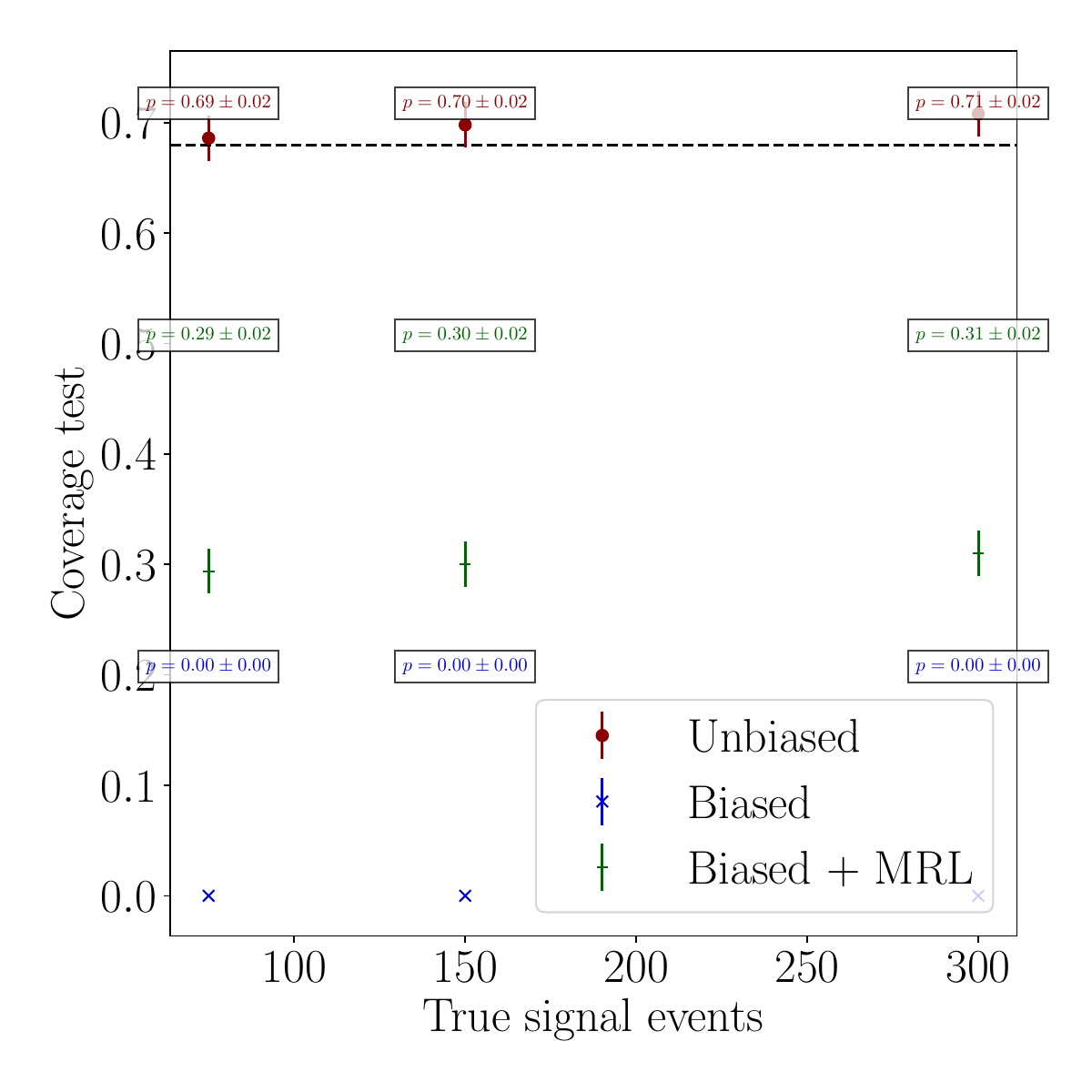}
    \caption{500 pseudo-experiment runs for different signal injection for the 2D example under the three benchmarks. Left:  Maximum Likelihood Estimates of the signal strength as a function of the true expected signal events, with uncertainties. Right: Coverage derived from the confidence interval as a function of the true expected signal events. 
    }
    \label{fig:toy_signal_injection_study_exp_2d}
\end{figure}

We perform the multiple pseudo-experiments for different signal injections in Fig.~\ref{fig:toy_signal_injection_study_exp_2d}. By inspecting the MLE and coverage for the nominal case, we observe how the bias is significantly but not completely corrected, as was the case for the 1D Gaussian nominal case. Moreover,the performance degrades as the injected signal gets smaller, with the MLE estimate increasing its deviation from one. For the low statistic example, both the MLE estimates and the coverage are consistent with the unbiased analysis across all signal injections (although the central values of the MLE do deviate from one at an albeit smaller degree and with larger errors than for the nominal case), signalling how the impact of the background bias and the resulting \method performance is dependent on the sample statistics.
The ``Unlikely'' example shows explicitly the dependence on the type of signal, since both the quality of the MLE and the coverage degrade significantly, with \method removing many signal events and thus underestimating the amount of signal present in the sample (in this case, even more than before the \method was applied), with all signal injections yielding estimates close to zero.

\section{A realistic example: \texttt{HI-SIGMA}}
\label{sec:hi-sigma}

As a realistic application, we consider the \texttt{HI-SIGMA} strategy introduced in Ref.~\cite{Amram:2025vqw} and study a slightly simplified version of it applied to the same di-Higgs test dataset. \texttt{HI-SIGMA} is an ideal testing case because we have unbinned, multi-dimensional background and signal models that can both be evaluated and also used to generate datasets. Moreover, while Ref.~\cite{Amram:2025vqw} showed how for large enough signal injections using \texttt{HI-SIGMA} yields a similar performance to state-of-the art ML-analyses with the benefit of providing a data-driven background model, it also showed that for low signal injections there is a remaining bias due to background mismodelling. In \cref{subsec:hisigma_details}, we detail the relevant features of \texttt{HI-SIGMA} and the test dataset, and in \cref{subsec:hisigma_results} we present the results of applying the \method method.

\subsection{The di-Higgs dataset and \texttt{HI-SIGMA}}
\label{subsec:hisigma_details}

In this section, we give a brief introduction to the di-Higgs dataset and the \texttt{HI-SIGMA} technique for estimating the number of di-Higg events in a non-resonant background. To keep the discussion focused, we concentrate on the relevant details and we refer the reader to Ref.~\cite{Amram:2025vqw} for a more complete description.

The di-Higgs dataset was used in Ref.~\cite{Amram:2025vqw} to showcase the power of \texttt{HI-SIGMA}. It was selected because of the high importance of di-Higgs measurement in present and future colliders~\cite{Dawson:2022zbb} and because the $bb\gamma\gamma$ final state, although very clean due to the diphoton mass resolution, has the intrinsic difficulty of searching for a small signal hidden among data-driven backgrounds, which requires a careful unbinned analysis. The dataset itself consists of synthetic events obtained by simulating both $hh\to bb\gamma\gamma$ and its main irreducible non-resonant background with \verb|MadGraph_aMC@NLO|~\cite{Alwall:2014hca} v3.5.7, \verb|Pythia8|~\cite{Sjostrand:2014zea,Bierlich:2022pfr} and \verb|Delphes|~\cite{deFavereau:2013fsa}. We select events with 2 $b$-tagged jets and two isolated photons which are grouped into two Higgs candidates. The event is then reduced to five features
\begin{equation}
   x\equiv\{m_{\gamma\gamma},p_{T}^{bb},p_{T}^{\gamma\gamma}/m_{\gamma \gamma}, \Delta R_{bb}, \Delta R_{\gamma\gamma}\}\,.
\end{equation}
Due to the good photon momentum resolution, the $m_{\gamma \gamma}$ shows a very sharp resonance for the signal on top of a smoothly decaying, non-resonant background. Thus, we use it to define a Signal Region by selecting events where $m_{\gamma\gamma}\in[90,180)\text{GeV}$. 

Although $m_{\gamma\gamma}$ already provides good discriminatory power between signal and background, and we can model its distribution for each process successfully with known parametric functions, the smallness of the di-Higgs signal motivates us to use the additional four features, denoted as $\vec{x}'$, in an unbinned fit. However, these distributions cannot be modelled using known parametric distributions and we need to estimate them using Machine Learning-based techniques. Moreover, the background simulations are not sufficiently accurate to model the true background and thus its distribution estimation needs to be data-driven.

To do the data-driven fits, \texttt{HI-SIGMA} splits the Signal Region into two disjoint subsets, the ``training'' and ``testing'' datasets.\footnote{In a more involved implementation, cross-validation can be implemented to better use all available data, as was done in Ref.~\cite{Amram:2025vqw}.} The training dataset will be used to infer the data-driven background distribution, and the testing dataset to infer the number of di-Higgs signal events. The two datasets are further subdivided in central and sideband regions,
\begin{equation}
    \label{eq:sr_sb_def}
    \begin{split}
        \text{central} &: m_{\gamma\gamma} \in [115, 135)\text{ GeV}\\
        \text{sideband (SB)} &: m_{\gamma\gamma} \in [90,115)\cup [135,180)\text{ GeV}\\
    \end{split}
\end{equation}
Although we expect the di-Higgs signal to be concentrated in the central region, we consider the complete central and sidebands in the statistical fit performed on the test dataset in order to better constrain the background mass distribution. We note that the definition of these regions, and of the signal region in general, is arbitrary, and in a realistic analysis a proper assesment of the impact of all definitions should be pursued.

To perform \method, we need a Calibration Region to determine the minimum likelihood \likelihood{\text{min}}. In this work, the Calibration Region is obtained by further splitting the testing dataset into a two disjoint subsets. Although it reduces the size of the final testing dataset, this allows us to avoid any biases in its background distribution, and we take advantage of the synthetic nature of the dataset to remove all signal events. In a more realistic application, the Calibration Region can be defined using the same mass variables but implementing a complementary veto that separates it from the true Signal Region. In the di-Higgs example, this could be a ``0 $b$-jets'' requirement. Care should be taken, however, since this would require us to account for the correlation between flavor-tagger and kinematics and for the presence of different resonant processes. The former could be corrected with a transfer function, such as that implemented in Ref.~\cite{ATLAS:2018rnh} and trained at a different mass region, which ensures that the kinematics are approximately morphed between regions at the expense of additional uncertainties. Regarding the difference in resonant processes, one should remember that the \method method is introduced to calibrate the data-driven estimation of the non-resonant backgrounds, mainly $bb \gamma\gamma$, $cc \gamma\gamma$ and $jj \gamma\gamma$. Thus, the presence of other resonant backgrounds such as $ZH$ and $ttH$ for the Signal Region and VBF and $(H \to \gamma\gamma) jj$ for the ``0 $b$-jets'' Calibration Region, which is neglected in this work, could be accounted for explicitly using simulation-trained probability densities, as we do for the signal. This, of course, assumes that simulations can be trusted to a higher degree for all resonant backgrounds.

To model the kinematic distributions, in this work  we consider a slightly simplified version of \texttt{HI-SIGMA} where we use a single conditional normalizing flow (cNF) without ensembling both for signal and background. The signal cNF $p_{s}(\vec{x}'|m_{\gamma\gamma},\theta_s)$ is trained on a large set of simulated events, while the background cNF $p_{b}(\vec{x}'|m_{\gamma\gamma},\theta_b)$ is trained on the sideband region of the training data.  
For the mass distributions, we use a double Crystal Ball function \cite{Oreglia:1980cs,Gaiser:1982yw} for the signal, and fit it to the same signal simulated events used to estimate $p_{s}(\vec{x}'|m_{\gamma\gamma},\theta_s)$. For the background shape we use an exponential distribution of the form
\begin{equation}
    p_b(m_{\gamma\gamma}|\theta_{b}) = p_0 e^{-p_1\bar{m} + p_2\bar{m}^2}\,,
\end{equation}
where the $p_i$ are floating free parameters, and $\bar{m}$ is just a rescaled version of $m_{\gamma\gamma}$, scaled to be in the range [0,1]. We fit the free parameters in the Calibration Region and treat them as nuisance parameters in the final fit on the testing data, and add an additional, learnable background normalization parameter. We define nominal parameters, instead of fitting them directly in the testing data as in Ref.~\cite{Amram:2025vqw}, because it allows us to define a full likelihood 
\begin{equation}
    p_{b}(x|\theta^{\text{nom.}}_{b})=p_{b}(m_{\gamma\gamma}|\theta^{\text{nom.}}_{b})p_{b}(\vec{x}'|m_{\gamma\gamma},\theta^{\text{nom.}}_{b})\,,\nonumber
\end{equation}
from which to sample toys and obtain $\likelihood{\text{min}}$ from the Calibration Region. In this work, the toy dataset, the calibration region and the testing datasets are approximately of the same size,  $\mathcal{O}(5\times 10^{4})$ events.

\subsection{Results}
\label{subsec:hisigma_results}

\begin{figure}[h!]
    \centering
    \includegraphics[width=0.45\linewidth]{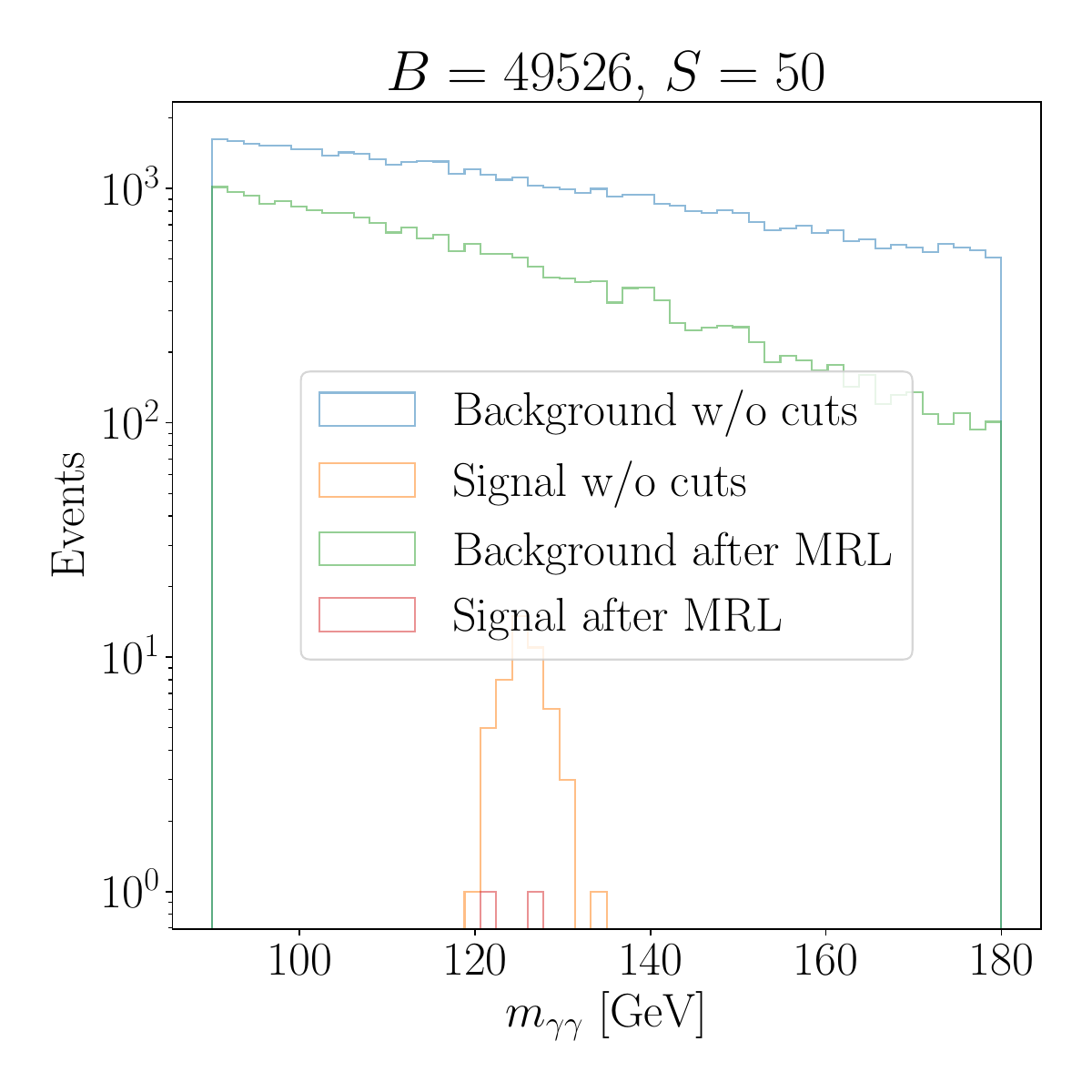}
    \includegraphics[width=0.45\linewidth]{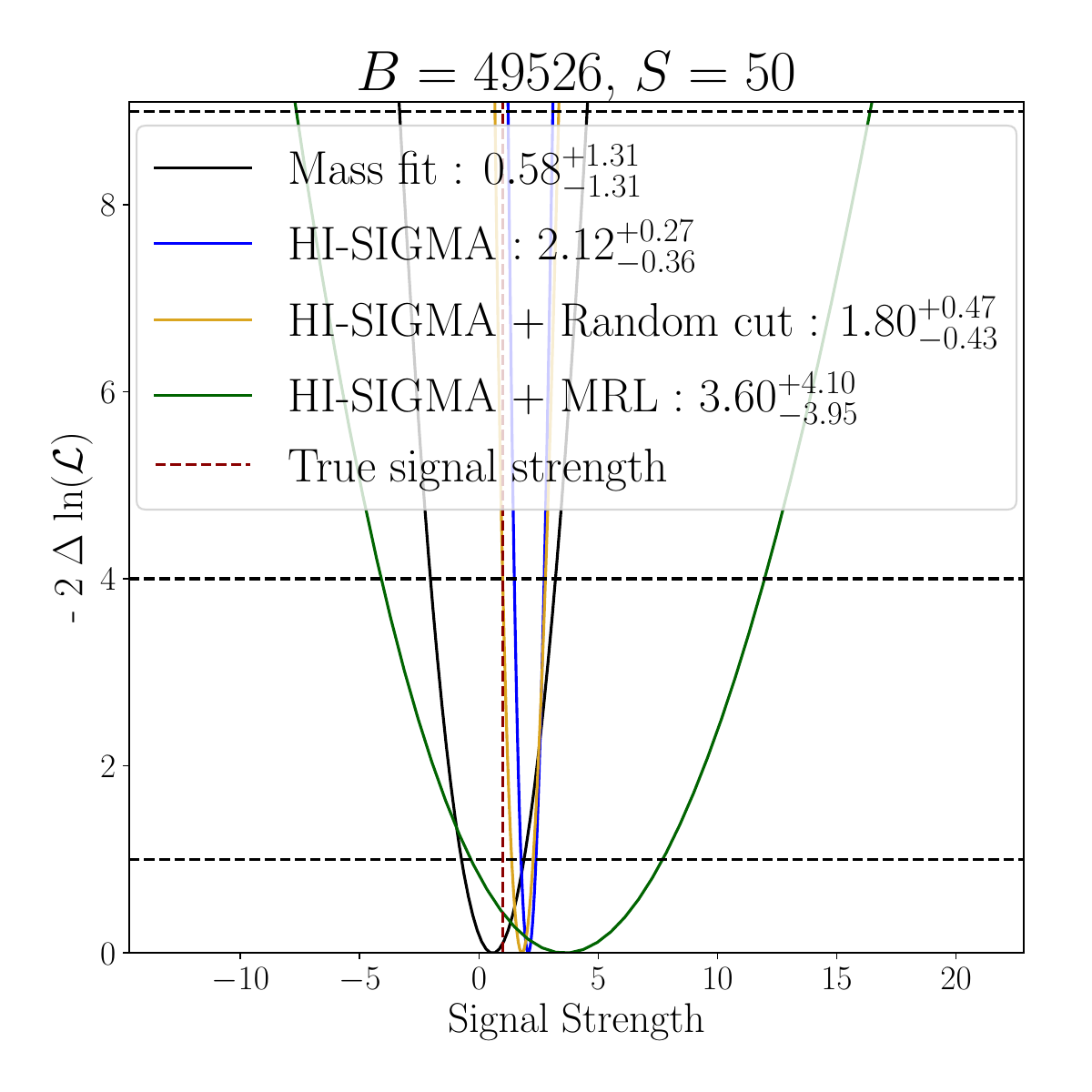}\\
    \includegraphics[width=0.45\linewidth]{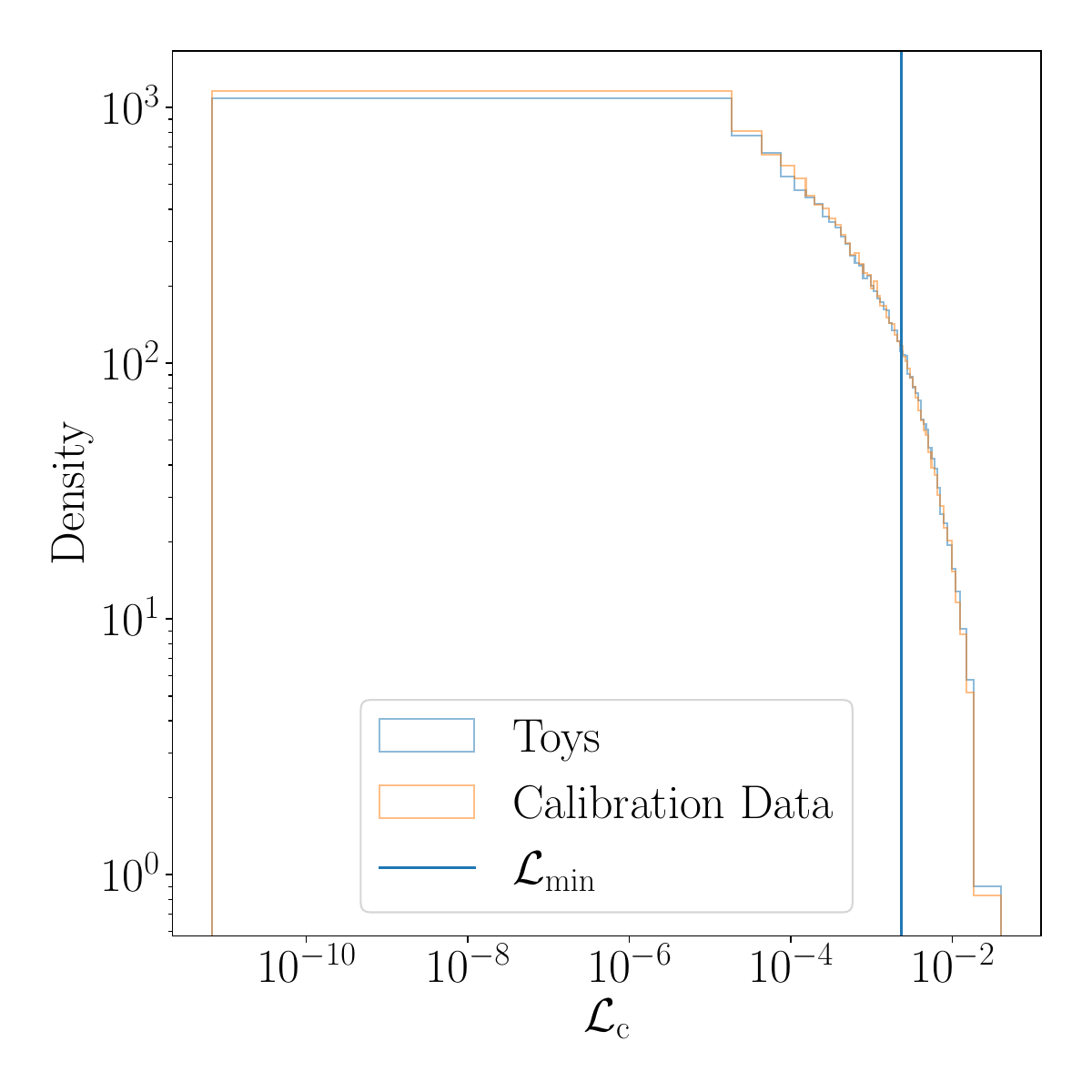}
    \includegraphics[width=0.45\linewidth]{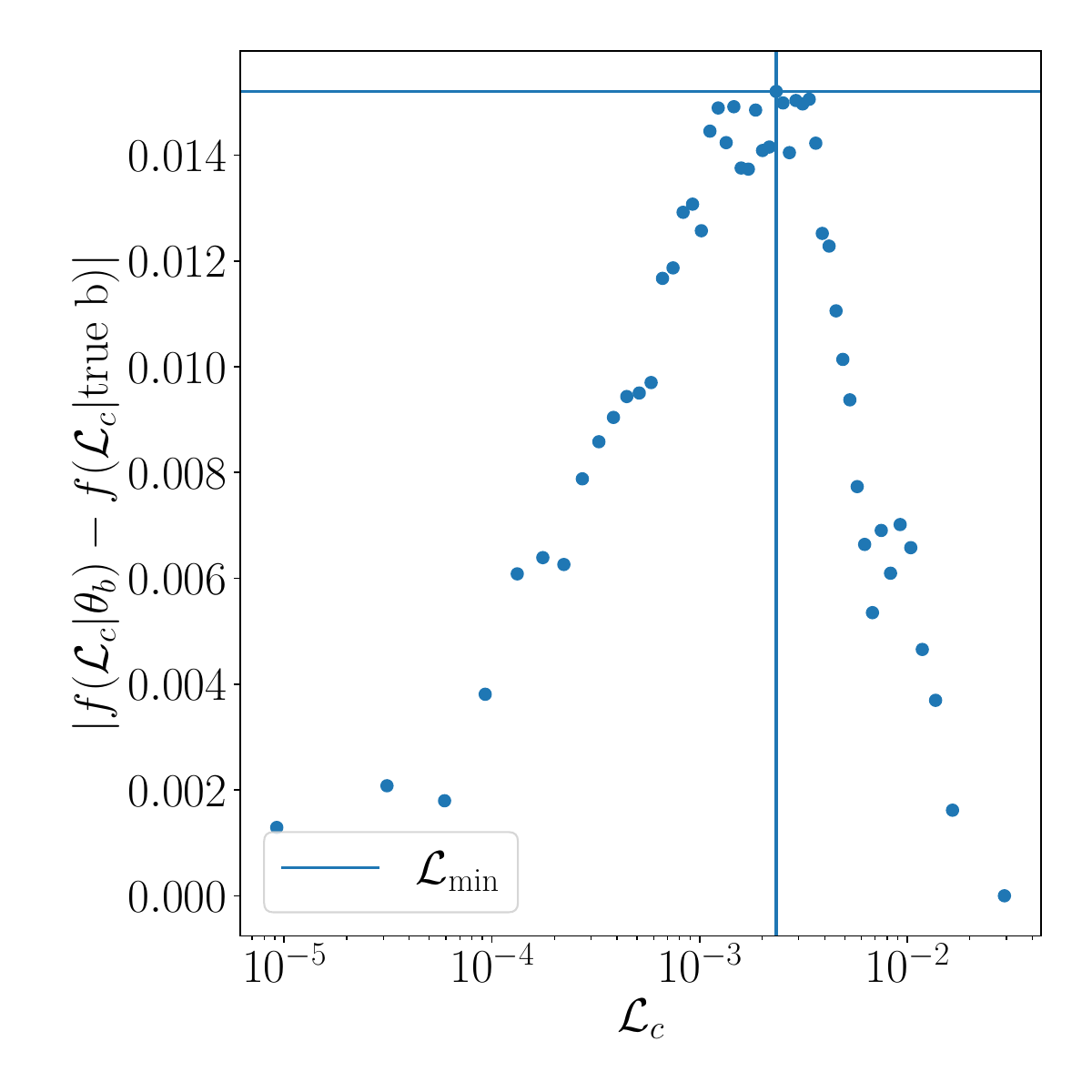}
    \caption{A single run of \texttt{HI-SIGMA+\method}, where $p(\vec{x}|m\in\text{SB})$ has not been masked. The figures show, in clockwise order: the $m_{\gamma\gamma}$ distribution for the Signal Region and when applying the Fiducial Signal Region cuts, the signal strength test statistic landscape using four diffferent strategies, the fraction difference distribution as a function of the critical likelihood \likelihood{c} obtained from comparing the toy and Calibration datasets, and the probability distribution of the model likelihood under the background model and the true background. We observe how $\likelihood{\text{min}}$ produces a very conservative fit where most signal events are lost and with very large uncertainties.}
    \label{fig:hi_sigma_example_no_mask}
\end{figure}

\begin{figure}[h!]
    \centering
    \includegraphics[width=0.45\linewidth]{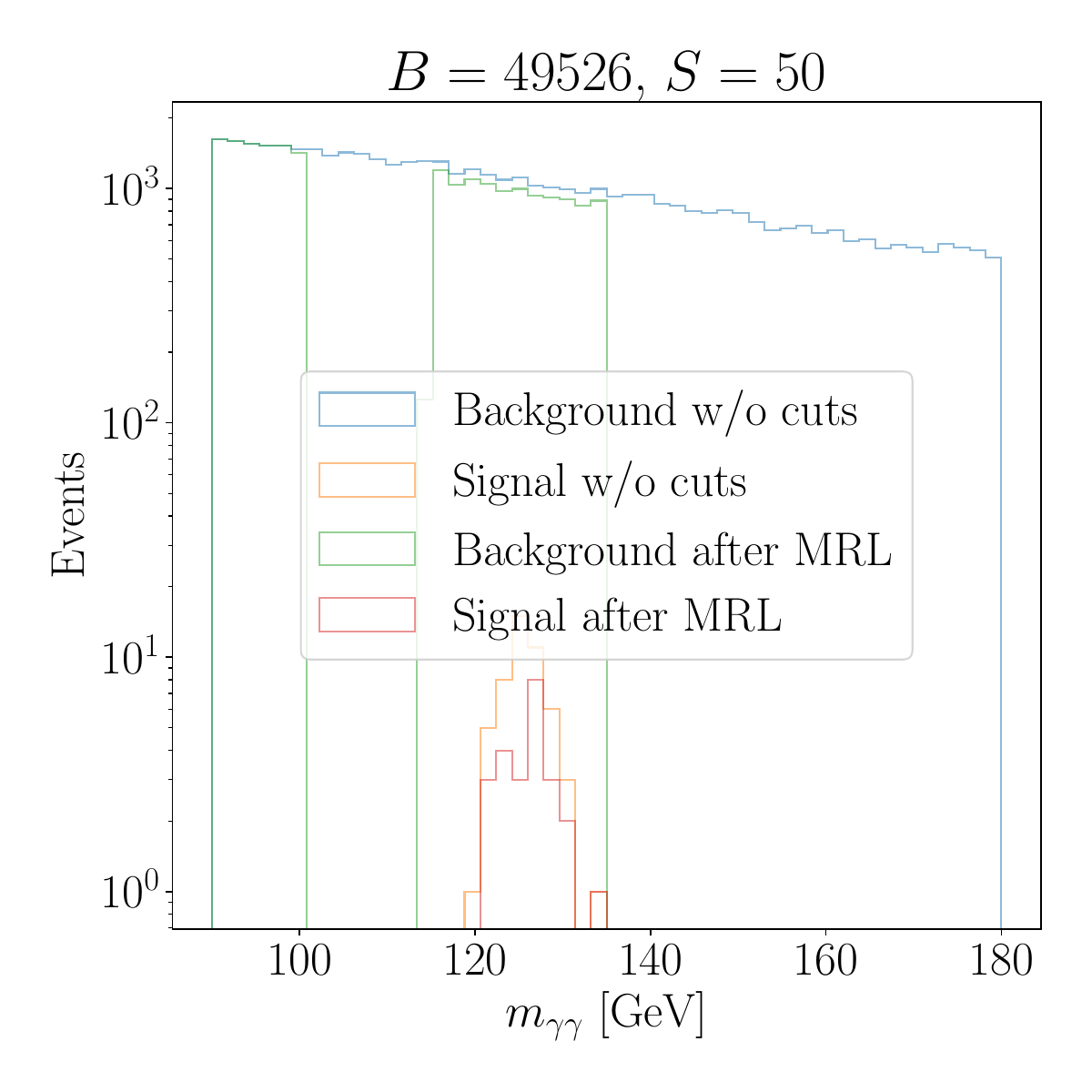}
    \includegraphics[width=0.45\linewidth]{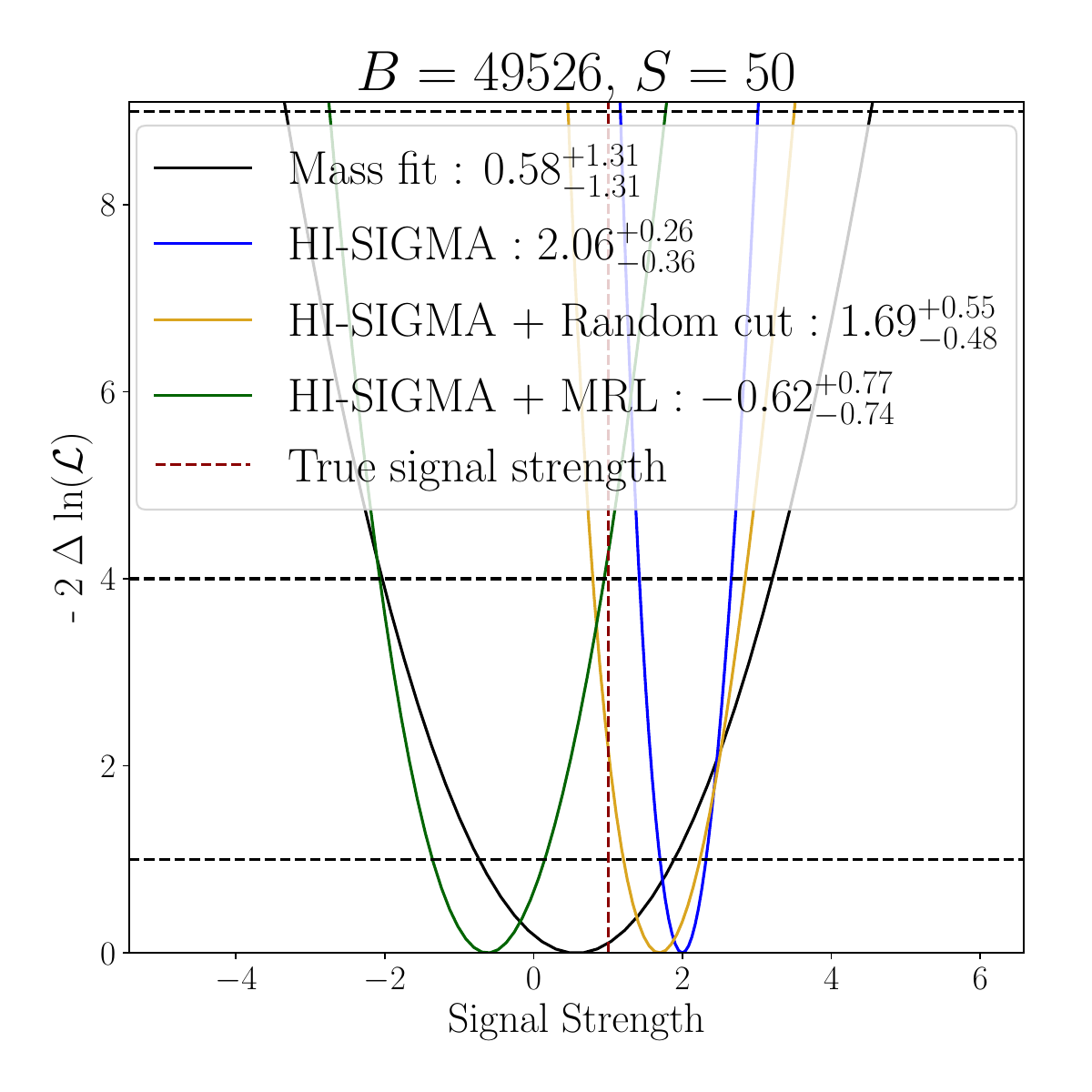}\\
    \includegraphics[width=0.45\linewidth]{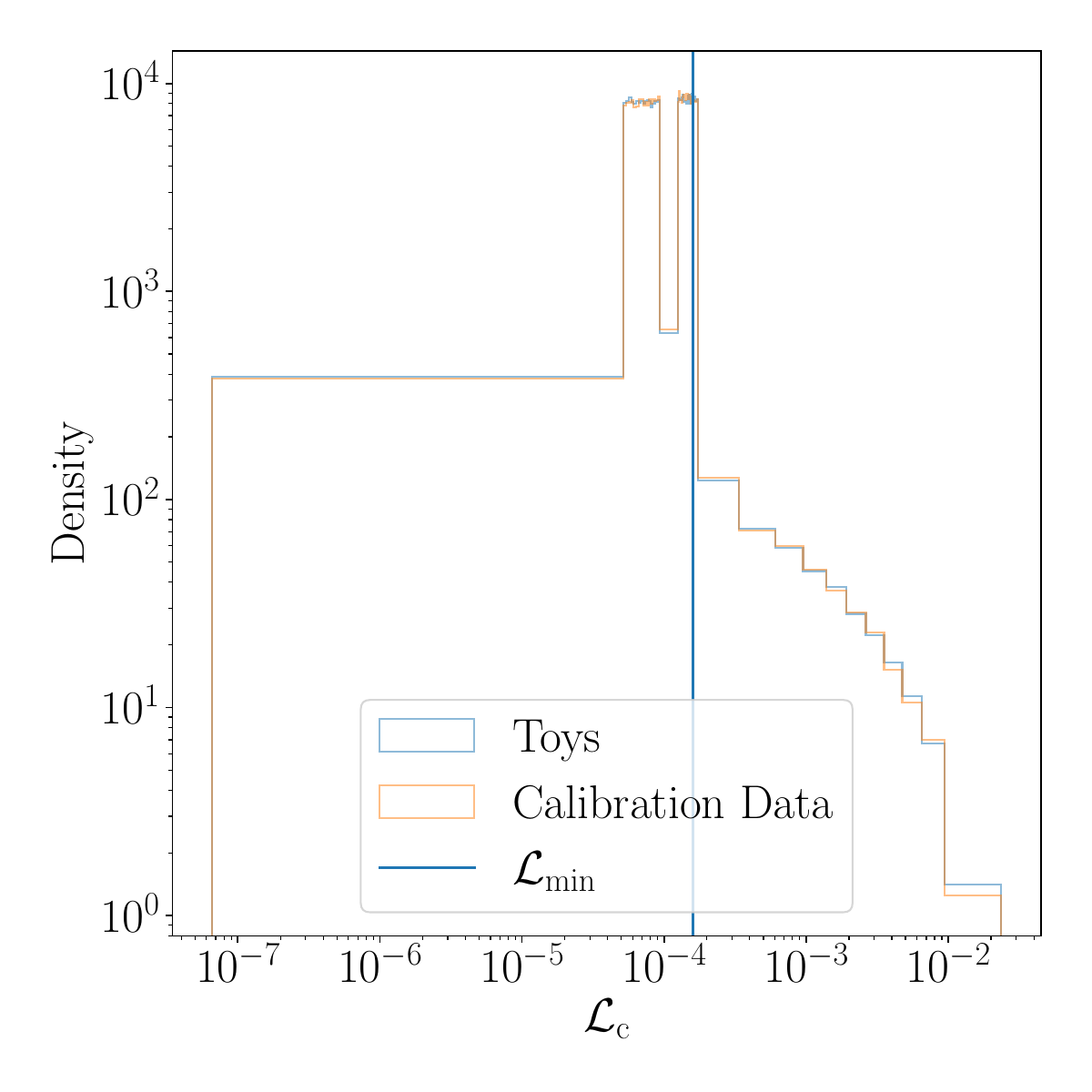}
    \includegraphics[width=0.45\linewidth]{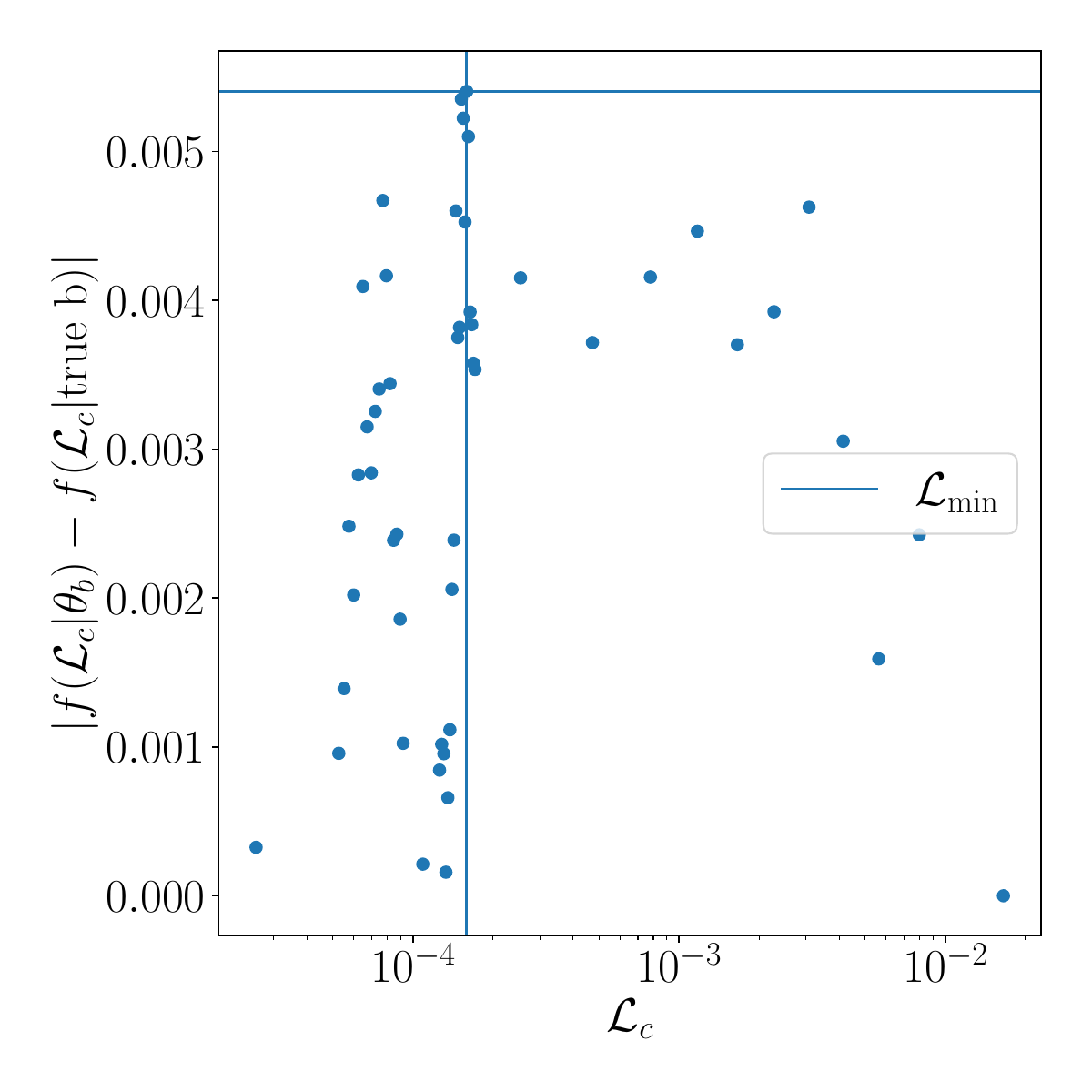}
    \caption{A single run of \texttt{HI-SIGMA+\method}, where $p(\vec{x}|m\in\text{SB})$ has been masked. The figures show, in clockwise order: the $m_{\gamma\gamma}$ distribution for the Signal Region and when applying the Fiducial Signal Region cuts, the signal strength test statistic landscape using four diffferent strategies, the fraction difference distribution as a function of the critical likelihood \likelihood{c} obtained from comparing the toy and Calibration datasets, and the probability distribution of the model likelihood under the background model and the true background. We observe how $\likelihood{\text{min}}$ effectively removes most of the sideband data, providing a fit with a reduced bias at the expense of increased uncertainty.}
    \label{fig:hi_sigma_example}
\end{figure}

In this section, we apply the \method to \texttt{HI-SIGMA}, producing similar results as in \cref{sec:toys}. For this example, we fix $\epsilon=10^{-4}$, which was found to yield reasonable results, without doing a more in-depth study via two-sample test. Additionally, and as a showcase of the usefulness of \method as a diagnostic tool for model bias, we also explore a problem already noticed in Ref.~\cite{Amram:2025vqw} where the authors found it necessary to clip the likelihoods, removing events that are outliers, and to include a sideband masking of the cNFs
\begin{equation*}
    p(\vec{x}'|m_{\gamma\gamma}\in\text{SB}) \to \frac{1}{L^d}\,,
\end{equation*}
where $L$ is a cutoff scale and $d$ the dimension of auxiliary features $\vec{x}$. The masking alters the likelihood landscape, and provides better results. The reason for this can be explored using the \method method and we show in Figs.~\ref{fig:hi_sigma_example_no_mask},\ref{fig:hi_sigma_example} the results for a particular run of \texttt{HI-SIGMA} without and with masking\footnote{For these examples, the uncertainties are estimated using \texttt{MINOS} routine in \texttt{iminuit} since the Hessian approximation fails.}.  We list in both cases the true number of background $B$ and signal events $S$ in the sample instead of the expected rates $\lambda_b,\lambda_s$ since we are not sampling them explicitly from a Poisson distribution.

Figs.~\ref{fig:hi_sigma_example_no_mask},\ref{fig:hi_sigma_example} show four fits to the data: using only the $m_{\gamma\gamma}$ distribution, using \texttt{HI-SIGMA} on the complete Signal Region, using \texttt{HI-SIGMA} on the Fiducial Signal Region, and using \texttt{HI-SIGMA} on a reduced Signal Region using a random subset of the Signal Region of the same size as the Fiducial Signal Region. Although the Fiducial Signal Region is defined on the five dimensional feature space, we find it useful to showcase how it looks on the $m_{\gamma\gamma}$ distribution in particular, where we can interpret the resulting cuts easily.\footnote{A similar analysis could be done using a classifier score, as in Ref.~\cite{Amram:2025vqw}.} 

We observe that the inference with no masking shows a similar distribution of fraction differences as the toy examples. However, the resulting cut is too conservative and actually captures the inverse crossing than intended, where we go from overestimating to underestimating the background. This, along with the signal being excluded from the cut, results in an overly conservative analysis where no signal can be found and with uncertainties on the signal strength that are larger than those of the mass-only fit. If we introduce the masking, the fraction differences see two flat regions appear, corresponding to both sides of the sideband. The resulting cut, however, is less strict and retains a larger fraction of the signal strength since it effectively removes part of the sideband from the fit (although events in the central region are removed as well). The resulting uncertainties in the signal strength are smaller than the mass only fit (but still larger than the standard \texttt{HI-SIGMA}). In both cases, we also show a ``Random cut'' which consists of a fit with restricted statistics, where we select a subset of the data with the same size as the FSR, but chosen at random. This aims to highlight how \method selects a non-trivial subset of data. 

The performance of \method with and without masking shows as well how masking, which was introduced original in Ref.~\cite{Amram:2025vqw} to reduce the impact of mismodelling, is actually a source of bias. This indicates that in order to take advantage of the sidebands better models should be developped. A more nuanced version of \method, which does not completely throw away events, or is at least aware of how nuisance parameters could reduce the disagreement between model and truth, would also avoid the excessive throwing away of events. We leave such developments for future work.

\begin{figure}[h!]
    \centering
    \includegraphics[width=0.45\linewidth]{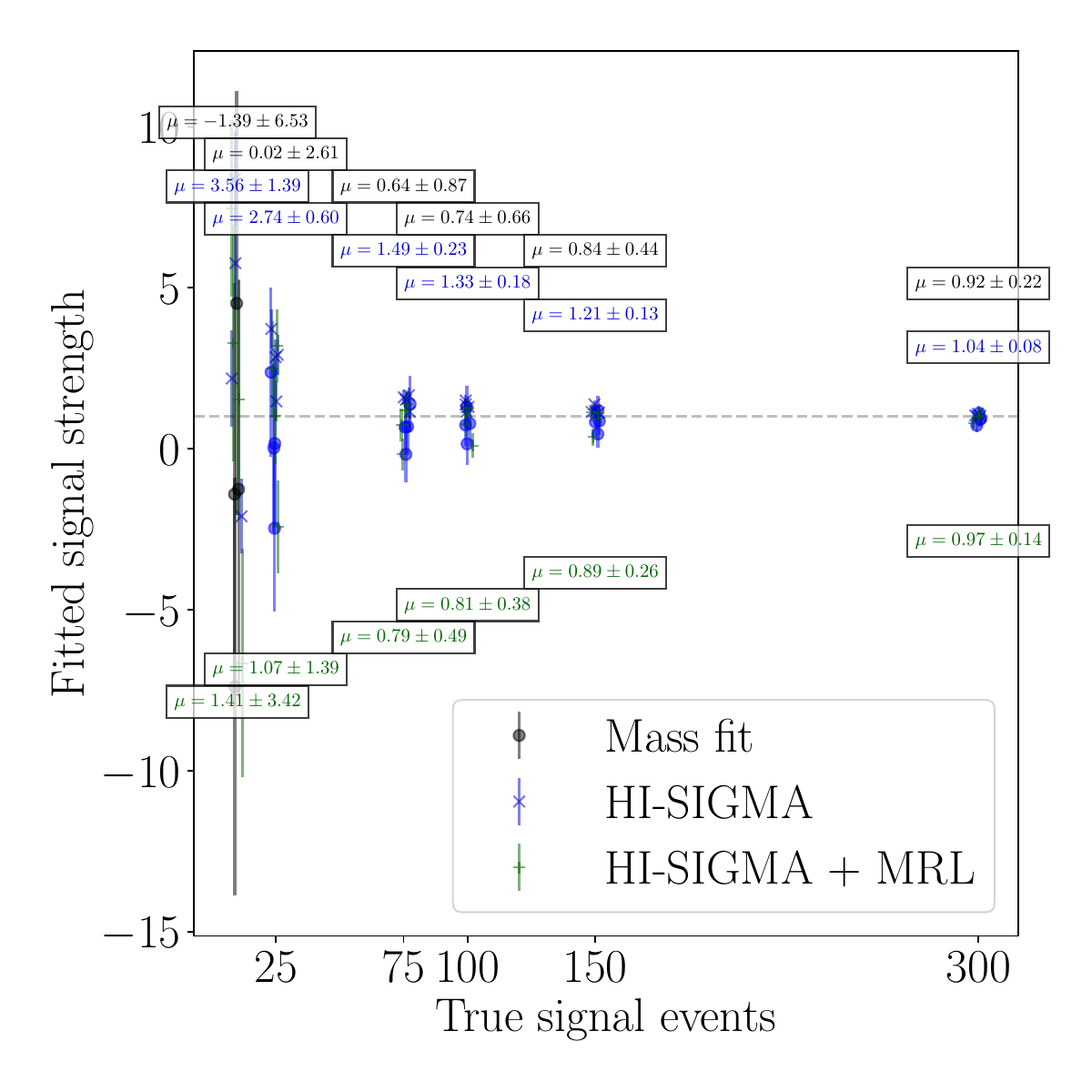}
    \includegraphics[width=0.45\linewidth]{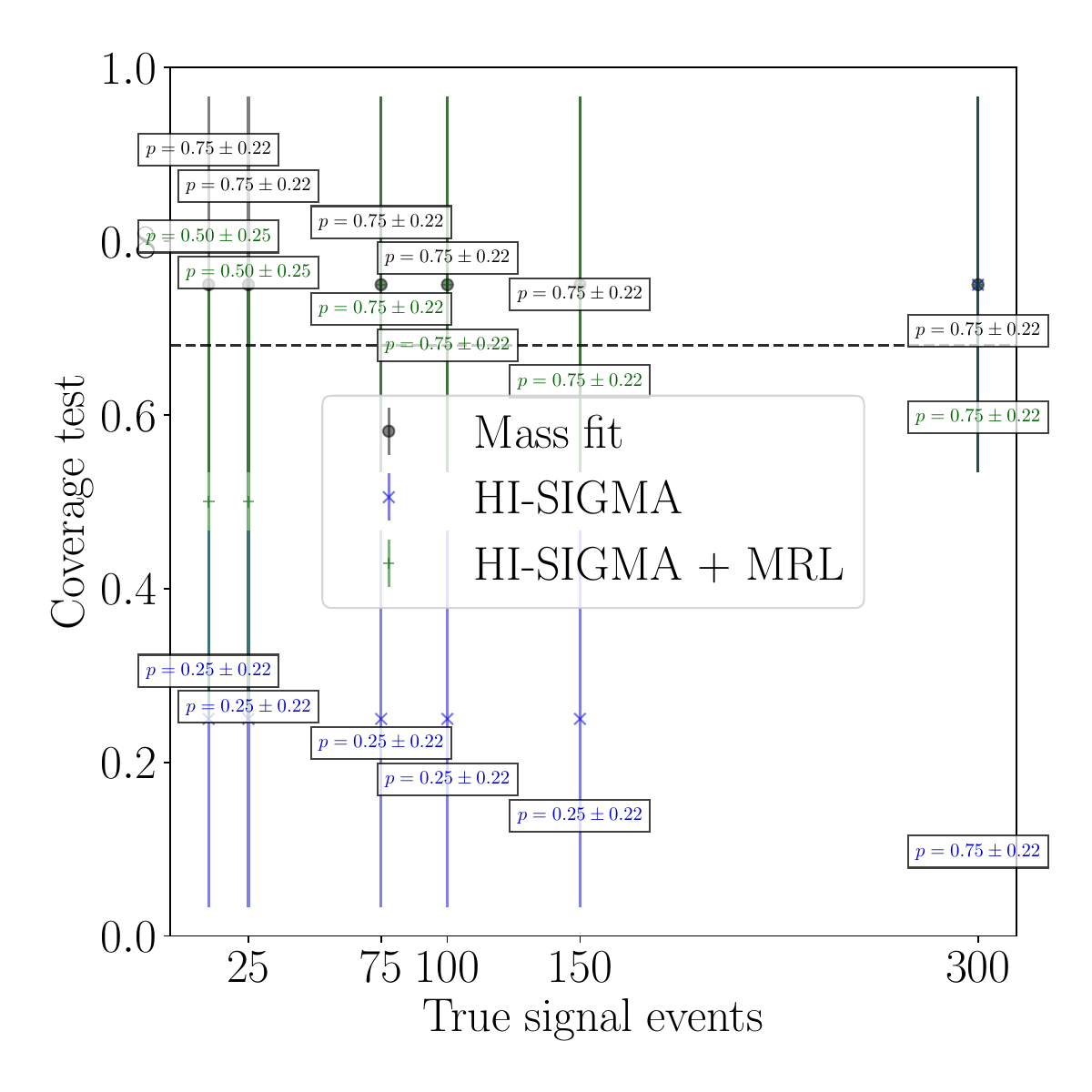}
    \caption{4 pseudo-experiment runs for the di-Higgs dataset, with sideband masking.  Left:  Maximum Likelihood Estimates of the signal strength as a function of the true expected signal events, with uncertainties. Right: Coverage derived from the confidence interval as a function of the true expected signal events. 
    We observe how \texttt{HI-SIGMA}+\method produces signal strength estimates with reduced bias and improved coverage.}
    \label{fig:hi_sigma_scan}
\end{figure}

We highlight as well that the added complexity of the model when compared to the toy examples render the estimated $\likelihood{\text{min}}$ noisier, even for similar dataset sizes. This motivates the consideration of larger toy samples and Calibration Regions, if available. Nevertheless, even if the statistics of the datasets and the quality of the cNFs can be improved substantially, this result already shows the power of the method as a diagnostic and selection tool. This is reinforced by the pseudo-experiments shown in Fig.~\ref{fig:hi_sigma_scan}. The statistics are again fairly small (we only have enough data for four pseudo-experiments, and the toys + Calibration Regions are of the same size as the data itself), but we observe how \method improves the quality of the estimated signal strength and its coverage, although it remains hard to assess the performance.

Even with low statistics, however, we observe the behavior that motivates the introduction of \method. For high enough signal injections, the impact of \method is minimal, while for low enough signal injections, \method reduces the impact of the bias substantially. In this case, and even with masking, the application of the \method method results in underestimation of the signal strength due to a large enough exclusion of signal samples from the FSR.

\section{Outlook}
\label{sec:outlook}

In this work, we have introduced a method, termed the Minimum Resolution Likelihood, to define a Fiducial Signal Region for unbinned, high-dimensional analyses where the background model bias may otherwise result in signal over-estimation. We have shown the necessary conditions for \method to be applicable and useful, and exemplified it using both toy examples and a more realistic analysis based on \texttt{HI-SIGMA} applied to di-Higgs searches.  In all cases, we have found that the Fiducial Signal Region turns at least part of the systematic effects into statistical uncertainties and that the resulting signal strength estimation has a reduced bias and improved coverage, as desired.

The method as presented here is conservative by definition, and thus shows many areas for improvement. One could be to avoid the hard cuts of a Fiducial Signal Region definition and instead improve on the masking introduced in \texttt{HI-SIGMA} by combining it with the estimation of the remaining likelihood volume below the threshold. Additionally, one should further test this method in realistic examples with high statistics.  All of these tests could also explore alternative conditions of disagreement between toy samples and the Calibration Region, or even a different definitions of $\likelihood{\text{min}}$ for \cref{eq:condition}. Although motivated and shown to be succesful, comparing the survival functions is by no means the only possible choice.

\method is envisioned to work generally, provided the set of conditions enumerated in this work are met. For example, another avenue to explore is its potential utility for Anomaly Detection, where data-driven background techniques are necessary almost by definition and introduce hard to quantify model biases. 

Beyond the realm of data-driven background models, one could still use \method as a diagnostic when background emulators are considered, such as ML-models for Simulation Based Inference analyses~\cite{Brehmer:2018kdj, Brehmer:2018eca, Cranmer_sbi_review,Ghosh_jrjc_proc, GomezAmbrosio:2022mpm, Bahl:2021dnc, Barrue:2023ysk, Schofbeck:2024zjo, Chai:2024zyl, nsbi_dihiggs, Benato:2025rgo}. Although we expect background bias effects to be subleading given the careful calibration of the models in realistic applications, as in Refs.~\cite{ATLAS_SBI_methods, ATLAS_SBI_measurement}, \method can be useful as a diagnostic, given its malleability and compatibility with the treatment of other systematic uncertainties. In particular, the scaling of $\likelihood{\text{min}}$ with training dataset size can help characterize the scaling laws of the relevant generative models, including foundation models~\cite{Amram:2026zzv}.

\section*{Code availability}
Public code can be found in
\href{https://github.com/ManuelSzewc/MinimumResolutionLikelihood/tree/main}{Github}. The toy examples can be run in a self-contained manner, while the \texttt{HI-SIGMA} example requires linking to the \href{https://github.com/OzAmram/HI-SIGMA/tree/main}{\texttt{HI-SIGMA} repository} and downloading the datasets from \href{https://doi.org/10.5281/zenodo.15587841}{Zenodo}.

\section*{Acknowledgments}
The author would like to thank Ezequiel Alvarez, Oz Amram, Prasanth Shyamsundar and Nicholas Smith for their detailed feedback and helpful suggestions. 
The author also expresses his gratitude to the public universities and the state research organizations of Argentina for their enduring commitment in the face of on-going challenges.

\appendix

\section{A hand-wavy argument to understand \method}
\label{app:eft}

A particular example of a model for an analysis is an effective field theory parameterization of the data, defined at a given energy scale $\Lambda$. Intuitively, one knows not to trust events with energy $E\gtrsim \Lambda$ to be well-described by the model. Since the likelihood can be described in terms of an energy density, $\mathcal{L}= \frac{g(E) e^{-\beta E}}{\int dE\,g(E) e^{-\beta E}}$, $\Lambda$ defines an energy scale but also a likelihood scale under a given model. We are simply redefining our model by setting a maximum energy such that 
\begin{align}
    \mathcal{L}'&= \frac{g(E)e^{-\beta E}\Theta(\Lambda - E)}{\int_{0}^{\Lambda}dE\,g(E)e^{-\beta E}}\,,\nonumber\\
    &=\mathcal{L}\Theta(\Lambda -E)\frac{\int_{0}^{\infty}dE\,g(E)e^{-\beta E}}{\int_{0}^{\Lambda}dE\,g(E)e^{-\beta E}}\,,\nonumber\\
    &=\mathcal{L}\frac{\Theta(\Lambda -E)}{\epsilon(\likelihood{c})}\nonumber\,.
\end{align}

\section{Interpreting $\likelihood{\text{min}}$}
\label{app:derivation}
To interpret \cref{eq:condition}, we can study its maximum by finding the roots of its derivative

\begin{align}
    \frac{d|\Delta(t)|}{dt}&= \text{sign}(\Delta(t))\frac{d \Delta(t)}{dt}\nonumber\\
    &=\text{sign}(\Delta(t))\frac{d}{dt}[\int dx \, \Theta(p_{b}(x|\theta_{b})-\likelihood{c})(p_{b}(x|\theta_{b})-p_{\text{true b}}(x)) ]\nonumber\\
    &=-\text{sign}(\Delta(t))\int dx \,\delta(p_{b}(x|\theta_{b})-\likelihood{c})(p_{b}(x|\theta_{b})-p_{\text{true b}}(x))\nonumber\\
    &=-\text{sign}(\Delta(t))\likelihood{c}(\sum_{i}\int_{S_i}d\sigma_{i}(x) \frac{(1-\frac{p_{\text{true b}}(x)}{\likelihood{c}})}{|\nabla p_{b}(x|\theta_{b})|})\,,
\end{align}
where $S_{i}$ are all hyper-surfaces over which $p_{b}(x|\theta_b)=\likelihood{c}$. There is a trivial solution to this equation, which  is $\Delta(t) = 0$ and that implies equal contained volumes for all $\likelihood{c}$ (including the strongest claim, that $p_{b}(x|\theta_{b}) = p_{\text{true b}}(x)$ for all $x$ such that $p_{b}(x|\theta_{b}) \geq \likelihood{c} $). However, since $|\Delta(t))|\geq 0$, this is not the solution we are interested in, as it corresponds to $\likelihood{\text{min}}\in\{0,\likelihood{\max}\}$.  The other term, which averages over all hypersurfaces where $p_{b}(x|\theta_{b}) = \likelihood{c} $, is null either if $p_{b}(x|\theta_{b}) = p_{\text{true b}}(x)$ for each hypersurface or if $\likelihood{\text{min}}$ is the average likelihood given by the implicit equation

\begin{align}
    \likelihood{\text{min}}&=\frac{\sum_{i}\int_{S_i} \frac{d\sigma_{i}(x)\, p_{\text{true b}}(x)}{|\nabla p(x|\theta_{b})|}}{\sum_{i}\int_{S_i} \frac{d\sigma_{i}(x)}{|\nabla p(x|\theta_{b})|}}\nonumber\\
    &=\sum_{i}\int_{S_i} \frac{1}{\sum_{k}\int_{S_k} \frac{d\sigma_{k}(x)}{|\nabla p(x|\theta_{b})|}}\frac{d\sigma_{i}(x)\, p_{\text{true b}}(x)}{|\nabla p(x|\theta_{b})|}\nonumber\\
    &\approx\sum_{i}w_{i}\int_{S} \,du \, p_{\text{true b}}(x=f_i(u))\nonumber\\
    &= \mathbb{E}_{u,i}[p_{\text{true b}}(x=f_{i}(u))]\,.
\end{align}

In other words, $\likelihood{\text{min}}$ finds the crossing between the regions where the background model underestimates and overestimates the likelihood surfaces. We emphasize that this is a weaker condition than ensuring point-by-point equality between truth and model since we are comparing the integrated likelihoods, a procedure which smooths some differences away. Moreover, this statement depends on the specific parameterization of the feature space, and thus we assume a domain-expert defined choice has been made. 

\bibliography{biblio}

\begin{thebibliography}{10}
\providecommand{\url}[1]{\texttt{#1}}
\providecommand{\urlprefix}{URL }
\expandafter\ifx\csname urlstyle\endcsname\relax
  \providecommand{\doi}[1]{doi:\discretionary{}{}{}#1}\else
  \providecommand{\doi}{doi:\discretionary{}{}{}\begingroup
  \urlstyle{rm}\Url}\fi
\providecommand{\eprint}[2][]{\url{#2}}

\bibitem{Behnke:2013pga}
O.~Behnke, K.~Kr{\"o}ninger, T.~Sch{\"o}rner-Sadenius and G.~Schott, eds.,
\newblock \emph{{Data analysis in high energy physics}: {A practical guide to
  statistical methods}},
\newblock Wiley-VCH, Weinheim, Germany,
\newblock ISBN 978-3-527-41058-3, 978-3-527-65344-7, 978-3-527-65343-0 (2013).

\bibitem{ATLAS:2018rnh}
M.~Aaboud \emph{et~al.},
\newblock \emph{{Search for pair production of Higgs bosons in the
  $b\bar{b}b\bar{b}$ final state using proton-proton collisions at $\sqrt{s} =
  13$ TeV with the ATLAS detector}},
\newblock JHEP \textbf{01}, 030 (2019),
\newblock \doi{10.1007/JHEP01(2019)030},
\newblock \eprint{1804.06174}.

\bibitem{CMS:2024tdk}
A.~Hayrapetyan \emph{et~al.},
\newblock \emph{{Search for ${\text {Z}{}{}} {\text {Z}{}{}} $ and ${\text
  {Z}{}{}} {\text {H}{}{}} $ production in the ${\text {b}{}{}} {\bar{{\text
  {b}{}{}}}{}{}} {\text {b}{}{}} {\bar{{\text {b}{}{}}}{}{}} $ final state
  using proton-proton collisions at $\sqrt{s}=13\,\text
  {Te}\hspace{-.08em}\text {V} $}},
\newblock Eur. Phys. J. C \textbf{84}(7), 712 (2024),
\newblock \doi{10.1140/epjc/s10052-024-13021-z},
\newblock \eprint{2403.20241}.

\bibitem{CMS:2026znu}
A.~Hayrapetyan \emph{et~al.},
\newblock \emph{{Improved results on Higgs boson pair production in the 4b
  final state}}  (2026),
\newblock \eprint{2604.27044}.

\bibitem{Hallin:2021wme}
A.~Hallin, J.~Isaacson, G.~Kasieczka, C.~Krause, B.~Nachman, T.~Quadfasel,
  M.~Schlaffer, D.~Shih and M.~Sommerhalder,
\newblock \emph{{Classifying anomalies through outer density estimation}},
\newblock Phys. Rev. D \textbf{106}(5), 055006 (2022),
\newblock \doi{10.1103/PhysRevD.106.055006},
\newblock \eprint{2109.00546}.

\bibitem{Raine:2022hht}
J.~A. Raine, S.~Klein, D.~Sengupta and T.~Golling,
\newblock \emph{{CURTAINs for your Sliding Window: Constructing Unobserved
  Regions by Transforming Adjacent Intervals}},
\newblock Front.Big Data \textbf{6}, 899345 (2022),
\newblock \doi{10.3389/fdata.2023.899345},
\newblock \eprint{2203.09470}.

\bibitem{Golling:2022nkl}
T.~Golling, S.~Klein, R.~Mastandrea and B.~Nachman,
\newblock \emph{{Flow-enhanced transportation for anomaly detection}},
\newblock Phys. Rev. D \textbf{107}(9), 096025 (2023),
\newblock \doi{10.1103/PhysRevD.107.096025},
\newblock \eprint{2212.11285}.

\bibitem{Hallin:2022eoq}
A.~Hallin, G.~Kasieczka, T.~Quadfasel, D.~Shih and M.~Sommerhalder,
\newblock \emph{{Resonant anomaly detection without background sculpting}},
\newblock Phys. Rev. D \textbf{107}(11), 114012 (2023),
\newblock \doi{10.1103/PhysRevD.107.114012},
\newblock \eprint{2210.14924}.

\bibitem{Sengupta:2023xqy}
D.~Sengupta, S.~Klein, J.~A. Raine and T.~Golling,
\newblock \emph{{CURTAINs flows for flows: Constructing unobserved regions with
  maximum likelihood estimation}},
\newblock SciPost Phys. \textbf{17}(2), 046 (2024),
\newblock \doi{10.21468/SciPostPhys.17.2.046},
\newblock \eprint{2305.04646}.

\bibitem{Das:2023bcj}
R.~Das, G.~Kasieczka and D.~Shih,
\newblock \emph{{Residual ANODE}}  (2023),
\newblock \eprint{2312.11629}.

\bibitem{ATLAS:2020ocz}
\emph{{Recommendations for the Modeling of Smooth Backgrounds}},
\newblock Tech. rep., CERN, Geneva,
\newblock All figures including auxiliary figures are available at
  https://atlas.web.cern.ch/Atlas/GROUPS/PHYSICS/PUBNOTES/ATL-PHYS-PUB-2020-028
  (2020).

\bibitem{Dauncey:2014xga}
P.~D. Dauncey, M.~Kenzie, N.~Wardle and G.~J. Davies,
\newblock \emph{{Handling uncertainties in background shapes}: {the discrete
  profiling method}},
\newblock JINST \textbf{10}(04), P04015 (2015),
\newblock \doi{10.1088/1748-0221/10/04/P04015},
\newblock \eprint{1408.6865}.

\bibitem{Haussmann:2026gbi}
M.~Hau{\ss}mann, R.~Winterhalder and M.~Ubiali,
\newblock \emph{{Uncertainty in Physics and AI: Taxonomy, Quantification, and
  Validation}}  (2026),
\newblock \eprint{2605.10378}.

\bibitem{Brehmer:2018kdj}
J.~Brehmer, K.~Cranmer, G.~Louppe and J.~Pavez,
\newblock \emph{{Constraining Effective Field Theories with Machine Learning}},
\newblock Phys. Rev. Lett. \textbf{121}(11), 111801 (2018),
\newblock \doi{10.1103/PhysRevLett.121.111801},
\newblock \eprint{1805.00013}.

\bibitem{Brehmer:2018eca}
J.~Brehmer, K.~Cranmer, G.~Louppe and J.~Pavez,
\newblock \emph{{A Guide to Constraining Effective Field Theories with Machine
  Learning}},
\newblock Phys. Rev. D \textbf{98}(5), 052004 (2018),
\newblock \doi{10.1103/PhysRevD.98.052004},
\newblock \eprint{1805.00020}.

\bibitem{Cranmer_sbi_review}
K.~Cranmer, J.~Brehmer and G.~Louppe,
\newblock \emph{The frontier of simulation-based inference},
\newblock Proceedings of the National Academy of Sciences \textbf{117}(48),
  30055 (2020),
\newblock \doi{10.1073/pnas.1912789117},
\newblock \eprint{https://www.pnas.org/doi/pdf/10.1073/pnas.1912789117}.

\bibitem{Ghosh_jrjc_proc}
A.~Ghosh,
\newblock \emph{Measuring quantum interference in the off-shell higgs to four
  leptons process with machine learning},
\newblock Proceedings of the Journ{\'e}es de Rencontre des Jeunes Chercheurs
  2019 (JRJC 2019),
\newblock
  \url{https://hal.archives-ouvertes.fr/hal-02971995v3/file/JRJC2019_Proceedings_v3.pdf}
  (2020).

\bibitem{GomezAmbrosio:2022mpm}
R.~Gomez~Ambrosio, J.~ter Hoeve, M.~Madigan, J.~Rojo and V.~Sanz,
\newblock \emph{{Unbinned multivariate observables for global SMEFT analyses
  from machine learning}},
\newblock JHEP \textbf{03}, 033 (2023),
\newblock \doi{10.1007/JHEP03(2023)033},
\newblock \eprint{2211.02058}.

\bibitem{Bahl:2021dnc}
H.~Bahl and S.~Brass,
\newblock \emph{{Constraining $ \mathcal{CP} $-violation in the Higgs-top-quark
  interaction using machine-learning-based inference}},
\newblock JHEP \textbf{03}, 017 (2022),
\newblock \doi{10.1007/JHEP03(2022)017},
\newblock \eprint{2110.10177}.

\bibitem{Barrue:2023ysk}
R.~Barru\'e, P.~Conde-Mu\'\i{}\~no, V.~Dao and R.~Santos,
\newblock \emph{{Simulation-based inference in the search for CP violation in
  leptonic WH production}},
\newblock JHEP \textbf{04}, 014 (2024),
\newblock \doi{10.1007/JHEP04(2024)014},
\newblock \eprint{2308.02882}.

\bibitem{Schofbeck:2024zjo}
R.~Sch\"ofbeck,
\newblock \emph{{Refinable modeling for unbinned SMEFT analyses}},
\newblock Mach. Learn. Sci. Tech. \textbf{6}(1), 015007 (2025),
\newblock \doi{10.1088/2632-2153/ad9fd1},
\newblock \eprint{2406.19076}.

\bibitem{Chai:2024zyl}
S.~Chai, J.~Gu and L.~Li,
\newblock \emph{{From optimal observables to machine learning: an
  effective-field-theory analysis of e$^{+}$e$^{-}$\textrightarrow{}
  W$^{+}$W$^{-}$ at future lepton colliders}},
\newblock JHEP \textbf{05}, 292 (2024),
\newblock \doi{10.1007/JHEP05(2024)292},
\newblock \eprint{2401.02474}.

\bibitem{nsbi_dihiggs}
R.~Mastandrea, B.~Nachman and T.~Plehn,
\newblock \emph{{Constraining the Higgs potential with neural simulation-based
  inference for di-Higgs production}},
\newblock Phys. Rev. D \textbf{110}(5), 056004 (2024),
\newblock \doi{10.1103/PhysRevD.110.056004},
\newblock \eprint{2405.15847}.

\bibitem{Benato:2025rgo}
L.~Benato, C.~Giordano, C.~Krause, A.~Li, R.~Sch\"ofbeck, D.~Schwarz,
  M.~Shooshtari and D.~Wang,
\newblock \emph{{Unbinned inclusive cross-section measurements with
  machine-learned systematic uncertainties}}  (2025),
\newblock \eprint{2505.05544}.

\bibitem{Diefenbacher:2026kki}
S.~Diefenbacher, S.~Palacios~Schweitzer and G.~Kasieczka,
\newblock \emph{{Generative Models and Statistical Validation}}  (2026),
\newblock \eprint{2605.30453}.

\bibitem{Amram:2025vqw}
O.~Amram and M.~Szewc,
\newblock \emph{{Data-driven high-dimensional statistical inference with
  generative models}},
\newblock JHEP \textbf{11}, 129 (2025),
\newblock \doi{10.1007/JHEP11(2025)129},
\newblock \eprint{2506.06438}.

\bibitem{Cranmer:2014lly}
K.~Cranmer,
\newblock \emph{{Practical Statistics for the LHC}},
\newblock In \emph{{2011 European School of High-Energy Physics}}, pp.
  267--308,
\newblock \doi{10.5170/CERN-2014-003.267} (2014), \eprint{1503.07622}.

\bibitem{Buchner_2023}
J.~Buchner,
\newblock \emph{Nested sampling methods},
\newblock Statistics Surveys \textbf{17}(none) (2023),
\newblock \doi{10.1214/23-ss144}.

\bibitem{Read:1999kh}
A.~L. Read,
\newblock \emph{{Linear interpolation of histograms}},
\newblock Nucl. Instrum. Meth. A \textbf{425}, 357 (1999),
\newblock \doi{10.1016/S0168-9002(98)01347-3}.

\bibitem{cranmer:2012sba}
K.~Cranmer, G.~Lewis, L.~Moneta, A.~Shibata and W.~Verkerke,
\newblock \emph{{HistFactory: A tool for creating statistical models for use
  with RooFit and RooStats}}  (2012).

\bibitem{iminuit}
H.~Dembinski and P.~O. et~al.,
\newblock \emph{scikit-hep/iminuit}  (2020),
\newblock \doi{10.5281/zenodo.3949207}.

\bibitem{Dawson:2022zbb}
S.~Dawson \emph{et~al.},
\newblock \emph{{Report of the Topical Group on Higgs Physics for Snowmass
  2021: The Case for Precision Higgs Physics}},
\newblock In \emph{{Snowmass 2021}} (2022), \eprint{2209.07510}.

\bibitem{Alwall:2014hca}
J.~Alwall, R.~Frederix, S.~Frixione, V.~Hirschi, F.~Maltoni, O.~Mattelaer,
  H.~S. Shao, T.~Stelzer, P.~Torrielli and M.~Zaro,
\newblock \emph{{The automated computation of tree-level and next-to-leading
  order differential cross sections, and their matching to parton shower
  simulations}},
\newblock JHEP \textbf{07}, 079 (2014),
\newblock \doi{10.1007/JHEP07(2014)079},
\newblock \eprint{1405.0301}.

\bibitem{Sjostrand:2014zea}
T.~Sj\"ostrand, S.~Ask, J.~R. Christiansen, R.~Corke, N.~Desai, P.~Ilten,
  S.~Mrenna, S.~Prestel, C.~O. Rasmussen and P.~Z. Skands,
\newblock \emph{{An introduction to PYTHIA 8.2}},
\newblock Comput. Phys. Commun. \textbf{191}, 159 (2015),
\newblock \doi{10.1016/j.cpc.2015.01.024},
\newblock \eprint{1410.3012}.

\bibitem{Bierlich:2022pfr}
C.~Bierlich \emph{et~al.},
\newblock \emph{{A comprehensive guide to the physics and usage of PYTHIA
  8.3}},
\newblock SciPost Phys. Codeb. \textbf{2022}, 8 (2022),
\newblock \doi{10.21468/SciPostPhysCodeb.8},
\newblock \eprint{2203.11601}.

\bibitem{deFavereau:2013fsa}
J.~de~Favereau, C.~Delaere, P.~Demin, A.~Giammanco, V.~Lema\^\i{}tre,
  A.~Mertens and M.~Selvaggi,
\newblock \emph{{DELPHES 3, A modular framework for fast simulation of a
  generic collider experiment}},
\newblock JHEP \textbf{02}, 057 (2014),
\newblock \doi{10.1007/JHEP02(2014)057},
\newblock \eprint{1307.6346}.

\bibitem{Oreglia:1980cs}
M.~J. Oreglia,
\newblock \emph{{A study of the reactions $\psi^\prime \to \gamma \gamma
  \psi$}},
\newblock Ph.D. thesis, Stanford University,
\newblock SLAC-R-236 (1980).

\bibitem{Gaiser:1982yw}
J.~E. Gaiser,
\newblock \emph{{Charmonium Spectroscopy From Radiative Decays of the $J/\psi$
  and $\psi^\prime$}},
\newblock Ph.D. thesis, Stanford University,
\newblock SLAC-R-255 (1982).

\bibitem{ATLAS_SBI_methods}
G.~Aad \emph{et~al.},
\newblock \emph{{An implementation of neural simulation-based inference for
  parameter estimation in ATLAS}}  (2024),
\newblock \eprint{2412.01600}.

\bibitem{ATLAS_SBI_measurement}
G.~Aad \emph{et~al.},
\newblock \emph{{Measurement of off-shell Higgs boson production in the
  $H^*\rightarrow ZZ\rightarrow 4\ell$ decay channel using a neural
  simulation-based inference technique in 13\,TeV pp collisions with the ATLAS
  detector}},
\newblock Rept. Prog. Phys. \textbf{88}(5), 057803 (2025),
\newblock \doi{10.1088/1361-6633/adcd9a},
\newblock \eprint{2412.01548}.

\bibitem{Amram:2026zzv}
O.~Amram, D.~A. Faroughy, T.~Gerdes, A.~Hallin, G.~Kasieczka, M.~Kr{\"a}mer,
  H.~Reyes-Gonzalez and D.~Shih,
\newblock \emph{{Neural Scaling Laws for Jet Generation}}  (2026),
\newblock \eprint{2605.28940}.

\end{thebibliography}

\end{document}